\documentclass[reprint,
 amsmath,amssymb,
aps, prd,
 longbibliography,
 superscriptaddress,
 nofootinbib
]{revtex4-2}

\usepackage{macros}
\usepackage{graphicx}
\usepackage{dcolumn}
\usepackage{bm}
\usepackage{hyperref}
\usepackage{color,units}
\usepackage{footmisc}
\usepackage[dvipsnames]{xcolor} 

\begin{document}

\title{Directly inferring cosmology and the neutron-star equation of state from gravitational-wave mergers}

\author{Spencer J. Magnall}
\thanks{Electronic Address: spencer.magnall@monash.edu}
\affiliation{School of Physics and Astronomy, Monash University, VIC 3800, Australia}
\affiliation{OzGrav: The ARC Centre of Excellence for Gravitational-Wave Discovery, Clayton, VIC 3800, Australia}
\author{Simon R. Goode}
\affiliation{School of Physics and Astronomy, Monash University, VIC 3800, Australia}
\affiliation{OzGrav: The ARC Centre of Excellence for Gravitational-Wave Discovery, Clayton, VIC 3800, Australia}
\author{Nikhil Sarin}
\affiliation{Oskar Klein Centre for Cosmoparticle Physics, Department of Physics,
Stockholm University, AlbaNova, Stockholm SE-106 91, Sweden}
\affiliation{Nordita, Stockholm University and KTH Royal Institute of Technology \\
Hannes Alfvéns väg 12, SE-106 91 Stockholm, Sweden}
\author{Paul D. Lasky}
\affiliation{School of Physics and Astronomy, Monash University, VIC 3800, Australia}
\affiliation{OzGrav: The ARC Centre of Excellence for Gravitational-Wave Discovery, Clayton, VIC 3800, Australia}

\begin{abstract}
Upgrades to existing gravitational-wave observatories have the potential to simultaneously constrain the nuclear equation of state and Hubble's constant $H_0$ to percent level with merging neutron star binaries. 
In practice, performing simultaneous inference of $H_0$ and the equation of state is limited computationally by the requirement to solve the equations of general-relativistic hydrostatic equilibrium millions of times. 
We develop a machine-learning model to solve the Tolman-Oppenheimer-Volkoff equations in less than a millisecond, and demonstrate its utility by performing direct inference of both equation of state and Hubble's constant for synthetic neutron star merger signals with LIGO-Virgo-KAGRA operating at A+ 
sensitivities.
We show that a population of fifteen mergers observed with A+ allows for the radius of a $1.4\,M_{\odot}$ neutron star and $H_0$ to be constrained to $R_{1.4} = 11.74^{+0.35}_{-0.28}$ km
and $H_0 = \unit[68^{+17}_{-13}]{km\,s^{-1}\,Mpc^{-1}}$, at 90\% credible interval and 68\% credible interval respectively. 
These constraints utilise only the gravitational-wave information to infer cosmological parameters; such numbers will be further improved with the addition of electromagnetic counterparts and/or galaxy catalogues.
\end{abstract}

\maketitle

\section{Introduction} \label{sec:intro}
Binary neutron star mergers are the site of both extreme violence and rich astrophysics. The simultaneous detection of gravitational waves and an electromagnetic counterpart to the event GW170817 \citep{GW170817_gw,GW170817_em,GW170817_grb,2017FERMI,2017INTEGRALGRB,2017SSS17a,AT2017gfo_Valenti} provided a wealth of information about gamma-ray burst astrophysics, tests of general relativity, $r$-process nucleosynthesis, and the cold nuclear equation of state at densities far exceeding nuclear saturation~\citep[e.g.,][]{GW170817_eos1,GW170817_eos2}. 

The merger GW170817 also provided a unique opportunity to measure the expansion of the Universe with gravitational waves~\citep{GW170817_bright_cosmo}. Mergers of compact binaries act as ``standard sirens'' analogous to standard candles, albeit completely independent of the cosmic distance ladder~\citep{1986Schutz,Holz_Huges_standard_sirens}. Standard sirens therefore provide complementary measurements of the expansion of the Universe compared to traditional methods.  

Broadly, gravitational-wave events can be split between bright and dark sirens, which have detectable and non-detectable electromagnetic counterparts, respectively. Regardless of this class, one needs to determine the redshift to make cosmological measurements. For example, the identification of the electromagnetic counterpart to GW170817, allowed the redshift of the host galaxy NGC4993 to be used to calculate a value of Hubble's constant $H_0 = 70^{+12}_{-8} \rm{km} \ \rm{s}^{-1} {\rm Mpc}^{-1}$~\citep{GW170817_bright_cosmo}. This constraint was further improved to $H_0 = 68.9^{+4.7}_{-4.6} \ \rm{km} \ \rm{s}^{-1} {\rm Mpc}^{-1}$ using a combination of electromagnetic, radio, and gravitational-wave measurements \citep{2019Hotokezaka}.
Measurements of Hubble's constant are not limited to single events; data from multiple events can be combined to further constrain $H_0$. The most recent estimate from the LIGO-Virgo-KAGRA collaboration~\citep{2015AdvancedLigo,2015AdvancedVirgo,2021KAGRA} combined 47 dark sirens, including information about their potential host galaxies from galaxy catalogues, with cosmological results from GW170817 to yield $H_0 = 68^{+6}_{-8} \ \rm{km} \ \rm{s}^{-1} {\rm Mpc}^{-1}$ \citep{2023GWTC3_cosmo}.

There are other ways to determine the redshift of dark sirens, either by statistically using features in the mass spectrum at a population level~\citep{2019Farr_Fishbach+,2022Ezquiag_Holz_Spectral}, or by utilising equation-of-state information in neutron-star mergers to break the so-called mass-redshift degeneracy~\citep{Messenger&Read2012}. In this Paper, we focus on the latter.

\
Measurements of tidal deformability from neutron star mergers thus far are relatively uninformative, only ruling out the ``stiffest'' equations of state ~\citep{GW170817_eos1,GW170817_eos2,AnzacDay,2021NSBH,GW230529}. As such, it is necessary to combine tidal deformability measurements from multiple neutron star mergers to constrain the equation of state~\cite[e.g.,][]{2013DelPozzo,2015LackeyWade,2015Agathos,2019Hernandez,2022AGolomb_Talbot,Walker2024}.

Since the equation of state is still relatively unknown, most gravitational-wave cosmology forecasts either 1)~assume the equation of state will be well constrained, and then combine $H_0$ posterior distributions~\cite[e.g.,][]{2021Chaterjee}, or 2) determine both the tidal deformability and $H_0$ hierarchically using a parameterised model for the equation of state~\cite[e.g.,][]{2022Ghosh,ghosh2024jointinferencepopulationcosmology}. We introduce methodology to perform simultaneous and direct Bayesian inference on the equation of state and $H_0$ for individual neutron star mergers. This allows for straight-forward hierarchical inference on the equation of state, Hubble's constant, and binary parameters in a way that alleviates the need for complicated likelihood interpolations~\cite[e.g.,][]{2019Hernandez,2022AGolomb_Talbot,Walker2024}. 

Our methodology requires significantly speeding up solutions of the equations of hydrostatic equilibrium---so-called Tolman-Oppenheimer-Volkoff (TOV) equations---in order to calculate the gravitational waveform inside our Bayesian-inference pipeline. To that end, we develop a machine-learning algorithm $\mu$-TOV (pronounced `microtov') that determines the tidal deformability and mass of a neutron star from the input equation-of-state parameters in $\approx\unit[500]{\mu s}$.\footnote{We refer the reader to similar machine-learning models by \cite{Soma2022,2024Brandes,2024McGinn, 2024Tiwari,2024Reed+}} We integrate $\mu$-TOV with the \textsc{bilby} parameter-estimation software~\citep{Bilby_Ashton, Bilby_Romero-Shaw} to perform direct equation-of-state inference on individual binary neutron star events, while simultaneously inferring $H_0$.

This Paper is laid out as follows: we describe and benchmark $\mu$-TOV in Section~\ref{sec:Molotov_NN}. We detail the methodology for performing simultaneous direct inference of the equation of state and Hubble's constant and validate our methodology with a synthetic injection of a binary neutron star event in Section~\ref{sec:Direct Inference}. We study a synthetic population of binary neutron star mergers forecasting constraints on the equation of state and $H_0$ during the $A{+}$
era in Section \ref{sec:popstudy}. We provide concluding remarks in Section~\ref{ref:conclusion}.

\section{\texorpdfstring{$\mu$-TOV solver}{mu-TOV solver}}\label{sec:Molotov_NN}
We have three key requirements for our machine-learning algorithm to solve the TOV equations:
\begin{enumerate}
    \item an error on tidal deformability that is less than the width of the tidal-deformability posterior obtained via parameter estimation,
    \item non-biased estimates of the tidal deformability, and
    \item prediction speed of the same order as a single likelihood evaluation, as to not be a bottleneck for inference. 
\end{enumerate}
The third requirement implies we require an evaluation speed of $\mathcal O(1 \ \rm ms)$, as this is the speed of a single likelihood evaluation for a long-duration binary neutron star merger using reduced-order methods~\citep{Smith+2016}. We stress that our methodology for performing direct inference of the equation of state and $H_0$ is solver independent and can easily be replaced with a more accurate and efficient solver if and when such a solver becomes available. 

\subsection{Piecewise Polytropic Equation of state parameterisation}
For simplicity, we utilise a piecewise polytropic equation of state~\citep{Read2009}, although we note that the equation of state can be better described using a spectral parameterisation \citep{Lindbolm2010} or non-parametric representations~\citep[e.g.,][]{2019Landry,2020Essick,2022Legred,Sarin2023}. The piecewise polytrope paramterisation of \citeauthor{Read2009} is not differentiable at dividing densities, and consequently sound speed is not continuous. Piecewise polytropes are also prone to developing posteriors with unphysical `kinks', and to implicit correlations \citep{Geert2018,Legred2022}. It is also not clear how the parameterisation may handle potential phase transitions and exotic physics of real neutron stars. 
We discuss the impact of this choice of paramaterisation on our synthetic population studies (Section~\ref{sec:popstudy}) further in Appendix~\ref{sec: PP_Bias}. We use the crust of the SLy equation of state \citep{SLY2001,SLY2004} at low densities, which we stitch smoothly to the piecewise polytrope at a density $\rho_0$ between $10^{12}$ and $10^{14.7} \rm g/cm^3$. We adopt a three piece polytropic parameterisation above this stitch point, which has four free parameters: the adiabatic indices $\Gamma_1$, $\Gamma_2$, $\Gamma_3$, and $p_1$ which defines the pressure at which $\Gamma_1$ and $\Gamma_2$ are continuous. Consequently, $p_1$ also determines the polytropic constants $K_i$ via 
\begin{equation}
    K_{i+1} = \frac{p(\rho_i)}{\rho^{\Gamma_{i+1}}_{i}}. 
\end{equation}

\begin{table}
    \renewcommand{\arraystretch}{1.5}
    \resizebox{0.8\linewidth}{!}{
    \begin{tabular}{ccc}
        \toprule
         Parameter & Unit & Range \\
        \hline\
         $\log_{10}(p_1)$  & $\log_{10}(\rm dyne/cm^2)$ & $\rm Uniform(33.5,35.5)$ \\
         $\Gamma_1$ & - &  $\rm Uniform(1.5, 5)$ \\
         $\Gamma_2$ & - &$\rm Uniform(1.5,4)$ \\
         $\Gamma_3$ & - &$\rm Uniform(1.5,4)$ \\
        $M$ & $M_\odot$ & $\rm Uniform(1.0,2.0)$ \\
    \end{tabular}}
    \caption{Ranges and distributions for mass $M$, first transition pressure $p_1$, and adiabatic indices $\Gamma_1$, $\Gamma_2$, $\Gamma_3$ used when constructing training data. Our equation-of-state ranges are informed via polytropic parameterisations \citep{Read2009} of tabulated equations of state. The term `Uniform($a$,$b$)' implies we draw training-data samples from a uniform distribution bounded by $a$ and $b$.}
    \label{tab:priors_trainingset}
\end{table}
\begin{figure*}
    \centering
    \includegraphics[width=\linewidth]{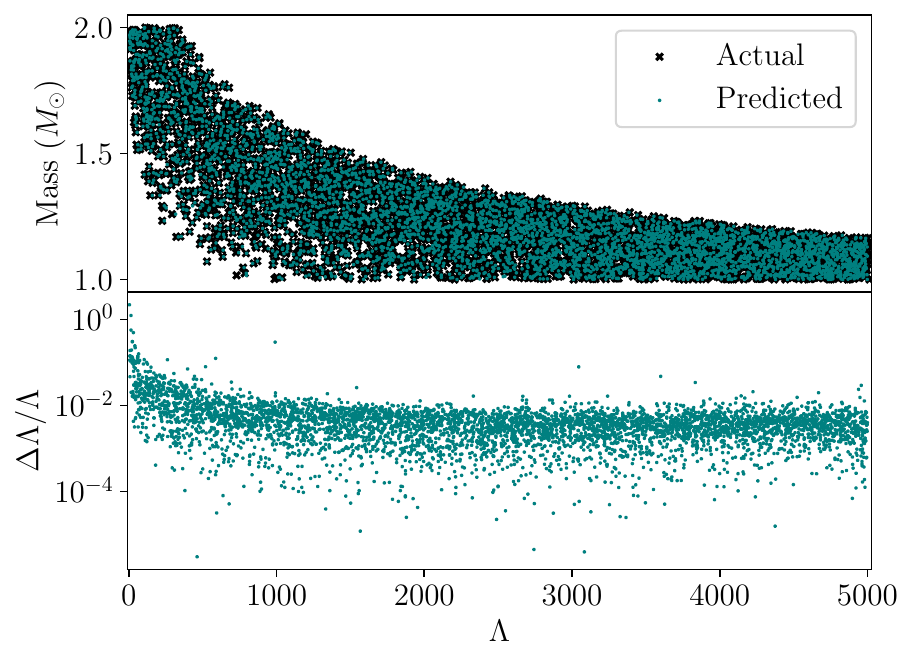}
    \caption{Top panel: a comparison between predicted values (teal dots) of tidal deformability using the $\mu$-TOV model, to the values obtained via numerical integration of the TOV equations (black crosses) for a set of randomly drawn equations of state and masses. Bottom panel: The relative error on tidal deformability for the model as a function of tidal deformability.}
    \label{fig:ML_delta_lambda_model_master}
\end{figure*}
We generate training data by drawing equations of state uniformly from our range in $\Gamma_i$ and $p_1$ (summarised in Table \ref{tab:priors_trainingset}). We enforce a minimum constraint on the maximum mass of a non-rotating neutron star $M_{\rm TOV} > 2.01 M_\odot$ for each equation of state draw, consistent with measurements of PSR J0384+0432 \citep{J0384+0432}. We also discard equation of state draws for which $\Gamma_1$ and $\log_{10}(p_1)$ are incompatible, i.e., where the slope of $\log_{10}(p)$ vs $\log_{10}(\rho)$ is too shallow to intersect the polytrope at the value of $\log_{10}(p_1)$.

For each equation of state draw, we solve the TOV and Love-number equations numerically using the \textsc{lalsimulation}~\citep{lalsuite} software package, to obtain the mass-tidal deformability ($M$ - $\Lambda$) curve. That is, we solve

\begin{equation}  
 \frac{\mathrm{d}m}{\mathrm{d}r} = 4\pi\rho r^2, 
\end{equation}
    
\begin{equation}
        \frac{\mathrm{d} P}{\mathrm{d} r} = -\frac{(\rho +P)(m+4\pi r^3P)}{r(r-2m)},
\end{equation}
\begin{equation}
    \frac{\mathrm{d}\Phi}{\mathrm{d}r} = - \frac{1}{\rho + P} \frac{\mathrm{d} P}{\mathrm{d} r},
\end{equation}
\begin{equation}
    k_2 = k_2(C,y),
\end{equation}
where $m$, $P$ and $\rho$ are the mass, pressure, and density, respectively as functions of radius from the center of the neutron star $r$, and
$\Phi(r)$ is a metric coefficient such that
\begin{equation}
    ds^2 = -e^{-2\Phi(r)}dt^2 + \frac{dr^2}{1-2m(r)/r} + r^2(d\theta^2 + \sin^2\theta d\phi^2).
\end{equation}
Here, $k_2$ is the gravitational Love number, which we compute numerically using the compactness $C\equiv M/R$ 
, and $y$ a function that depends on the value of the metric and its derivative with respect to $r$; for details see~\citet{2008Hinderer}.
After solving these equations, we compute the dimensionless tidal deformability $\Lambda$ via 
\begin{equation}
    \label{eq:dimensionless_tidal}
    \Lambda = \frac{2}{3}k_2 C^{-5}.
\end{equation}
We then randomly draw points on the mass-tidal deformability curve, and add these to our training data. Since the solver is focused on providing accurate estimates of tidal deformability for inference, we restrict our parameter space in mass to be $1.0$ - $2.0 M_\odot$, which covers the parameter space of neutron star masses expected for most low-spin ($a < 0.05$) binary neutron star mergers. 
We also only consider equations of state with tidal deformabilities in the range $\Lambda = 0$ to 5000, consistent with the typical priors employed in parameter estimation for gravitational-wave data analysis~\cite[e.g.,][]{GW170817_eos1,GW170817_eos2}. 

\subsection{Neural Network}
We utilise a simple neural network developed and trained using the \textsc{keras} framework as part of the \textsc {tensorflow} software library for machine learning and artificial intelligence~\citep{tensorflow2015-whitepaper}. We detail the network architecture, training and hyper-parameter tuning in Appendix \ref{sec:NN_appendix}. 

Once trained, we quantify the prediction accuracy of our model by generating a set of $\approx 10^6$ new equations of state, masses and resultant tidal deformabilties. We then compare the true tidal deformability to the model prediction.
Figure \ref{fig:ML_delta_lambda_model_master} shows the prediction error of our model on this data set. The top panel, shows a comparison between predicted values of tidal deformability (teal dots), to the values obtained by numerically integrating the TOV equations (black cross) in mass-tidal deformability space. The bottom panel in Figure \ref{fig:ML_delta_lambda_model_master} shows the relative error in tidal deformability prediction,
\begin{equation}
    \label{eq:tidal_rel_error}
    \frac{\Delta \Lambda}{\Lambda} = \frac{\Lambda_{\rm ML} - \Lambda_{\rm TOV}}{\Lambda_{\rm TOV}},
\end{equation}
where $\Lambda_{\rm ML}$ is the predicted tidal deformability from the machine learning model, and $\Lambda_{\rm TOV}$ is the `true' tidal deformability obtained from our TOV solver. We plot the relative error in the model as a function of $\Lambda_{\rm TOV}$. 

We see that for the majority of the parameter space, the machine learning model does an excellent job in predicting the tidal deformability, with mean relative errors of $\Delta \Lambda/\Lambda\leq 1\%$.
We see larger relative errors for smaller values of tidal deformability, however we note this is because the denominator in $\Delta\Lambda/\Lambda$ is larger, implying larger fractional errors. In other words, the relative error $\Delta\Lambda$ is largely constant across the range of parameter space.  

We also note occasional outliers with relatively large ($>10\%$) errors across the entire parameter space. These rare false model estimates are negligible when performing parameter estimation, as the large number of live points in either Markov chain Monte Carlo or nested sampling algorithms de-values single likelihood outliers.
In rare cases, our model predicts an unphysical negative tidal deformability, which would cause errors during parameter estimation as waveform models are not defined. When this happens, we simply set the tidal deformability to zero.

\begin{figure}
    \centering
        \includegraphics[width=\linewidth]{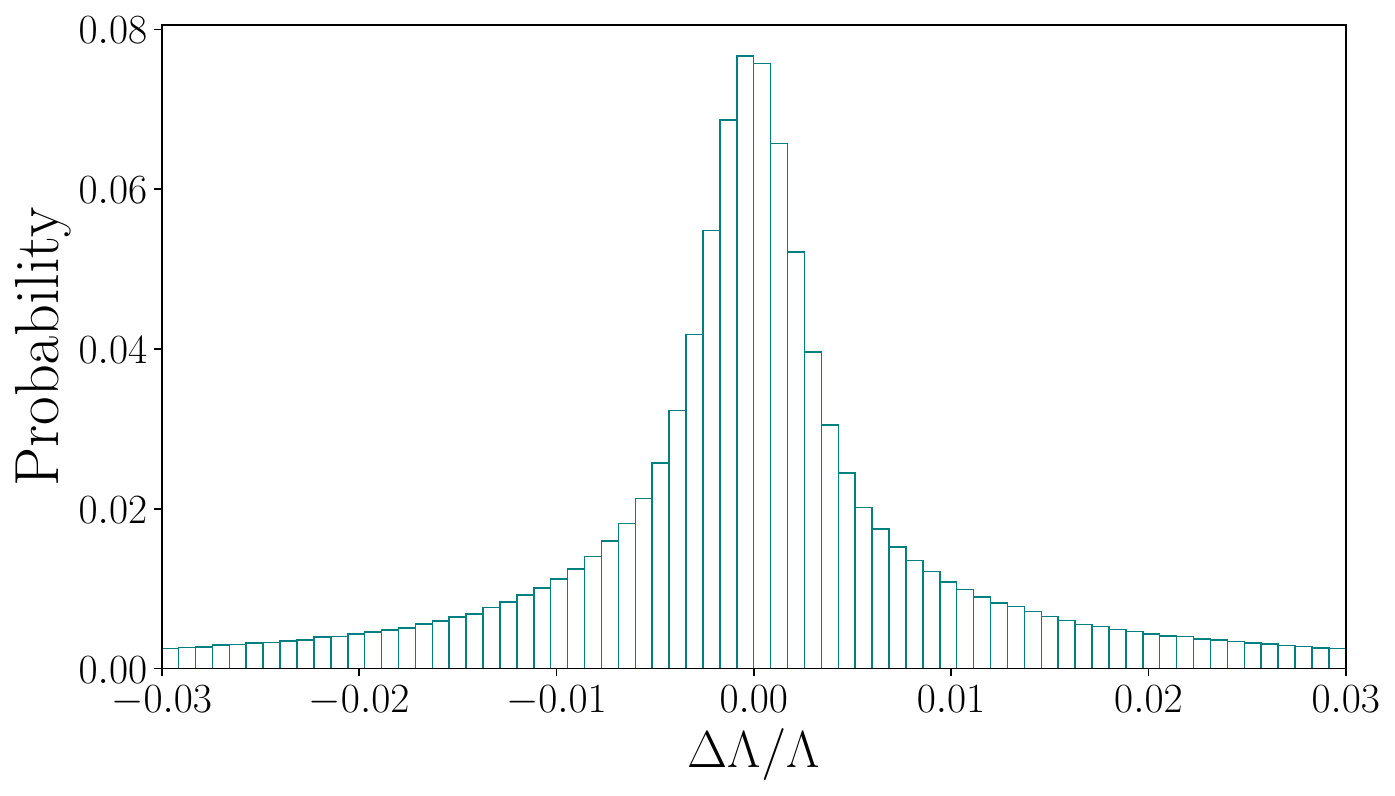}
    \caption{Histogram of relative error in prediction of the $\mu$-TOV model compared to the exact solutions for a random set of equations of state and masses. We see a symmetric distribution with a slight bias towards under-prediction at a  $< 1 \%$ level.}
    \label{fig:delta_lambda_bias}
\end{figure}

We quantify the overall bias of our model by drawing randomly from our set of equations of state and masses. We then calculate the relative error $\Delta \Lambda/\Lambda$ in the prediction of tidal deformability from the machine learning model compared to the TOV solver using Equation \ref{eq:tidal_rel_error} and plot a histogram of the results in Figure~\ref{fig:delta_lambda_bias}.
 
For a non-biased model, we expect this histogram to be symmetric centered at zero. Figure \ref{fig:delta_lambda_bias} indeed shows an almost symmetric distribution with a slight bias towards under-prediction (corresponding to a negative $\Delta \Lambda/\Lambda$) at a $< 1\%$ level. We are satisfied that this bias is small enough to be negligible when performing parameter estimation of binary neutron star events, although we caution that it will need to be considered if we are combining posterior distributions from many tens of events in the future.

A single prediction of tidal deformability with $\mu$-TOV takes approximately $500 \ \rm \mu{\rm s}$, which is negligible when compared to a single likelihood evaluation for parameter estimation of a binary neutron star signal using reduced-order quadrature models in third-generation gravitational wave observatories~\citep[e.g.,][]{Smith+2021}.

Given the above, we are satisfied that $\mu$-TOV satisfies our three requirements for performing parameter estimation set out at the beginning of this Section.

\section{Direct inference}\label{sec:Direct Inference}
\subsection{EOS inference}
We perform Bayesian inference of a binary neutron star signal using the parallel implementation~\citep{pbilby_paper} of the \textsc{bilby}  Bayesian inference library~\citep{Bilby_Ashton,Bilby_Romero-Shaw}. 
Instead of sampling directly over tidal deformability as is usually done in binary neutron star parameter estimation, we sample over the equation-of-state parameters directly. For each likelihood calculation, new equation of state and mass parameters are drawn, and $\mu$-TOV is used to calculate the tidal deformability. The full procedure for calculating the likelihood with $\mu$-TOV for each sample draw in \textsc{bilby} is:
\begin{enumerate}
    \item For each sample, pass the 15 binary neutron star parameters ($m_1$, $m_2$, $D_L$, etc.), equation of state parameters ($\log_{10} (p_1)$, $\Gamma_1$, $\Gamma_2$, $\Gamma_3$), and redshift ($z$) into a custom frequency-domain source model. 
    \item Using the redshift sample, calculate the source-frame masses $m_1$ and $m_2$ via  $m_{\rm source} = {m_z}/1+z$. where $m_z$ is the lab-frame mass.   
    \item Calculate the tidal deformability for each neutron star using $\mu$-TOV. 
    \item Pass the computed tidal deformabilities along with the other 15 binary neutron star parameters into the regular frequency-domain source model to evaluate the waveform, and hence calculate the likelihood. 
\end{enumerate}
We use the standard likelihood used for Bayesian inference of gravitational-wave events; e.g., see Refs.~\cite{veitch2015, Bilby_Romero-Shaw}

\subsection{\texorpdfstring{$H_0$ inference}{H0 inference}} \label{sec:H0}
To infer Hubble's constant directly from the gravitational-wave signal, we require luminosity distance  and redshift. If the neutron star equation of state is well constrained, then the source frame mass $m_{\rm source}$ can be determined directly from the tidal deformability measurement \citep{Messenger&Read2012}. Thus, redshift can be obtained from the definition of detector frame mass
\begin{equation}
    \label{eq:Redshifted mass}
    m_z = (1+z) m_{\rm source}.
\end{equation}
In practice, we do not know the equation of state exactly and instead sample over redshift, detector frame mass and equation of state parameters as described above. For a flat, $\Lambda \rm CDM$ universe, luminosity distance $d_L$, redshift $z$, and Hubble's constant $H_0$ are related via 
\begin{equation}
    d_L = \frac{c(1+z)}{H_0} \int^{z}_0 \frac{dz'}{\sqrt{(1+z')^3\Omega_M+\Omega_\Lambda}},
\end{equation}
where $\Omega_M$ and $\Omega_\Lambda$ are the matter density, and dark energy density, respectively.  
For $z\ll1$, this reduces to the approximation 
\begin{equation}
    d_L \approx \frac{cz}{H_0}.
\end{equation} 

We can chose to either sample directly in $H_0$, or sample in redshift, generating the $H_0$ posterior distribution in post-processing. We choose to sample directly in $H_0$ so that we can explicitly enforce a prior, which we take to be uniform. 
\label{sec:GW170817}
We construct a synthetic injection of a future  GW170817-like event, GW270817.
We inject a binary neutron star signal using the \textsc{IMRPhenomPv2\_NRTidal} waveform approximate~\citep{Dietrich2019_NRtidal2} into a three-detector network consisting of LIGO Hanford, LIGO Livingston, and Virgo at $A{+}$ sensitivity\footnote{We use sensitivity curves for $A+$ and AdV from https://dcc.ligo.org/LIGO-T1500293.}. 

We utilise the \textsc{dynesty} \citep{Dynesty} nested sampler for inference. The priors for inference are summarised in Table \ref{tab:priors_BNSinference}. In addition to these priors, we have constraints on allowed equations of state due to causality. 
We enforce these limits in post-processing by discarding samples that do not fall within this range. 

\begin{figure*}
    \centering
    \includegraphics[width=\linewidth]
    {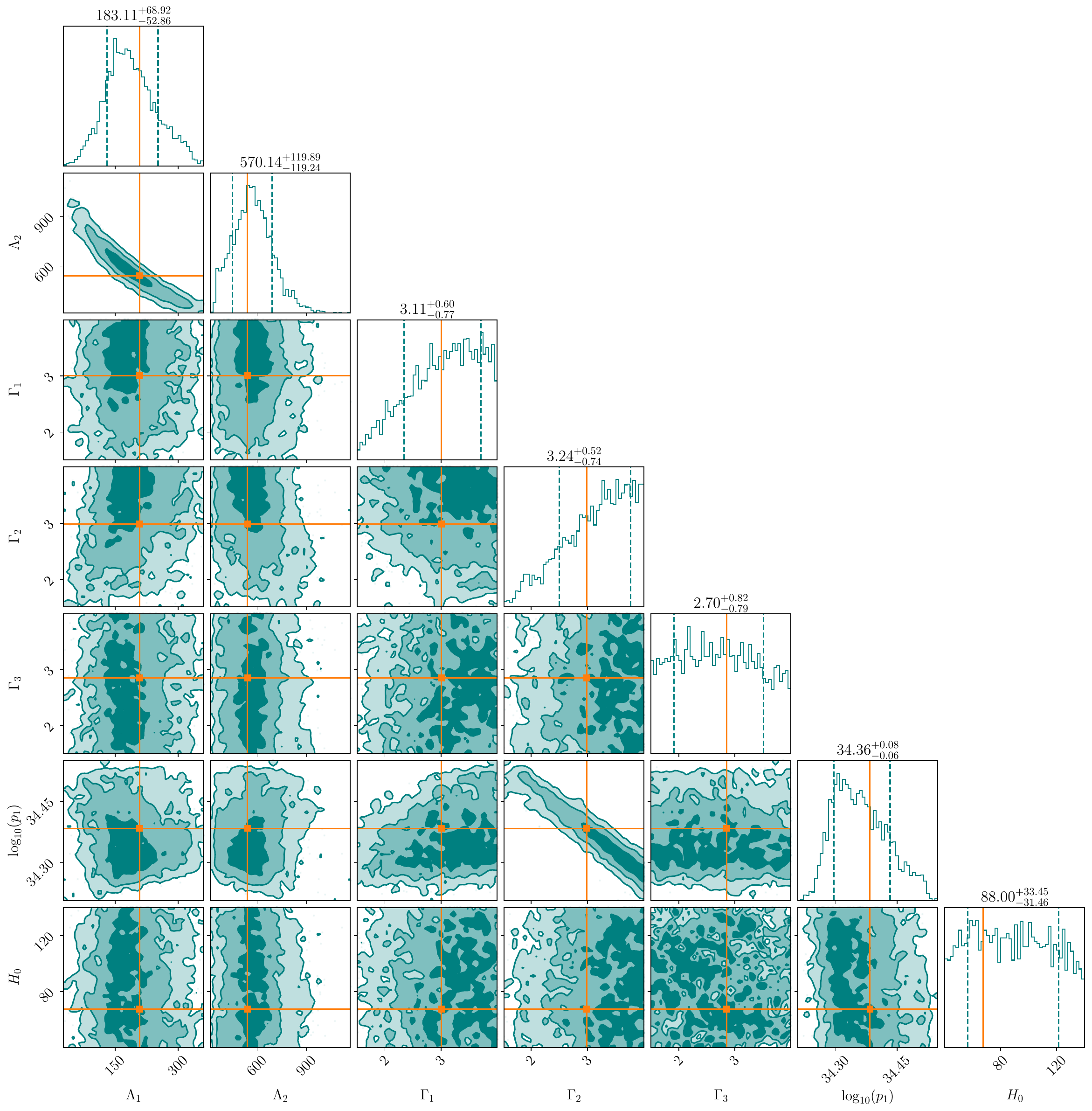}
    \caption{Marginalised posterior distribution for the injection of a GW170817 like signal with tidal deformabilities consistent with the SLy equation of state. Posterior distributions are shown in teal, while the injected values are shown by the orange lines. The first transition pressure, $p_1$ has units of $\rm dyne/cm^2$, while Hubble's constant $H_0$ has standard units of $\rm km \ s^{-1} \ Mpc^{-1}$. Tidal deformabilites are reconstructed in post processing using our machine learning model $\mu$-TOV. We see good recovery of all parameters, including the tidal deformabilities.}
    \label{fig:GW170817-like}
\end{figure*}

Figure \ref{fig:GW170817-like} shows the resultant marginalised posterior distribution, with the corresponding injected values being represented with orange lines. We see good recovery of our injected tidal deformabilities (which are calculated in post-processing), and posteriors for equation of state parameters consistent with the piecewise polytrope parameterisation of SLy. We note that we do not show the full posterior distribution here, but only the marginalised posterior for the equation of state parameters, $H_0$ and the tidal deformabilities calculated in post-processing. The posterior distribution for the other parameters are also consistent with their injected values. 

\begin{table}
    \centering
    \begin{tabular}{ccccc}
        \toprule
        variable & unit & prior & min & max \\
        \hline
        $m_{1,2}$ & $M_\odot$ & uniform & 1 & 2 \\
        $a_{1,2}$ & \_ & uniform & -0.05 & 0.05 \\ 
        $d_L$ & $\rm Mpc$ & comoving & 10 & 500 \\
        $\rm ra$ & $\rm rad.$ & uniform & 0 & $2\pi$ \\ 
        $\rm dec$ & $\rm rad.$ & cos & $-\pi/2$ & $\pi/2$ \\ 
        $\theta_{\rm JN}$ & $\rm rad.$ & sin & 0 & $\pi$ \\ 
        $\psi$ & $\rm rad.$ & uniform & 0 & $\pi$ \\ 
        $\phi_c$ & $\rm rad.$ & uniform & 0 & $2\pi$ \\ 
        $\Gamma_1$ & \_ & uniform & 1.5 & 5 \\ 
        $\Gamma_2$ & \_ & uniform & 1.5 & 4.5 \\ 
        $\Gamma_3$ & \_ & uniform & 1.5 & 4.5 \\
        $\log_{10}(p_1)$ & $\log_{10}(\rm dyne/cm^2)$ & uniform & 33.5 & 35.5 \\
        $\log_{10}(\rho_0)$ & $\log_{10}(\rm dyne/cm^2)$  & constraint & 12.5 & 14.7 \\
        $H_0$ & $\rm km \ s^{-1} \ Mpc^{-1}$  & uniform & 40 & 140
    \end{tabular}
    \caption{Priors for equation of state inference of our GW170817-like binary neutron star merger. Here, `comoving' implies a prior uniform in comoving volume, and `constraint' implies we constrain the denisty where our polytrope intersects the low-density crust.} 
    \label{tab:priors_BNSinference}
\end{table}

\begin{figure}
    \centering
    \includegraphics[width=\linewidth]{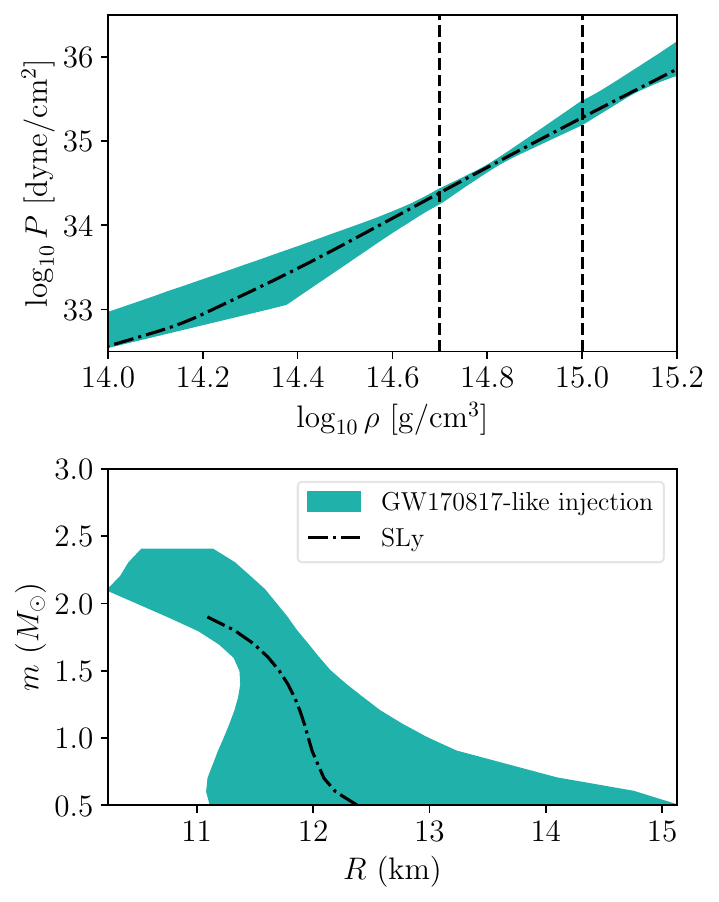}
    \caption{Equation of state reconstruction for a single GW170817-like event at $A+$ sensitivity. Top panel: The equation of state posterior predictive distribution for pressure $P$ as a function of density $\rho$. Bottom panel: The neutron star mass $m$ as a function of stellar radius $R$.  The dashed curves in the top panel correspond to the transition densities between different polytropic indices. We show the 90\% percent credible interval by the shaded teal region, while the dashed dotted curve represents the injected SLy equation of state in both panels.}
    \label{fig:GW270817_eos}
\end{figure}
In Figure~\ref{fig:GW270817_eos} we show the reconstructed pressure density diagram (top panel) and the neutron star mass-radius constraint (bottom panel) for our injected event GW270817. The 90\% credible intervals are shown with the aquamarine shaded region, while the injected equation of state (SLy) is shown with the black dotted dashed curve. We see that for a $1.4 M_\odot$ neutron star, the 90\% credible interval is constrained to within $900$ m.
\section{Synthetic Population Study}\label{sec:popstudy}
 We study a synthetic population of binary neutron star mergers in the $A{+}$ era.
 We again consider a three-detector network consisting of the LIGO Hanford, LIGO Livingston and Virgo interferometers.  
We consider fifteen events at $A{+}$ sensitivity,
which we take as representative of a pessimistic estimate of the number of expected binary neutron star observations during two years of $A{+}$~\citep{Floor_A+_Asharp_rates}.

We randomly draw parameters ($a$, $\rm ra $, $\rm dec$, etc.) for each event from the parameter ranges listed in Table \ref{tab:injectionset}, and masses from a Gaussian mixture distribution that describes the galactic neutron star population while providing support for GW190425~\citep{AnzacDay}. We assume this mass distribution takes the form
\begin{equation}
\label{eq:Mixtue}
 p(m) = (1-\epsilon) \mathcal{N}(\mu_1, \sigma_1) + \epsilon \mathcal{N}(\mu_2, \sigma_2),   
\end{equation}
where $p(m)$ is the source frame mass, $\epsilon=0.35$ is the mixing fraction, and the means and standard deviations are $\mu_1 =1.32 \,M_\odot$, $\sigma_1 =0.11\,M_\odot$, $\mu_{2} = 1.80\,M_\odot $, $\sigma_2 = 0.21\,M_\odot$ \citep[e.g.,][]{2018Alsing}.

\begin{table}
    \centering
    \begin{tabular}{ccccc}
        \toprule
        variable & unit & distribution & min & max \\
        \hline
        $q$ & - & uniform & 0.25 & 1 \\
        $a_{1,2}$ & \_ & uniform & -0.05 & 0.05 \\ 
        $d_L$ & $\rm Mpc$ & comoving & 10 & 600 \\
        $\rm ra$ & $\rm rad.$ & uniform & 0 & $2\pi$ \\ 
        $\rm dec$ & $\rm rad.$ & cos & $-\pi/2$ & $\pi/2$ \\ 
        $\iota$ & $\rm rad.$ & sin & 0 & $\pi$ \\ 
        $\psi$ & $\rm rad.$ & uniform & 0 & $\pi$ \\ 
        $\phi_c$ & $\rm rad.$ & uniform & 0 & $2\pi$ \\ 
       \\
    \end{tabular}
    \caption{Parameter ranges for the injection set of our synthetic binary neutron star population study. 
    }
    \label{tab:injectionset}
\end{table}

We create synthetic signals using an equation of state consistent with the SLy \citep{2001SLy} for each event, and use redshift $z$ and luminosity-distance $d_L$ values consistent with a cosmology of $H_0 = 67.4 \ \rm km \ s^{-1}\,Mpc^{-1} $ \citep{2020A&A...641A...6P}.
We additionally enforce a network signal-to-noise ratio cut, requiring all injections to have a network signal-to-noise ratio of $\rho_{\rm net} \geq 12$. 
Since binary neutron star mergers are expected to be in-band for minutes to hours, we use a reduced-order quadrature waveform that accelerates inference by removing redundant waveform calculations via reduced-order quadrature integration \citep{Smith+2016,Smith+2021,2023MorisakiROQ}. 
\

\begin{figure}
    \centering
    \includegraphics[width=\linewidth]{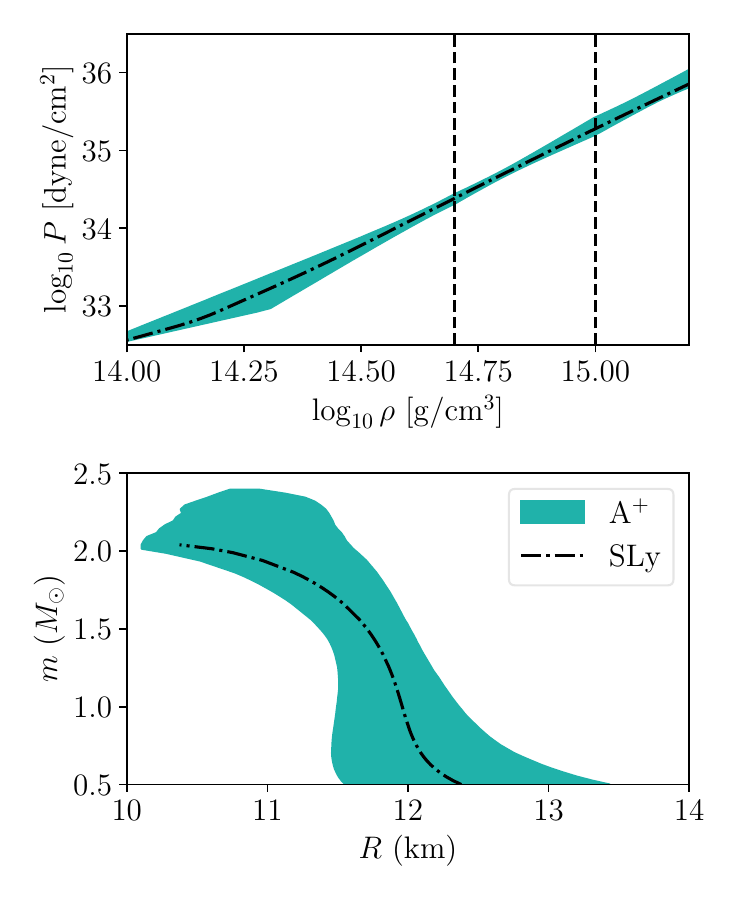}
    \caption{Top panel:  The equation of state reconstruction for pressure $P$ as a function of density $\rho$. Bottom panel: The neutron star mass $m$ as a function of stellar radius $R$.  The dashed curves in the top panel correspond to the transition densities between different polytropic indices. We show the 90\% percent credible interval by the teal shaded region, while the dashed dotted curves represent the injected SLy equation of state in both panels.}
    \label{fig:EOS_constraints}
\end{figure}

 We obtain `stacked' posteriors on equation-of-state parameters and $H_0$ by combining individual event posteriors with the methodology outlined in Appendix \ref{sec:Stacking}. 
Our results for the equation of state constraints are shown in Figure \ref{fig:EOS_constraints} for pressure as a function of density (top panel), and for mass as a function of radius (bottom panel). The 90\% credible intervals are shown with the teal shaded region, 
while the dotted dashed curves correspond to the injected equation of state.  We see that for a $1.4 M_\odot$ NS, the 90\% credible  interval of the radius will be constrained to within 700 m.

\begin{figure}
    \centering
    \includegraphics[width=\linewidth]{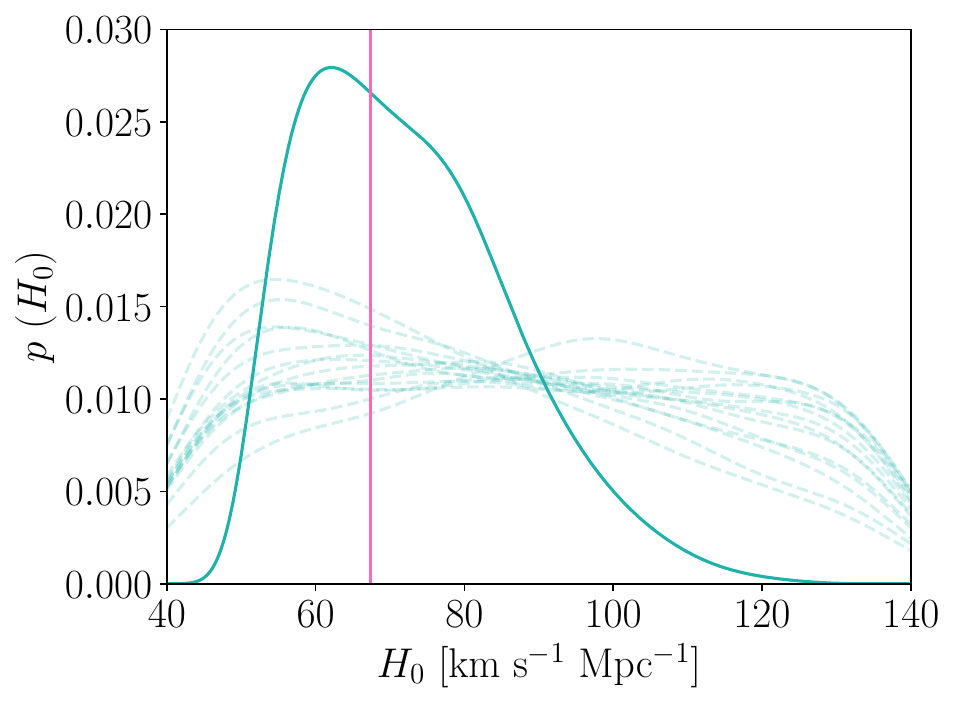}
    \caption{Combined posterior distribution of $H_0$ obtained from our synthetic injections (solid teal). Individual event $H_0$ posterior distributions are shown with the dashed teal curves. The pink solid line represents the injected cosmology $H_0 = 67.4 \ \rm km \ s^{-1} \ Mpc^{-1}$.}
    \label{fig:H_0_constraints}
\end{figure}

Figure \ref{fig:H_0_constraints} shows posterior distributions of Hubble's constant for the $A{+}$ era,
with the teal dashed curves showing the individual posteriors of $H_0$ from each event. Note that due to the luminosity distance-inclination angle degeneracy, most of the individual posteriors are peaked away from the true value. However, through combining posteriors, they eventually converge to the true value. We constrain Hubble's constant to $H_0 = 68^{+17}_{-13}  \ \rm km \ s^{-1}\, Mpc^{-1}$ (68\% CI) in the $A+$ era. 

\section{Discussion and Conclusion}\label{ref:conclusion}

Direct inference of the neutron star equation of state from gravitational-wave signals, such that inference of the equation of state is performed simultaneously with all other binary parameters, can be somewhat limited by the computation time required to solve the general relativistic equations of hydrostatic equilibrium. In this work, we develop a machine-learning algorithm, $\mu$-TOV, that solves these equations in $\approx\unit[500]{\mu s}$; faster than a single likelihood evaluation for most gravitational-wave calculations. We train $\mu$-TOV on piecewise polytropes in this work, however we emphasise that we can trivially use more complex equation-of-state parameterisations.

We implement $\mu$-TOV in the \textsc{Bilby} software suite to perform inference of the equation of state for gravitational-wave mergers. Using the method of~\citet{Messenger&Read2012}, we show how to simultaneously measure the equation of state and Hubble's constant ($H_0$) independent of the cosmic distance ladder. As we are performing inference directly on the equation of state and $H_0$, the combination of posterior distributions from multiple observations is rendered rather trivial.

By performing synthetic observations and inference calculations, we show that the O5 (i.e., A+)
observing run of LIGO/Virgo can constrain the 90\% CI radius of a 1.4 $M_\odot$  neutron star to within $\unit[700]{m}$.
We also demonstrate that $H_0$ can be constrained to $H_0 = 68^{+17}_{-13}  \ \rm km \ s^{-1} Mpc^{-1}$ 
in the $A+$ era
from the gravitational-wave signal alone. These are fractional uncertainties of 3\% and 20\% for the neutron star radius and Hubble's constant, respectively. 

Our results are conservative; we do not make use of galaxy catalogues to refine the possible redshift information~\citep{2023Galaxy_Catalogue}, we do not assume any electromagnetic counterpart to these sources has been found, and we do not use prior knowledge of the equation of state from gravitational-wave (e.g., \citealt{GW170817_eos1,GW170817_eos2}), kilonovae (e.g., \citealt{2023Pang}) or NICER measurements (e.g., \citealt{2019RileyNICERView}). The inclusion of any of these elements would improve the constraints on $H_0$ and the equation of state. 

There are a few caveats to our analyses. For example, a potential source of error is the choice of the piecewise polytrope parameterisation. Such a parameterisation may introduce implicit correlations~\citep{Geert2018,2022Legred}, and future work should investigate the use of spectral \citep{Lindbolm2010} or non-parametric \citep{2019Landry,2020Essick,2022Legred,Sarin2023} equations of state. We again stress that our trained machine-learning algorithm $\mu$-TOV should not have any additional difficulties learning those equation of state parameterisations compared to the piecewise polytropes.

In this work, we have also not considered the impact of selection effects on our synthetic population. For example, since we use a network signal to noise ratio cut of $\rho_{\rm net} \geq 12 $ 
and we choose our neutron star mass distribution to be a Gaussian mixture (with means $\mu_1 = 1.32\, M_{\odot}$, and $\mu_2 = 1.80\, M_{\odot}$; see Section~\ref{sec:popstudy}) we have a preference towards observing larger (and therefore louder) neutron star mergers. Gravitational-wave signals are also louder for face-on, rather than edge-on, systems, implying a potential bias in the inferred value of $H_0$.
We leave detailed analysis on the impact of these selection effects and more on both the equation of state and Hubble's constant to future work. 

\
In this Paper, we restrict our analyses to binary neutron star mergers as we expect them to contain more information about tidal deformability than neutron star-black hole mergers. In principle, we could also perform equation-of-state and $H_0$ inference on neutron star-black hole mergers with the same methodology, however, we do not expect those to better constrain the nuclear equation of state for individual mergers~\citep{2023Biscoveanu,2023Clarke}. The merger rates of both neutron star-black hole and binary neutron star mergers still have large uncertainties; given the former are detectable out to larger distances, they may directly compete with binary neutron star mergers when constraining the equation of state~\citep{Sarin2023}. Consequently, their impact on constraining Hubble's constant may also be significant.

\
Our forecast constraints on Hubble's constant are broadly consistent with recent work~\citep{2021Chaterjee,2022Ghosh,ghosh2024jointinferencepopulationcosmology}, albeit with slightly broader confidence intervals. We attribute this to the use of an uninformative equation of state prior and/or lack of additional constraints from the neutron star population. During the $A+$ era, we find the 90\% credible interval of a $1.4 M_\odot$ neutron star to within $700 \ \rm  m$ of the injected value. This is consistent with work by \cite{2020Landry}, who constrain the radius to within $1 \ \rm  km$. However, a direct comparison is difficult owing to differences in network sensitivity, synthetic population, equation-of-state parameterisation, and injected equation-of-state values. 

\
In this Work, we present a method that allows for direct inference of the equation of state and Hubble's constant using tidal information from individual binary neutron star mergers. Using a machine-learning model to approximate a TOV solver, we sample directly over equation-of-state parameters and $H_0$ in our binary neutron star inference scheme. 
\
We validate our method by performing inference of the equation of state and Hubble's constant with a GW170817-like event. 
We further demonstrate our method by performing a synthetic study of a population of binary neutron star mergers, consisting of fifteen events at $A{+}$ sensitivity.
We show that the radius of a $1.4 M_\odot$ neutron star can be constrained to within $700 \rm m$ at the 90\% credible  level. We obtain a constraint on Hubble's constant of $H_0 = 68^{+17}_{-13}  \ \rm km \ s^{-1} Mpc^{-1}$ from the gravitational-wave signals alone. 

\section{Acknowledgments}
The authors thank Leslie Wade for comments on the manuscript. 
S. J. M. receives support from the Australian Government Research Training Program. N.S. is supported by a Nordita Fellowship and also acknowledges support from the Knut and Alice Wallenberg foundation through the ``Gravity Meets Light" project. Nordita is supported in part by NordForsk. The authors are supported through Australian Research Council (ARC) Centre of Excellences CE170100004 and CE230900016, Discovery Projects DP220101610 and DP230103088, and LIEF Project LE210100002. We also acknowledge CPU time on OzSTAR, funded by Swinburne University and the Australian Government
\appendix

\section{Neural Network architecture and training}\label{sec:NN_appendix}
\begin{figure}
    \centering
    \includegraphics[width=\linewidth]{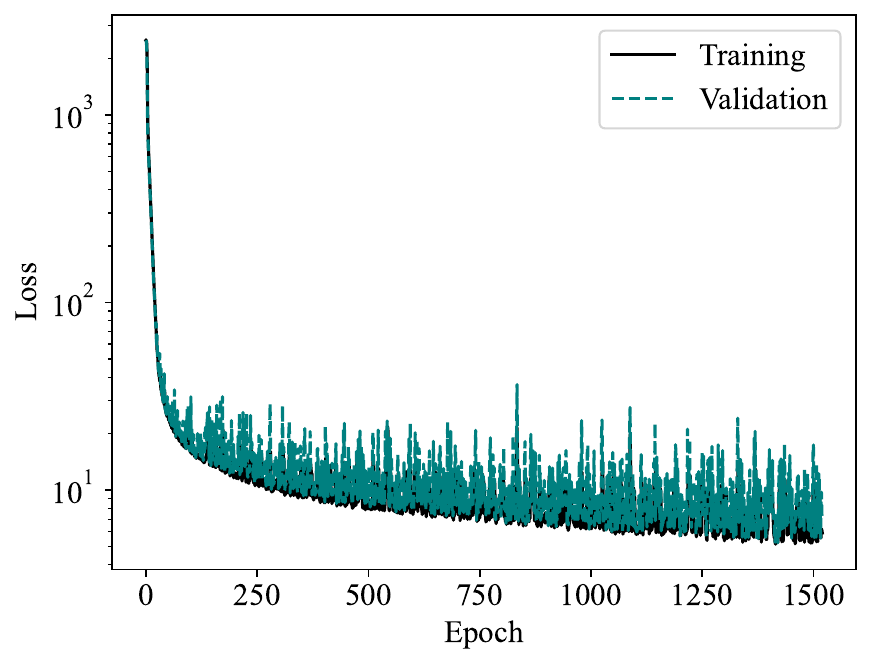}
    \caption{Loss curves for $\mu$-TOV. The training loss is shown with the black solid curve, while the validation loss is shown with the dashed teal curve. We see no significant divergence between the training and validation curves, which indicatives the model is not being over-fit.}
    \label{fig:loss_curves}
\end{figure}

We construct an artificial neural network consisting of an input layer of five neurons corresponding to the equation-of-state parameters and mass, four hidden layers with 128, 64, 32, and 16 neurons respectively, and an output layer with a single neuron corresponding to tidal deformability. The ReLU (rectified linear unit; \cite{Fukushima1969VisualFE}) activation function is used for all hidden layers, while the output neuron has no activation function. 
We choose a learning rate of $\eta = 0.001$, with the ADAM optimiser \citep{ADAM}, and mean absolute error as the loss function. The dataset used for training and testing this model consists of 100,000 samples of equation-of-state parameters and their corresponding tidal deformability, evenly sampled across $\Lambda$ as to avoid bias in the model. The training set consists of 90,000 samples, with 5,000 each reserved for the validation and test sets. A batch size of 2,812 is used to ensure each epoch contains 32 training steps. We utilise early stopping for regularization, which leads to our model being trained for $\approx 1500$ epochs. The combination of large batch size and many epochs ensures the model captures the complex relationship between the equation-of-state parameters and the tidal deformability quickly and effectively. Figure \ref{fig:loss_curves} shows the training and validation loss curves from the model and demonstrates no signs of under-fitting or over-fitting.

\section{Combining posterior distributions}\label{sec:Stacking}
We derive the equations for combining posterior distributions of both the cold neutron star equation of state and Hubble's constant directly from Bayes' theorem
\begin{equation}
    p(\theta|d) = \frac{\mathcal{L}(d|\theta)\pi(\theta)}{\mathcal{Z}},
\end{equation}
where $\theta$ are the model parameters, $d$ is the data, $p(\theta|d)$ is the posterior distribution, $\mathcal{L}(d|\theta)$ is the likelihood, $\pi(\theta)$ is the prior, and $\mathcal{Z}$ is a normalisation factor known as the evidence. 


We consider first the probability of obtaining data from two events (which we name $d_1$ and $d_2$) for a given equation of state. If $d_1$ and $d_2$ are independent then 
\begin{equation}
\label{eq:combined_likelihood}
    \mathcal{L}(d_1,d_2|\theta) = \mathcal{L}(d_1|\theta) \ \mathcal{L}(d_2|\theta).
\end{equation}
In practice, since we have posterior distributions for each event and not likelihoods, we use Bayes' theorem to write 
\begin{equation}
\label{eq:joint_like}
    \mathcal{L}(d_1,d_2|\theta) \propto \frac{p(\theta|d_1)}{\pi_2(\theta)} \frac{p(\theta|d_2)}{\pi_1(\theta)},
\end{equation}
where $\pi_1(\theta)$, and $\pi_2(\theta)$ are the priors for the first and second events, respectively. 
Since we are only interested in the equation-of-state parameters, we consider the marginalised posterior distribution
\begin{equation}
    p(\theta_i|d) = \int \left(\prod_{k\ne i} d\theta_k\right)p(\theta|d),
\end{equation}
where we have marginalised over all parameters except for $\theta_i = {\log_{10}({p_1}),\Gamma_1,\Gamma_2,\Gamma_3}$. Equation \ref{eq:joint_like} may now be written in the form
\begin{equation}
    \mathcal{L}(d_1,d_2|\theta_i) \propto \frac{p(\theta_i|d_1)}{\pi(\theta_i)} \frac{p(\theta_i|d_2)}{\pi(\theta_i)},
\end{equation}
where $\pi(\theta_i)$ is the prior on equation-of-state parameters.  
Utilising Bayes theorem, the combined posterior distribution becomes 
\begin{equation}
     p(\theta|d_1,d_2) \propto \frac{p(\theta|d_1)}{\pi(\theta)} \frac{p(\theta|d_2)}{\pi(\theta)} \pi(\theta)=\frac{1}{\pi(\theta)} p(\theta|d_1) p(\theta|d_2).
\end{equation}
Generalising to $N$ events we have 
\begin{equation}
\label{eq:stacked_posterior}
     p(\theta_i|d_1,...,d_N) \propto \frac{1}{\pi^{N-1}(\theta_i)} \prod_{a=1}^{N}p(\theta_i|d_a).
\end{equation}

We combine posterior distributions via a multi-variate kernel-density-estimate 
\begin{equation}
     p(\theta_i|d_1,...,d_N) \approx  \frac{1}{\pi^{N-1}(\theta_i)}\prod_{a=1}^{N} \rm{KDE} \ (S_a),
\end{equation}
where $S_a$ are the marginalised posterior samples for each event.
Since we have uniform priors on equation-of-state parameters, we obtain
\begin{equation}
     \label{eq:KDE_stack}p(\theta_i|d_1,...,d_N) \propto \prod_{a=1}^{N} \rm{KDE} \ (S_a).
\end{equation}

Similarly, posteriors on $H_0$ are combined via
\begin{equation}
    p(H_0|d_1,...,d_N) \propto \frac{1}{\pi^{N-1}(H_0)} \prod_{a=1}^{N}p(H_0|d_a),
\end{equation}
where $\pi(H_0)$ is the prior on $H_0$, and $p(H_0|d_a)$ is the marginalised posterior of $H_0$ for event $a$.  
 We choose a uniform prior on $H_0$ of $P(H_0) = \rm Uniform \ [40,140] \ \rm km \ s^{-1} \ Mpc^{-1} $ for all analyses in this work. 
Like the equation of state inference, we calculate and take the product of these posteriors distributions via a kernel density estimate~(Equation \ref{eq:KDE_stack}). 

\section{Piecewise Polytrope Bias}\label{sec: PP_Bias}
\begin{figure}
    \centering
    \includegraphics[width=\linewidth]{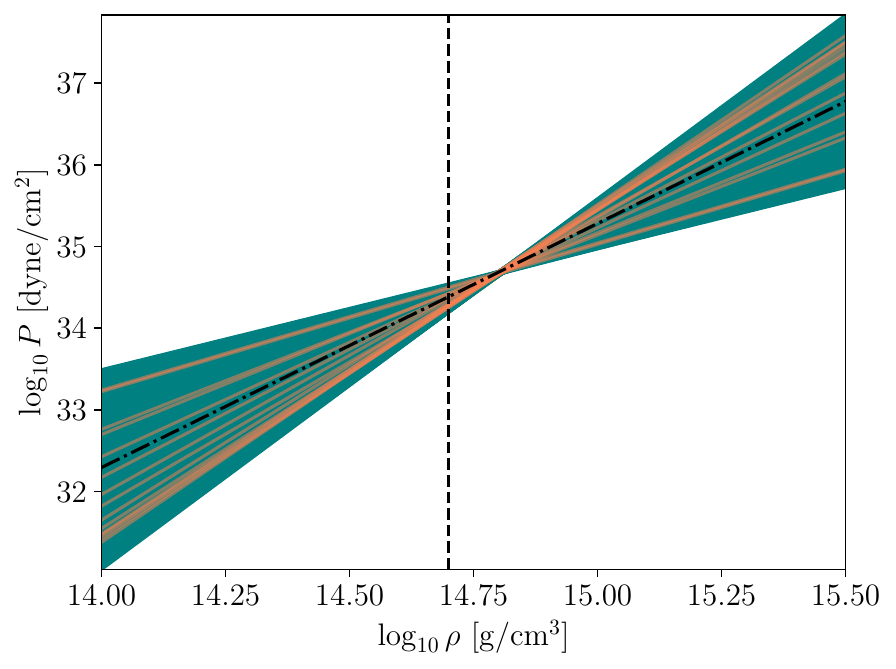}
    \caption{Illustration of the covariance between $\Gamma_2$ and $p_1$. We consider a simplified EOS, which is parameterised entirely by $\Gamma_2$, and $\log(p_1)$ (i.e linear in log space). Individual draws are shown with the coral curves. Due to the linearity of the parameterisation, we see an artificial squeezing of the EOS at the density probed by the NS mergers ($\rho = 10^{14.8} \rm g \ cm^{-3} $). 
    The injected EOS (dotted line) may be fit at this density by either a small $\Gamma_2$ and large value of $\log(p_1)$ or conversely large values of $\Gamma_2$ and small values of $\log(p_1)$. 
    }
    \label{fig:PP_Bias}
\end{figure}
Throughout this work, we utilise the piecewise-polytropic equation of state paramterisation of \citealt{Read2009} due to its simplicity. However, we note that directly sampling using this parameterisation can lead to significant systematic biases when combining a large number of posterior distributions, or if sampling a loud (signal-to-noise ratio$\gtrsim100$) event. We argue below that this can be mitigated by using aggressive sampler settings in our nested-sampling algorithm (e.g., a large number of live points or number of autocorrelation times, however the computational cost becomes prohibitive for these loud events, and is therefore considered beyond the scope of this work, as described below.

To illustrate this bias, we consider only the second adiabatic index $\Gamma_2$,  and the pressure anchor point $p_1$. These variables are highly covariant and probe the density regime expected during binary neutron star mergers.
For example, we may fit the same point in density with either a high value $\log_{10}(p_1)$ and a small value of $\Gamma_2$, or conversely a small value of $\log_{10}(p_1)$ and a large value of $\Gamma_2$. Since we require a monotonically increasing equation of state, $\Gamma_2$ must always be positive. Figure \ref{fig:PP_Bias} shows an illustration of the phenomenon. We see that we can fit more curves, with a large value of $\Gamma_2$ and a small value of $\log_{10}(p_1)$ than vice versa. 

\begin{figure}
    \centering
    \includegraphics[width=\linewidth]{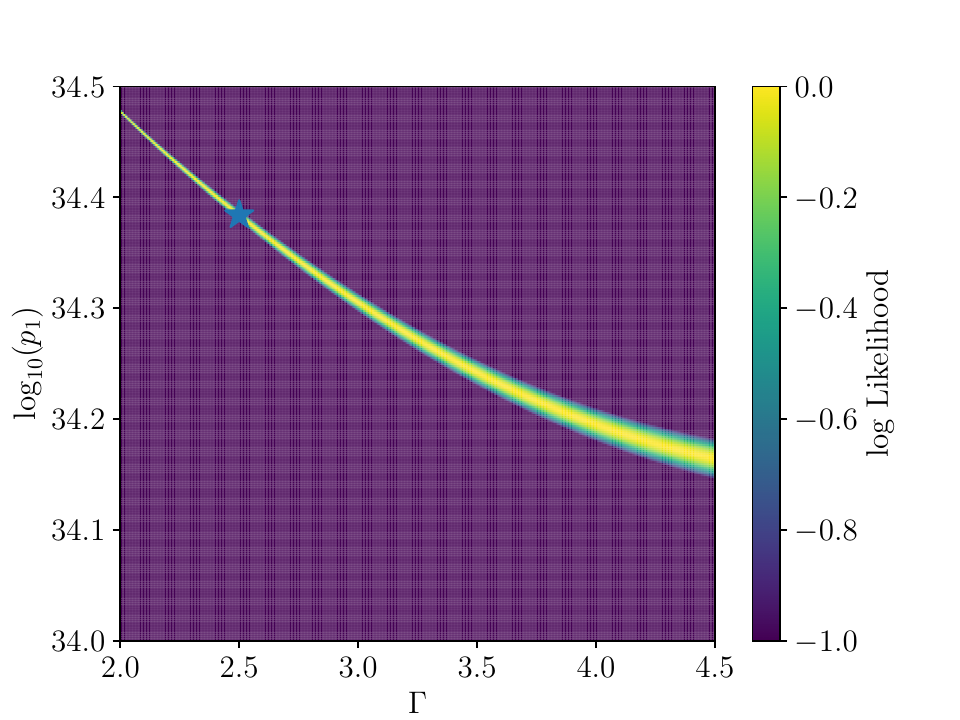}
    \caption{The likelihood surface for $\Gamma_2$ and $\log_{10}(p_1)$. The `true' equation of state injected, is shown with the blue star. We see that the EOS parameterisation has a likelihood surface `wedge' that becomes more finely peaked as $\Gamma_2$ is decreased and $\log_{10}(p_1)$ is increased. }
    \label{fig:likelihood_wedge}
\end{figure}
This degeneracy between adiabatic index and anchor pressure point, combined with the requirement for monotonicity, results in a wedge shaped likelihood surface that becomes increasingly sharp as pressure is increased and adiabatic index is decreased. We show this wedge-shaped region in Fig.~\ref{fig:likelihood_wedge} where we plot the likelihood heat map as a function of the two parameters of interest. One can see here that the true value (at the star) is located at the thin edge of the wedge. This wedge is so finely peaked near the true value that the sampler has trouble finding the peak without aggressive sampler settings, but finds it easier to locate the broad edge of the wedge.  As a consequence, the posterior distributions tend to be peaked away from the injected values leading to a bias. 
Since the radius of a neutron star is highly correlated with the pressure of the equation of state near nuclear saturation \citep{2001Lattimer_Prakash} this may lead to a significant bias in the inferred constraints on neutron star radius. 

We emphasise that this bias occurs because of our choice of piecewise polytropic parameterisation. For other reasons described in the body of the text, this is not an optimal choice for the equation of state. In future, we will move to better parameterisations, that we also anticipate will not suffer from the problem described above.


\begin{thebibliography}{73}%
\makeatletter
\providecommand \@ifxundefined [1]{%
 \@ifx{#1\undefined}
}%
\providecommand \@ifnum [1]{%
 \ifnum #1\expandafter \@firstoftwo
 \else \expandafter \@secondoftwo
 \fi
}%
\providecommand \@ifx [1]{%
 \ifx #1\expandafter \@firstoftwo
 \else \expandafter \@secondoftwo
 \fi
}%
\providecommand \natexlab [1]{#1}%
\providecommand \enquote  [1]{``#1''}%
\providecommand \bibnamefont  [1]{#1}%
\providecommand \bibfnamefont [1]{#1}%
\providecommand \citenamefont [1]{#1}%
\providecommand \href@noop [0]{\@secondoftwo}%
\providecommand \href [0]{\begingroup \@sanitize@url \@href}%
\providecommand \@href[1]{\@@startlink{#1}\@@href}%
\providecommand \@@href[1]{\endgroup#1\@@endlink}%
\providecommand \@sanitize@url [0]{\catcode `\\12\catcode `\$12\catcode `\&12\catcode `\#12\catcode `\^12\catcode `\_12\catcode `\%12\relax}%
\providecommand \@@startlink[1]{}%
\providecommand \@@endlink[0]{}%
\providecommand \url  [0]{\begingroup\@sanitize@url \@url }%
\providecommand \@url [1]{\endgroup\@href {#1}{\urlprefix }}%
\providecommand \urlprefix  [0]{URL }%
\providecommand \Eprint [0]{\href }%
\providecommand \doibase [0]{https://doi.org/}%
\providecommand \selectlanguage [0]{\@gobble}%
\providecommand \bibinfo  [0]{\@secondoftwo}%
\providecommand \bibfield  [0]{\@secondoftwo}%
\providecommand \translation [1]{[#1]}%
\providecommand \BibitemOpen [0]{}%
\providecommand \bibitemStop [0]{}%
\providecommand \bibitemNoStop [0]{.\EOS\space}%
\providecommand \EOS [0]{\spacefactor3000\relax}%
\providecommand \BibitemShut  [1]{\csname bibitem#1\endcsname}%
\let\auto@bib@innerbib\@empty
\bibitem [{\citenamefont {Abbott}\ \emph {et~al.}(2017{\natexlab{a}})\citenamefont {Abbott} \emph {et~al.}}]{GW170817_gw}%
  \BibitemOpen
  \bibfield  {author} {\bibinfo {author} {\bibfnamefont {B.~P.}\ \bibnamefont {Abbott}} \emph {et~al.},\ }\bibfield  {title} {\bibinfo {title} {{GW170817: Observation of Gravitational Waves from a Binary Neutron Star Inspiral}},\ }\href {https://doi.org/10/2017} {\bibfield  {journal} {\bibinfo  {journal} {Physical Review Letters}\ }\textbf {\bibinfo {volume} {119}},\ \bibinfo {pages} {161101} (\bibinfo {year} {2017}{\natexlab{a}})}\BibitemShut {NoStop}%
\bibitem [{\citenamefont {Abbott}\ \emph {et~al.}(2017{\natexlab{b}})\citenamefont {Abbott} \emph {et~al.}}]{GW170817_em}%
  \BibitemOpen
  \bibfield  {author} {\bibinfo {author} {\bibfnamefont {B.~P.}\ \bibnamefont {Abbott}} \emph {et~al.},\ }\bibfield  {title} {\bibinfo {title} {{Multi-messenger Observations of a Binary Neutron Star Merger}},\ }\href {https://doi.org/10.3847/2041-8213/aa91c9} {\bibfield  {journal} {\bibinfo  {journal} {\apjl}\ }\textbf {\bibinfo {volume} {848}},\ \bibinfo {eid} {L12} (\bibinfo {year} {2017}{\natexlab{b}})},\ \Eprint {https://arxiv.org/abs/1710.05833} {arXiv:1710.05833 [astro-ph.HE]} \BibitemShut {NoStop}%
\bibitem [{\citenamefont {{Abbott}}\ \emph {et~al.}(2017{\natexlab{a}})\citenamefont {{Abbott}} \emph {et~al.}}]{GW170817_grb}%
  \BibitemOpen
  \bibfield  {author} {\bibinfo {author} {\bibfnamefont {B.~P.}\ \bibnamefont {{Abbott}}} \emph {et~al.},\ }\bibfield  {title} {\bibinfo {title} {{Gravitational Waves and Gamma-Rays from a Binary Neutron Star Merger: GW170817 and GRB 170817A}},\ }\href {https://doi.org/10.3847/2041-8213/aa920c} {\bibfield  {journal} {\bibinfo  {journal} {\apjl}\ }\textbf {\bibinfo {volume} {848}},\ \bibinfo {eid} {L13} (\bibinfo {year} {2017}{\natexlab{a}})},\ \Eprint {https://arxiv.org/abs/1710.05834} {arXiv:1710.05834 [astro-ph.HE]} \BibitemShut {NoStop}%
\bibitem [{\citenamefont {{Goldstein}}\ \emph {et~al.}(2017)\citenamefont {{Goldstein}}, \citenamefont {{Veres}}, \citenamefont {{Burns}}, \citenamefont {{Briggs}}, \citenamefont {{Hamburg}}, \citenamefont {{Kocevski}}, \citenamefont {{Wilson-Hodge}}, \citenamefont {{Preece}}, \citenamefont {{Poolakkil}}, \citenamefont {{Roberts}}, \citenamefont {{Hui}}, \citenamefont {{Connaughton}}, \citenamefont {{Racusin}}, \citenamefont {{von Kienlin}}, \citenamefont {{Dal Canton}}, \citenamefont {{Christensen}}, \citenamefont {{Littenberg}}, \citenamefont {{Siellez}}, \citenamefont {{Blackburn}}, \citenamefont {{Broida}}, \citenamefont {{Bissaldi}}, \citenamefont {{Cleveland}}, \citenamefont {{Gibby}}, \citenamefont {{Giles}}, \citenamefont {{Kippen}}, \citenamefont {{McBreen}}, \citenamefont {{McEnery}}, \citenamefont {{Meegan}}, \citenamefont {{Paciesas}},\ and\ \citenamefont {{Stanbro}}}]{2017FERMI}%
  \BibitemOpen
  \bibfield  {author} {\bibinfo {author} {\bibfnamefont {A.}~\bibnamefont {{Goldstein}}}, \bibinfo {author} {\bibfnamefont {P.}~\bibnamefont {{Veres}}}, \bibinfo {author} {\bibfnamefont {E.}~\bibnamefont {{Burns}}}, \bibinfo {author} {\bibfnamefont {M.~S.}\ \bibnamefont {{Briggs}}}, \bibinfo {author} {\bibfnamefont {R.}~\bibnamefont {{Hamburg}}}, \bibinfo {author} {\bibfnamefont {D.}~\bibnamefont {{Kocevski}}}, \bibinfo {author} {\bibfnamefont {C.~A.}\ \bibnamefont {{Wilson-Hodge}}}, \bibinfo {author} {\bibfnamefont {R.~D.}\ \bibnamefont {{Preece}}}, \bibinfo {author} {\bibfnamefont {S.}~\bibnamefont {{Poolakkil}}}, \bibinfo {author} {\bibfnamefont {O.~J.}\ \bibnamefont {{Roberts}}}, \bibinfo {author} {\bibfnamefont {C.~M.}\ \bibnamefont {{Hui}}}, \bibinfo {author} {\bibfnamefont {V.}~\bibnamefont {{Connaughton}}}, \bibinfo {author} {\bibfnamefont {J.}~\bibnamefont {{Racusin}}}, \bibinfo {author} {\bibfnamefont {A.}~\bibnamefont {{von Kienlin}}}, \bibinfo {author} {\bibfnamefont {T.}~\bibnamefont {{Dal
  Canton}}}, \bibinfo {author} {\bibfnamefont {N.}~\bibnamefont {{Christensen}}}, \bibinfo {author} {\bibfnamefont {T.}~\bibnamefont {{Littenberg}}}, \bibinfo {author} {\bibfnamefont {K.}~\bibnamefont {{Siellez}}}, \bibinfo {author} {\bibfnamefont {L.}~\bibnamefont {{Blackburn}}}, \bibinfo {author} {\bibfnamefont {J.}~\bibnamefont {{Broida}}}, \bibinfo {author} {\bibfnamefont {E.}~\bibnamefont {{Bissaldi}}}, \bibinfo {author} {\bibfnamefont {W.~H.}\ \bibnamefont {{Cleveland}}}, \bibinfo {author} {\bibfnamefont {M.~H.}\ \bibnamefont {{Gibby}}}, \bibinfo {author} {\bibfnamefont {M.~M.}\ \bibnamefont {{Giles}}}, \bibinfo {author} {\bibfnamefont {R.~M.}\ \bibnamefont {{Kippen}}}, \bibinfo {author} {\bibfnamefont {S.}~\bibnamefont {{McBreen}}}, \bibinfo {author} {\bibfnamefont {J.}~\bibnamefont {{McEnery}}}, \bibinfo {author} {\bibfnamefont {C.~A.}\ \bibnamefont {{Meegan}}}, \bibinfo {author} {\bibfnamefont {W.~S.}\ \bibnamefont {{Paciesas}}},\ and\ \bibinfo {author} {\bibfnamefont {M.}~\bibnamefont {{Stanbro}}},\
  }\bibfield  {title} {\bibinfo {title} {{An Ordinary Short Gamma-Ray Burst with Extraordinary Implications: Fermi-GBM Detection of GRB 170817A}},\ }\href {https://doi.org/10.3847/2041-8213/aa8f41} {\bibfield  {journal} {\bibinfo  {journal} {\apjl}\ }\textbf {\bibinfo {volume} {848}},\ \bibinfo {eid} {L14} (\bibinfo {year} {2017})},\ \Eprint {https://arxiv.org/abs/1710.05446} {arXiv:1710.05446 [astro-ph.HE]} \BibitemShut {NoStop}%
\bibitem [{\citenamefont {{Savchenko}}\ \emph {et~al.}(2017)\citenamefont {{Savchenko}}, \citenamefont {{Ferrigno}}, \citenamefont {{Kuulkers}}, \citenamefont {{Bazzano}}, \citenamefont {{Bozzo}}, \citenamefont {{Brandt}}, \citenamefont {{Chenevez}}, \citenamefont {{Courvoisier}}, \citenamefont {{Diehl}}, \citenamefont {{Domingo}}, \citenamefont {{Hanlon}}, \citenamefont {{Jourdain}}, \citenamefont {{von Kienlin}}, \citenamefont {{Laurent}}, \citenamefont {{Lebrun}}, \citenamefont {{Lutovinov}}, \citenamefont {{Martin-Carrillo}}, \citenamefont {{Mereghetti}}, \citenamefont {{Natalucci}}, \citenamefont {{Rodi}}, \citenamefont {{Roques}}, \citenamefont {{Sunyaev}},\ and\ \citenamefont {{Ubertini}}}]{2017INTEGRALGRB}%
  \BibitemOpen
  \bibfield  {author} {\bibinfo {author} {\bibfnamefont {V.}~\bibnamefont {{Savchenko}}}, \bibinfo {author} {\bibfnamefont {C.}~\bibnamefont {{Ferrigno}}}, \bibinfo {author} {\bibfnamefont {E.}~\bibnamefont {{Kuulkers}}}, \bibinfo {author} {\bibfnamefont {A.}~\bibnamefont {{Bazzano}}}, \bibinfo {author} {\bibfnamefont {E.}~\bibnamefont {{Bozzo}}}, \bibinfo {author} {\bibfnamefont {S.}~\bibnamefont {{Brandt}}}, \bibinfo {author} {\bibfnamefont {J.}~\bibnamefont {{Chenevez}}}, \bibinfo {author} {\bibfnamefont {T.~J.~L.}\ \bibnamefont {{Courvoisier}}}, \bibinfo {author} {\bibfnamefont {R.}~\bibnamefont {{Diehl}}}, \bibinfo {author} {\bibfnamefont {A.}~\bibnamefont {{Domingo}}}, \bibinfo {author} {\bibfnamefont {L.}~\bibnamefont {{Hanlon}}}, \bibinfo {author} {\bibfnamefont {E.}~\bibnamefont {{Jourdain}}}, \bibinfo {author} {\bibfnamefont {A.}~\bibnamefont {{von Kienlin}}}, \bibinfo {author} {\bibfnamefont {P.}~\bibnamefont {{Laurent}}}, \bibinfo {author} {\bibfnamefont {F.}~\bibnamefont {{Lebrun}}}, \bibinfo
  {author} {\bibfnamefont {A.}~\bibnamefont {{Lutovinov}}}, \bibinfo {author} {\bibfnamefont {A.}~\bibnamefont {{Martin-Carrillo}}}, \bibinfo {author} {\bibfnamefont {S.}~\bibnamefont {{Mereghetti}}}, \bibinfo {author} {\bibfnamefont {L.}~\bibnamefont {{Natalucci}}}, \bibinfo {author} {\bibfnamefont {J.}~\bibnamefont {{Rodi}}}, \bibinfo {author} {\bibfnamefont {J.~P.}\ \bibnamefont {{Roques}}}, \bibinfo {author} {\bibfnamefont {R.}~\bibnamefont {{Sunyaev}}},\ and\ \bibinfo {author} {\bibfnamefont {P.}~\bibnamefont {{Ubertini}}},\ }\bibfield  {title} {\bibinfo {title} {{INTEGRAL Detection of the First Prompt Gamma-Ray Signal Coincident with the Gravitational-wave Event GW170817}},\ }\href {https://doi.org/10.3847/2041-8213/aa8f94} {\bibfield  {journal} {\bibinfo  {journal} {\apjl}\ }\textbf {\bibinfo {volume} {848}},\ \bibinfo {eid} {L15} (\bibinfo {year} {2017})},\ \Eprint {https://arxiv.org/abs/1710.05449} {arXiv:1710.05449 [astro-ph.HE]} \BibitemShut {NoStop}%
\bibitem [{\citenamefont {{Coulter}}\ \emph {et~al.}(2017)\citenamefont {{Coulter}}, \citenamefont {{Foley}}, \citenamefont {{Kilpatrick}}, \citenamefont {{Drout}}, \citenamefont {{Piro}}, \citenamefont {{Shappee}}, \citenamefont {{Siebert}}, \citenamefont {{Simon}}, \citenamefont {{Ulloa}}, \citenamefont {{Kasen}}, \citenamefont {{Madore}}, \citenamefont {{Murguia-Berthier}}, \citenamefont {{Pan}}, \citenamefont {{Prochaska}}, \citenamefont {{Ramirez-Ruiz}}, \citenamefont {{Rest}},\ and\ \citenamefont {{Rojas-Bravo}}}]{2017SSS17a}%
  \BibitemOpen
  \bibfield  {author} {\bibinfo {author} {\bibfnamefont {D.~A.}\ \bibnamefont {{Coulter}}}, \bibinfo {author} {\bibfnamefont {R.~J.}\ \bibnamefont {{Foley}}}, \bibinfo {author} {\bibfnamefont {C.~D.}\ \bibnamefont {{Kilpatrick}}}, \bibinfo {author} {\bibfnamefont {M.~R.}\ \bibnamefont {{Drout}}}, \bibinfo {author} {\bibfnamefont {A.~L.}\ \bibnamefont {{Piro}}}, \bibinfo {author} {\bibfnamefont {B.~J.}\ \bibnamefont {{Shappee}}}, \bibinfo {author} {\bibfnamefont {M.~R.}\ \bibnamefont {{Siebert}}}, \bibinfo {author} {\bibfnamefont {J.~D.}\ \bibnamefont {{Simon}}}, \bibinfo {author} {\bibfnamefont {N.}~\bibnamefont {{Ulloa}}}, \bibinfo {author} {\bibfnamefont {D.}~\bibnamefont {{Kasen}}}, \bibinfo {author} {\bibfnamefont {B.~F.}\ \bibnamefont {{Madore}}}, \bibinfo {author} {\bibfnamefont {A.}~\bibnamefont {{Murguia-Berthier}}}, \bibinfo {author} {\bibfnamefont {Y.~C.}\ \bibnamefont {{Pan}}}, \bibinfo {author} {\bibfnamefont {J.~X.}\ \bibnamefont {{Prochaska}}}, \bibinfo {author} {\bibfnamefont {E.}~\bibnamefont
  {{Ramirez-Ruiz}}}, \bibinfo {author} {\bibfnamefont {A.}~\bibnamefont {{Rest}}},\ and\ \bibinfo {author} {\bibfnamefont {C.}~\bibnamefont {{Rojas-Bravo}}},\ }\bibfield  {title} {\bibinfo {title} {{Swope Supernova Survey 2017a (SSS17a), the optical counterpart to a gravitational wave source}},\ }\href {https://doi.org/10.1126/science.aap9811} {\bibfield  {journal} {\bibinfo  {journal} {Science}\ }\textbf {\bibinfo {volume} {358}},\ \bibinfo {pages} {1556} (\bibinfo {year} {2017})},\ \Eprint {https://arxiv.org/abs/1710.05452} {arXiv:1710.05452 [astro-ph.HE]} \BibitemShut {NoStop}%
\bibitem [{\citenamefont {{Valenti}}\ \emph {et~al.}(2017)\citenamefont {{Valenti}}, \citenamefont {{Sand}}, \citenamefont {{Yang}}, \citenamefont {{Cappellaro}}, \citenamefont {{Tartaglia}}, \citenamefont {{Corsi}}, \citenamefont {{Jha}}, \citenamefont {{Reichart}}, \citenamefont {{Haislip}},\ and\ \citenamefont {{Kouprianov}}}]{AT2017gfo_Valenti}%
  \BibitemOpen
  \bibfield  {author} {\bibinfo {author} {\bibfnamefont {S.}~\bibnamefont {{Valenti}}}, \bibinfo {author} {\bibfnamefont {D.~J.}\ \bibnamefont {{Sand}}}, \bibinfo {author} {\bibfnamefont {S.}~\bibnamefont {{Yang}}}, \bibinfo {author} {\bibfnamefont {E.}~\bibnamefont {{Cappellaro}}}, \bibinfo {author} {\bibfnamefont {L.}~\bibnamefont {{Tartaglia}}}, \bibinfo {author} {\bibfnamefont {A.}~\bibnamefont {{Corsi}}}, \bibinfo {author} {\bibfnamefont {S.~W.}\ \bibnamefont {{Jha}}}, \bibinfo {author} {\bibfnamefont {D.~E.}\ \bibnamefont {{Reichart}}}, \bibinfo {author} {\bibfnamefont {J.}~\bibnamefont {{Haislip}}},\ and\ \bibinfo {author} {\bibfnamefont {V.}~\bibnamefont {{Kouprianov}}},\ }\bibfield  {title} {\bibinfo {title} {{The Discovery of the Electromagnetic Counterpart of GW170817: Kilonova AT 2017gfo/DLT17ck}},\ }\href {https://doi.org/10.3847/2041-8213/aa8edf} {\bibfield  {journal} {\bibinfo  {journal} {\apjl}\ }\textbf {\bibinfo {volume} {848}},\ \bibinfo {eid} {L24} (\bibinfo {year} {2017})},\ \Eprint
  {https://arxiv.org/abs/1710.05854} {arXiv:1710.05854 [astro-ph.HE]} \BibitemShut {NoStop}%
\bibitem [{\citenamefont {{Abbott}}\ \emph {et~al.}(2018)\citenamefont {{Abbott}} \emph {et~al.}}]{GW170817_eos1}%
  \BibitemOpen
  \bibfield  {author} {\bibinfo {author} {\bibfnamefont {B.~P.}\ \bibnamefont {{Abbott}}} \emph {et~al.},\ }\bibfield  {title} {\bibinfo {title} {{GW170817: Measurements of Neutron Star Radii and Equation of State}},\ }\href {https://doi.org/10.1103/PhysRevLett.121.161101} {\bibfield  {journal} {\bibinfo  {journal} {\prl}\ }\textbf {\bibinfo {volume} {121}},\ \bibinfo {eid} {161101} (\bibinfo {year} {2018})},\ \Eprint {https://arxiv.org/abs/1805.11581} {arXiv:1805.11581 [gr-qc]} \BibitemShut {NoStop}%
\bibitem [{\citenamefont {{Abbott}}\ \emph {et~al.}(2019)\citenamefont {{Abbott}} \emph {et~al.}}]{GW170817_eos2}%
  \BibitemOpen
  \bibfield  {author} {\bibinfo {author} {\bibfnamefont {B.~P.}\ \bibnamefont {{Abbott}}} \emph {et~al.},\ }\bibfield  {title} {\bibinfo {title} {{Properties of the Binary Neutron Star Merger GW170817}},\ }\href {https://doi.org/10.1103/PhysRevX.9.011001} {\bibfield  {journal} {\bibinfo  {journal} {Physical Review X}\ }\textbf {\bibinfo {volume} {9}},\ \bibinfo {eid} {011001} (\bibinfo {year} {2019})},\ \Eprint {https://arxiv.org/abs/1805.11579} {arXiv:1805.11579 [gr-qc]} \BibitemShut {NoStop}%
\bibitem [{\citenamefont {{Abbott}}\ \emph {et~al.}(2017{\natexlab{b}})\citenamefont {{Abbott}} \emph {et~al.}}]{GW170817_bright_cosmo}%
  \BibitemOpen
  \bibfield  {author} {\bibinfo {author} {\bibfnamefont {B.~P.}\ \bibnamefont {{Abbott}}} \emph {et~al.},\ }\bibfield  {title} {\bibinfo {title} {{A gravitational-wave standard siren measurement of the Hubble constant}},\ }\href {https://doi.org/10.1038/nature24471} {\bibfield  {journal} {\bibinfo  {journal} {\nat}\ }\textbf {\bibinfo {volume} {551}},\ \bibinfo {pages} {85} (\bibinfo {year} {2017}{\natexlab{b}})},\ \Eprint {https://arxiv.org/abs/1710.05835} {arXiv:1710.05835 [astro-ph.CO]} \BibitemShut {NoStop}%
\bibitem [{\citenamefont {{Schutz}}(1986)}]{1986Schutz}%
  \BibitemOpen
  \bibfield  {author} {\bibinfo {author} {\bibfnamefont {B.~F.}\ \bibnamefont {{Schutz}}},\ }\bibfield  {title} {\bibinfo {title} {{Determining the Hubble constant from gravitational wave observations}},\ }\href {https://doi.org/10.1038/323310a0} {\bibfield  {journal} {\bibinfo  {journal} {\nat}\ }\textbf {\bibinfo {volume} {323}},\ \bibinfo {pages} {310} (\bibinfo {year} {1986})}\BibitemShut {NoStop}%
\bibitem [{\citenamefont {{Holz}}\ and\ \citenamefont {{Hughes}}(2005)}]{Holz_Huges_standard_sirens}%
  \BibitemOpen
  \bibfield  {author} {\bibinfo {author} {\bibfnamefont {D.~E.}\ \bibnamefont {{Holz}}}\ and\ \bibinfo {author} {\bibfnamefont {S.~A.}\ \bibnamefont {{Hughes}}},\ }\bibfield  {title} {\bibinfo {title} {{Using Gravitational-Wave Standard Sirens}},\ }\href {https://doi.org/10.1086/431341} {\bibfield  {journal} {\bibinfo  {journal} {\apj}\ }\textbf {\bibinfo {volume} {629}},\ \bibinfo {pages} {15} (\bibinfo {year} {2005})},\ \Eprint {https://arxiv.org/abs/astro-ph/0504616} {arXiv:astro-ph/0504616 [astro-ph]} \BibitemShut {NoStop}%
\bibitem [{\citenamefont {{Hotokezaka}}\ \emph {et~al.}(2019)\citenamefont {{Hotokezaka}}, \citenamefont {{Nakar}}, \citenamefont {{Gottlieb}}, \citenamefont {{Nissanke}}, \citenamefont {{Masuda}}, \citenamefont {{Hallinan}}, \citenamefont {{Mooley}},\ and\ \citenamefont {{Deller}}}]{2019Hotokezaka}%
  \BibitemOpen
  \bibfield  {author} {\bibinfo {author} {\bibfnamefont {K.}~\bibnamefont {{Hotokezaka}}}, \bibinfo {author} {\bibfnamefont {E.}~\bibnamefont {{Nakar}}}, \bibinfo {author} {\bibfnamefont {O.}~\bibnamefont {{Gottlieb}}}, \bibinfo {author} {\bibfnamefont {S.}~\bibnamefont {{Nissanke}}}, \bibinfo {author} {\bibfnamefont {K.}~\bibnamefont {{Masuda}}}, \bibinfo {author} {\bibfnamefont {G.}~\bibnamefont {{Hallinan}}}, \bibinfo {author} {\bibfnamefont {K.~P.}\ \bibnamefont {{Mooley}}},\ and\ \bibinfo {author} {\bibfnamefont {A.~T.}\ \bibnamefont {{Deller}}},\ }\bibfield  {title} {\bibinfo {title} {{A Hubble constant measurement from superluminal motion of the jet in GW170817}},\ }\href {https://doi.org/10.1038/s41550-019-0820-1} {\bibfield  {journal} {\bibinfo  {journal} {Nature Astronomy}\ }\textbf {\bibinfo {volume} {3}},\ \bibinfo {pages} {940} (\bibinfo {year} {2019})},\ \Eprint {https://arxiv.org/abs/1806.10596} {arXiv:1806.10596 [astro-ph.CO]} \BibitemShut {NoStop}%
\bibitem [{\citenamefont {{Aasi}}\ \emph {et~al.}(2015)\citenamefont {{Aasi}} \emph {et~al.}}]{2015AdvancedLigo}%
  \BibitemOpen
  \bibfield  {author} {\bibinfo {author} {\bibfnamefont {J.}~\bibnamefont {{Aasi}}} \emph {et~al.},\ }\bibfield  {title} {\bibinfo {title} {{Advanced LIGO}},\ }\href {https://doi.org/10.1088/0264-9381/32/7/074001} {\bibfield  {journal} {\bibinfo  {journal} {Classical and Quantum Gravity}\ }\textbf {\bibinfo {volume} {32}},\ \bibinfo {eid} {074001} (\bibinfo {year} {2015})},\ \Eprint {https://arxiv.org/abs/1411.4547} {arXiv:1411.4547 [gr-qc]} \BibitemShut {NoStop}%
\bibitem [{\citenamefont {{Acernese}}\ \emph {et~al.}(2015)\citenamefont {{Acernese}} \emph {et~al.}}]{2015AdvancedVirgo}%
  \BibitemOpen
  \bibfield  {author} {\bibinfo {author} {\bibfnamefont {F.}~\bibnamefont {{Acernese}}} \emph {et~al.},\ }\bibfield  {title} {\bibinfo {title} {{Advanced Virgo: a second-generation interferometric gravitational wave detector}},\ }\href {https://doi.org/10.1088/0264-9381/32/2/024001} {\bibfield  {journal} {\bibinfo  {journal} {Classical and Quantum Gravity}\ }\textbf {\bibinfo {volume} {32}},\ \bibinfo {eid} {024001} (\bibinfo {year} {2015})},\ \Eprint {https://arxiv.org/abs/1408.3978} {arXiv:1408.3978 [gr-qc]} \BibitemShut {NoStop}%
\bibitem [{\citenamefont {{Akutsu}}\ \emph {et~al.}(2021)\citenamefont {{Akutsu}} \emph {et~al.}}]{2021KAGRA}%
  \BibitemOpen
  \bibfield  {author} {\bibinfo {author} {\bibfnamefont {T.}~\bibnamefont {{Akutsu}}} \emph {et~al.},\ }\bibfield  {title} {\bibinfo {title} {{Overview of KAGRA: Detector design and construction history}},\ }\href {https://doi.org/10.1093/ptep/ptaa125} {\bibfield  {journal} {\bibinfo  {journal} {Progress of Theoretical and Experimental Physics}\ }\textbf {\bibinfo {volume} {2021}},\ \bibinfo {eid} {05A101} (\bibinfo {year} {2021})},\ \Eprint {https://arxiv.org/abs/2005.05574} {arXiv:2005.05574 [physics.ins-det]} \BibitemShut {NoStop}%
\bibitem [{\citenamefont {{Abbott}}\ \emph {et~al.}(2023)\citenamefont {{Abbott}} \emph {et~al.}}]{2023GWTC3_cosmo}%
  \BibitemOpen
  \bibfield  {author} {\bibinfo {author} {\bibfnamefont {R.}~\bibnamefont {{Abbott}}} \emph {et~al.},\ }\bibfield  {title} {\bibinfo {title} {{Constraints on the Cosmic Expansion History from GWTC{\textendash}3}},\ }\href {https://doi.org/10.3847/1538-4357/ac74bb} {\bibfield  {journal} {\bibinfo  {journal} {\apj}\ }\textbf {\bibinfo {volume} {949}},\ \bibinfo {eid} {76} (\bibinfo {year} {2023})},\ \Eprint {https://arxiv.org/abs/2111.03604} {arXiv:2111.03604 [astro-ph.CO]} \BibitemShut {NoStop}%
\bibitem [{\citenamefont {{Farr}}\ \emph {et~al.}(2019)\citenamefont {{Farr}}, \citenamefont {{Fishbach}}, \citenamefont {{Ye}},\ and\ \citenamefont {{Holz}}}]{2019Farr_Fishbach+}%
  \BibitemOpen
  \bibfield  {author} {\bibinfo {author} {\bibfnamefont {W.~M.}\ \bibnamefont {{Farr}}}, \bibinfo {author} {\bibfnamefont {M.}~\bibnamefont {{Fishbach}}}, \bibinfo {author} {\bibfnamefont {J.}~\bibnamefont {{Ye}}},\ and\ \bibinfo {author} {\bibfnamefont {D.~E.}\ \bibnamefont {{Holz}}},\ }\bibfield  {title} {\bibinfo {title} {{A Future Percent-level Measurement of the Hubble Expansion at Redshift 0.8 with Advanced LIGO}},\ }\href {https://doi.org/10.3847/2041-8213/ab4284} {\bibfield  {journal} {\bibinfo  {journal} {\apjl}\ }\textbf {\bibinfo {volume} {883}},\ \bibinfo {eid} {L42} (\bibinfo {year} {2019})},\ \Eprint {https://arxiv.org/abs/1908.09084} {arXiv:1908.09084 [astro-ph.CO]} \BibitemShut {NoStop}%
\bibitem [{\citenamefont {{Ezquiaga}}\ and\ \citenamefont {{Holz}}(2022)}]{2022Ezquiag_Holz_Spectral}%
  \BibitemOpen
  \bibfield  {author} {\bibinfo {author} {\bibfnamefont {J.~M.}\ \bibnamefont {{Ezquiaga}}}\ and\ \bibinfo {author} {\bibfnamefont {D.~E.}\ \bibnamefont {{Holz}}},\ }\bibfield  {title} {\bibinfo {title} {{Spectral Sirens: Cosmology from the Full Mass Distribution of Compact Binaries}},\ }\href {https://doi.org/10.1103/PhysRevLett.129.061102} {\bibfield  {journal} {\bibinfo  {journal} {\prl}\ }\textbf {\bibinfo {volume} {129}},\ \bibinfo {eid} {061102} (\bibinfo {year} {2022})},\ \Eprint {https://arxiv.org/abs/2202.08240} {arXiv:2202.08240 [astro-ph.CO]} \BibitemShut {NoStop}%
\bibitem [{\citenamefont {{Messenger}}\ and\ \citenamefont {{Read}}(2012)}]{Messenger&Read2012}%
  \BibitemOpen
  \bibfield  {author} {\bibinfo {author} {\bibfnamefont {C.}~\bibnamefont {{Messenger}}}\ and\ \bibinfo {author} {\bibfnamefont {J.}~\bibnamefont {{Read}}},\ }\bibfield  {title} {\bibinfo {title} {{Measuring a Cosmological Distance-Redshift Relationship Using Only Gravitational Wave Observations of Binary Neutron Star Coalescences}},\ }\href {https://doi.org/10.1103/PhysRevLett.108.091101} {\bibfield  {journal} {\bibinfo  {journal} {\prl}\ }\textbf {\bibinfo {volume} {108}},\ \bibinfo {eid} {091101} (\bibinfo {year} {2012})},\ \Eprint {https://arxiv.org/abs/1107.5725} {arXiv:1107.5725 [gr-qc]} \BibitemShut {NoStop}%
\bibitem [{\citenamefont {{Abbott}}\ \emph {et~al.}(2020)\citenamefont {{Abbott}} \emph {et~al.}}]{AnzacDay}%
  \BibitemOpen
  \bibfield  {author} {\bibinfo {author} {\bibfnamefont {B.~P.}\ \bibnamefont {{Abbott}}} \emph {et~al.},\ }\bibfield  {title} {\bibinfo {title} {{GW190425: Observation of a Compact Binary Coalescence with Total Mass {\ensuremath{\sim}} 3.4 M$_{{\ensuremath{\odot}}}$}},\ }\href {https://doi.org/10.3847/2041-8213/ab75f5} {\bibfield  {journal} {\bibinfo  {journal} {\apjl}\ }\textbf {\bibinfo {volume} {892}},\ \bibinfo {eid} {L3} (\bibinfo {year} {2020})},\ \Eprint {https://arxiv.org/abs/2001.01761} {arXiv:2001.01761 [astro-ph.HE]} \BibitemShut {NoStop}%
\bibitem [{\citenamefont {{Abbott}}\ \emph {et~al.}(2021)\citenamefont {{Abbott}} \emph {et~al.}}]{2021NSBH}%
  \BibitemOpen
  \bibfield  {author} {\bibinfo {author} {\bibfnamefont {R.}~\bibnamefont {{Abbott}}} \emph {et~al.},\ }\bibfield  {title} {\bibinfo {title} {{Observation of Gravitational Waves from Two Neutron Star-Black Hole Coalescences}},\ }\href {https://doi.org/10.3847/2041-8213/ac082e} {\bibfield  {journal} {\bibinfo  {journal} {\apjl}\ }\textbf {\bibinfo {volume} {915}},\ \bibinfo {eid} {L5} (\bibinfo {year} {2021})},\ \Eprint {https://arxiv.org/abs/2106.15163} {arXiv:2106.15163 [astro-ph.HE]} \BibitemShut {NoStop}%
\bibitem [{\citenamefont {{Abac}}\ \emph {et~al.}(2024)\citenamefont {{Abac}} \emph {et~al.}}]{GW230529}%
  \BibitemOpen
  \bibfield  {author} {\bibinfo {author} {\bibfnamefont {A.~G.}\ \bibnamefont {{Abac}}} \emph {et~al.},\ }\bibfield  {title} {\bibinfo {title} {{Observation of Gravitational Waves from the Coalescence of a 2.5{\textendash}4.5 M $_{{\ensuremath{\odot}}}$ Compact Object and a Neutron Star}},\ }\href {https://doi.org/10.3847/2041-8213/ad5beb} {\bibfield  {journal} {\bibinfo  {journal} {\apjl}\ }\textbf {\bibinfo {volume} {970}},\ \bibinfo {eid} {L34} (\bibinfo {year} {2024})},\ \Eprint {https://arxiv.org/abs/2404.04248} {arXiv:2404.04248 [astro-ph.HE]} \BibitemShut {NoStop}%
\bibitem [{\citenamefont {{Del Pozzo}}\ \emph {et~al.}(2013)\citenamefont {{Del Pozzo}}, \citenamefont {{Li}}, \citenamefont {{Agathos}}, \citenamefont {{Van Den Broeck}},\ and\ \citenamefont {{Vitale}}}]{2013DelPozzo}%
  \BibitemOpen
  \bibfield  {author} {\bibinfo {author} {\bibfnamefont {W.}~\bibnamefont {{Del Pozzo}}}, \bibinfo {author} {\bibfnamefont {T.~G.~F.}\ \bibnamefont {{Li}}}, \bibinfo {author} {\bibfnamefont {M.}~\bibnamefont {{Agathos}}}, \bibinfo {author} {\bibfnamefont {C.}~\bibnamefont {{Van Den Broeck}}},\ and\ \bibinfo {author} {\bibfnamefont {S.}~\bibnamefont {{Vitale}}},\ }\bibfield  {title} {\bibinfo {title} {{Demonstrating the Feasibility of Probing the Neutron-Star Equation of State with Second-Generation Gravitational-Wave Detectors}},\ }\href {https://doi.org/10.1103/PhysRevLett.111.071101} {\bibfield  {journal} {\bibinfo  {journal} {\prl}\ }\textbf {\bibinfo {volume} {111}},\ \bibinfo {eid} {071101} (\bibinfo {year} {2013})},\ \Eprint {https://arxiv.org/abs/1307.8338} {arXiv:1307.8338 [gr-qc]} \BibitemShut {NoStop}%
\bibitem [{\citenamefont {{Lackey}}\ and\ \citenamefont {{Wade}}(2015)}]{2015LackeyWade}%
  \BibitemOpen
  \bibfield  {author} {\bibinfo {author} {\bibfnamefont {B.~D.}\ \bibnamefont {{Lackey}}}\ and\ \bibinfo {author} {\bibfnamefont {L.}~\bibnamefont {{Wade}}},\ }\bibfield  {title} {\bibinfo {title} {{Reconstructing the neutron-star equation of state with gravitational-wave detectors from a realistic population of inspiralling binary neutron stars}},\ }\href {https://doi.org/10.1103/PhysRevD.91.043002} {\bibfield  {journal} {\bibinfo  {journal} {\prd}\ }\textbf {\bibinfo {volume} {91}},\ \bibinfo {eid} {043002} (\bibinfo {year} {2015})},\ \Eprint {https://arxiv.org/abs/1410.8866} {arXiv:1410.8866 [gr-qc]} \BibitemShut {NoStop}%
\bibitem [{\citenamefont {{Agathos}}\ \emph {et~al.}(2015)\citenamefont {{Agathos}}, \citenamefont {{Meidam}}, \citenamefont {{Del Pozzo}}, \citenamefont {{Li}}, \citenamefont {{Tompitak}}, \citenamefont {{Veitch}}, \citenamefont {{Vitale}},\ and\ \citenamefont {{Van Den Broeck}}}]{2015Agathos}%
  \BibitemOpen
  \bibfield  {author} {\bibinfo {author} {\bibfnamefont {M.}~\bibnamefont {{Agathos}}}, \bibinfo {author} {\bibfnamefont {J.}~\bibnamefont {{Meidam}}}, \bibinfo {author} {\bibfnamefont {W.}~\bibnamefont {{Del Pozzo}}}, \bibinfo {author} {\bibfnamefont {T.~G.~F.}\ \bibnamefont {{Li}}}, \bibinfo {author} {\bibfnamefont {M.}~\bibnamefont {{Tompitak}}}, \bibinfo {author} {\bibfnamefont {J.}~\bibnamefont {{Veitch}}}, \bibinfo {author} {\bibfnamefont {S.}~\bibnamefont {{Vitale}}},\ and\ \bibinfo {author} {\bibfnamefont {C.}~\bibnamefont {{Van Den Broeck}}},\ }\bibfield  {title} {\bibinfo {title} {{Constraining the neutron star equation of state with gravitational wave signals from coalescing binary neutron stars}},\ }\href {https://doi.org/10.1103/PhysRevD.92.023012} {\bibfield  {journal} {\bibinfo  {journal} {\prd}\ }\textbf {\bibinfo {volume} {92}},\ \bibinfo {eid} {023012} (\bibinfo {year} {2015})},\ \Eprint {https://arxiv.org/abs/1503.05405} {arXiv:1503.05405 [gr-qc]} \BibitemShut {NoStop}%
\bibitem [{\citenamefont {{Hernandez Vivanco}}\ \emph {et~al.}(2019)\citenamefont {{Hernandez Vivanco}}, \citenamefont {{Smith}}, \citenamefont {{Thrane}}, \citenamefont {{Lasky}}, \citenamefont {{Talbot}},\ and\ \citenamefont {{Raymond}}}]{2019Hernandez}%
  \BibitemOpen
  \bibfield  {author} {\bibinfo {author} {\bibfnamefont {F.}~\bibnamefont {{Hernandez Vivanco}}}, \bibinfo {author} {\bibfnamefont {R.}~\bibnamefont {{Smith}}}, \bibinfo {author} {\bibfnamefont {E.}~\bibnamefont {{Thrane}}}, \bibinfo {author} {\bibfnamefont {P.~D.}\ \bibnamefont {{Lasky}}}, \bibinfo {author} {\bibfnamefont {C.}~\bibnamefont {{Talbot}}},\ and\ \bibinfo {author} {\bibfnamefont {V.}~\bibnamefont {{Raymond}}},\ }\bibfield  {title} {\bibinfo {title} {{Measuring the neutron star equation of state with gravitational waves: The first forty binary neutron star merger observations}},\ }\href {https://doi.org/10.1103/PhysRevD.100.103009} {\bibfield  {journal} {\bibinfo  {journal} {\prd}\ }\textbf {\bibinfo {volume} {100}},\ \bibinfo {eid} {103009} (\bibinfo {year} {2019})},\ \Eprint {https://arxiv.org/abs/1909.02698} {arXiv:1909.02698 [gr-qc]} \BibitemShut {NoStop}%
\bibitem [{\citenamefont {{Golomb}}\ and\ \citenamefont {{Talbot}}(2022)}]{2022AGolomb_Talbot}%
  \BibitemOpen
  \bibfield  {author} {\bibinfo {author} {\bibfnamefont {J.}~\bibnamefont {{Golomb}}}\ and\ \bibinfo {author} {\bibfnamefont {C.}~\bibnamefont {{Talbot}}},\ }\bibfield  {title} {\bibinfo {title} {{Hierarchical Inference of Binary Neutron Star Mass Distribution and Equation of State with Gravitational Waves}},\ }\href {https://doi.org/10.3847/1538-4357/ac43bc} {\bibfield  {journal} {\bibinfo  {journal} {\apj}\ }\textbf {\bibinfo {volume} {926}},\ \bibinfo {eid} {79} (\bibinfo {year} {2022})},\ \Eprint {https://arxiv.org/abs/2106.15745} {arXiv:2106.15745 [astro-ph.HE]} \BibitemShut {NoStop}%
\bibitem [{\citenamefont {{Walker}}\ \emph {et~al.}(2024)\citenamefont {{Walker}}, \citenamefont {{Smith}}, \citenamefont {{Thrane}},\ and\ \citenamefont {{Reardon}}}]{Walker2024}%
  \BibitemOpen
  \bibfield  {author} {\bibinfo {author} {\bibfnamefont {K.}~\bibnamefont {{Walker}}}, \bibinfo {author} {\bibfnamefont {R.}~\bibnamefont {{Smith}}}, \bibinfo {author} {\bibfnamefont {E.}~\bibnamefont {{Thrane}}},\ and\ \bibinfo {author} {\bibfnamefont {D.~J.}\ \bibnamefont {{Reardon}}},\ }\bibfield  {title} {\bibinfo {title} {{Precision constraints on the neutron star equation of state with third-generation gravitational-wave observatories}},\ }\href {https://doi.org/10.48550/arXiv.2401.02604} {\bibfield  {journal} {\bibinfo  {journal} {arXiv e-prints}\ ,\ \bibinfo {eid} {arXiv:2401.02604}} (\bibinfo {year} {2024})},\ \Eprint {https://arxiv.org/abs/2401.02604} {arXiv:2401.02604 [astro-ph.HE]} \BibitemShut {NoStop}%
\bibitem [{\citenamefont {{Chatterjee}}\ \emph {et~al.}(2021)\citenamefont {{Chatterjee}}, \citenamefont {{Hegade K.~R.}}, \citenamefont {{Holder}}, \citenamefont {{Holz}}, \citenamefont {{Perkins}}, \citenamefont {{Yagi}},\ and\ \citenamefont {{Yunes}}}]{2021Chaterjee}%
  \BibitemOpen
  \bibfield  {author} {\bibinfo {author} {\bibfnamefont {D.}~\bibnamefont {{Chatterjee}}}, \bibinfo {author} {\bibfnamefont {A.}~\bibnamefont {{Hegade K.~R.}}}, \bibinfo {author} {\bibfnamefont {G.}~\bibnamefont {{Holder}}}, \bibinfo {author} {\bibfnamefont {D.~E.}\ \bibnamefont {{Holz}}}, \bibinfo {author} {\bibfnamefont {S.}~\bibnamefont {{Perkins}}}, \bibinfo {author} {\bibfnamefont {K.}~\bibnamefont {{Yagi}}},\ and\ \bibinfo {author} {\bibfnamefont {N.}~\bibnamefont {{Yunes}}},\ }\bibfield  {title} {\bibinfo {title} {{Cosmology with Love: Measuring the Hubble constant using neutron star universal relations}},\ }\href {https://doi.org/10.1103/PhysRevD.104.083528} {\bibfield  {journal} {\bibinfo  {journal} {\prd}\ }\textbf {\bibinfo {volume} {104}},\ \bibinfo {eid} {083528} (\bibinfo {year} {2021})},\ \Eprint {https://arxiv.org/abs/2106.06589} {arXiv:2106.06589 [gr-qc]} \BibitemShut {NoStop}%
\bibitem [{\citenamefont {{Ghosh}}\ \emph {et~al.}(2022)\citenamefont {{Ghosh}}, \citenamefont {{Biswas}},\ and\ \citenamefont {{Bose}}}]{2022Ghosh}%
  \BibitemOpen
  \bibfield  {author} {\bibinfo {author} {\bibfnamefont {T.}~\bibnamefont {{Ghosh}}}, \bibinfo {author} {\bibfnamefont {B.}~\bibnamefont {{Biswas}}},\ and\ \bibinfo {author} {\bibfnamefont {S.}~\bibnamefont {{Bose}}},\ }\bibfield  {title} {\bibinfo {title} {{Simultaneous inference of neutron star equation of state and the Hubble constant with a population of merging neutron stars}},\ }\href {https://doi.org/10.1103/PhysRevD.106.123529} {\bibfield  {journal} {\bibinfo  {journal} {\prd}\ }\textbf {\bibinfo {volume} {106}},\ \bibinfo {eid} {123529} (\bibinfo {year} {2022})},\ \Eprint {https://arxiv.org/abs/2203.11756} {arXiv:2203.11756 [astro-ph.CO]} \BibitemShut {NoStop}%
\bibitem [{\citenamefont {Ghosh}\ \emph {et~al.}(2024)\citenamefont {Ghosh}, \citenamefont {Biswas}, \citenamefont {Bose},\ and\ \citenamefont {Kapadia}}]{ghosh2024jointinferencepopulationcosmology}%
  \BibitemOpen
  \bibfield  {author} {\bibinfo {author} {\bibfnamefont {T.}~\bibnamefont {Ghosh}}, \bibinfo {author} {\bibfnamefont {B.}~\bibnamefont {Biswas}}, \bibinfo {author} {\bibfnamefont {S.}~\bibnamefont {Bose}},\ and\ \bibinfo {author} {\bibfnamefont {S.~J.}\ \bibnamefont {Kapadia}},\ }\href {https://arxiv.org/abs/2407.16669} {\bibinfo {title} {Joint inference of population, cosmology, and neutron star equation of state from gravitational waves of dark binary neutron stars}} (\bibinfo {year} {2024}),\ \Eprint {https://arxiv.org/abs/2407.16669} {arXiv:2407.16669 [gr-qc]} \BibitemShut {NoStop}%
\bibitem [{\citenamefont {{Soma}}\ \emph {et~al.}(2022)\citenamefont {{Soma}}, \citenamefont {{Wang}}, \citenamefont {{Shi}}, \citenamefont {{St{\"o}cker}},\ and\ \citenamefont {{Zhou}}}]{Soma2022}%
  \BibitemOpen
  \bibfield  {author} {\bibinfo {author} {\bibfnamefont {S.}~\bibnamefont {{Soma}}}, \bibinfo {author} {\bibfnamefont {L.}~\bibnamefont {{Wang}}}, \bibinfo {author} {\bibfnamefont {S.}~\bibnamefont {{Shi}}}, \bibinfo {author} {\bibfnamefont {H.}~\bibnamefont {{St{\"o}cker}}},\ and\ \bibinfo {author} {\bibfnamefont {K.}~\bibnamefont {{Zhou}}},\ }\bibfield  {title} {\bibinfo {title} {{Neural network reconstruction of the dense matter equation of state from neutron star observables}},\ }\href {https://doi.org/10.1088/1475-7516/2022/08/071} {\bibfield  {journal} {\bibinfo  {journal} {\jcap}\ }\textbf {\bibinfo {volume} {2022}},\ \bibinfo {eid} {071} (\bibinfo {year} {2022})},\ \Eprint {https://arxiv.org/abs/2201.01756} {arXiv:2201.01756 [hep-ph]} \BibitemShut {NoStop}%
\bibitem [{\citenamefont {{Brandes}}\ \emph {et~al.}(2024)\citenamefont {{Brandes}}, \citenamefont {{Modi}}, \citenamefont {{Ghosh}}, \citenamefont {{Farrell}}, \citenamefont {{Lindblom}}, \citenamefont {{Heinrich}}, \citenamefont {{Steiner}}, \citenamefont {{Weber}},\ and\ \citenamefont {{Whiteson}}}]{2024Brandes}%
  \BibitemOpen
  \bibfield  {author} {\bibinfo {author} {\bibfnamefont {L.}~\bibnamefont {{Brandes}}}, \bibinfo {author} {\bibfnamefont {C.}~\bibnamefont {{Modi}}}, \bibinfo {author} {\bibfnamefont {A.}~\bibnamefont {{Ghosh}}}, \bibinfo {author} {\bibfnamefont {D.}~\bibnamefont {{Farrell}}}, \bibinfo {author} {\bibfnamefont {L.}~\bibnamefont {{Lindblom}}}, \bibinfo {author} {\bibfnamefont {L.}~\bibnamefont {{Heinrich}}}, \bibinfo {author} {\bibfnamefont {A.~W.}\ \bibnamefont {{Steiner}}}, \bibinfo {author} {\bibfnamefont {F.}~\bibnamefont {{Weber}}},\ and\ \bibinfo {author} {\bibfnamefont {D.}~\bibnamefont {{Whiteson}}},\ }\bibfield  {title} {\bibinfo {title} {{Neural Simulation-Based Inference of the Neutron Star Equation of State directly from Telescope Spectra}},\ }\href {https://doi.org/10.48550/arXiv.2403.00287} {\bibfield  {journal} {\bibinfo  {journal} {arXiv e-prints}\ ,\ \bibinfo {eid} {arXiv:2403.00287}} (\bibinfo {year} {2024})},\ \Eprint {https://arxiv.org/abs/2403.00287} {arXiv:2403.00287 [astro-ph.HE]}
  \BibitemShut {NoStop}%
\bibitem [{\citenamefont {{McGinn}}\ \emph {et~al.}(2024)\citenamefont {{McGinn}}, \citenamefont {{Mukherjee}}, \citenamefont {{Irwin}}, \citenamefont {{Messenger}}, \citenamefont {{Williams}},\ and\ \citenamefont {{Heng}}}]{2024McGinn}%
  \BibitemOpen
  \bibfield  {author} {\bibinfo {author} {\bibfnamefont {J.}~\bibnamefont {{McGinn}}}, \bibinfo {author} {\bibfnamefont {A.}~\bibnamefont {{Mukherjee}}}, \bibinfo {author} {\bibfnamefont {J.}~\bibnamefont {{Irwin}}}, \bibinfo {author} {\bibfnamefont {C.}~\bibnamefont {{Messenger}}}, \bibinfo {author} {\bibfnamefont {M.~J.}\ \bibnamefont {{Williams}}},\ and\ \bibinfo {author} {\bibfnamefont {I.~S.}\ \bibnamefont {{Heng}}},\ }\bibfield  {title} {\bibinfo {title} {{Rapid neutron star equation of state inference with Normalising Flows}},\ }\href {https://doi.org/10.48550/arXiv.2403.17462} {\bibfield  {journal} {\bibinfo  {journal} {arXiv e-prints}\ ,\ \bibinfo {eid} {arXiv:2403.17462}} (\bibinfo {year} {2024})},\ \Eprint {https://arxiv.org/abs/2403.17462} {arXiv:2403.17462 [gr-qc]} \BibitemShut {NoStop}%
\bibitem [{\citenamefont {{Tiwari}}\ and\ \citenamefont {{Pai}}(2024)}]{2024Tiwari}%
  \BibitemOpen
  \bibfield  {author} {\bibinfo {author} {\bibfnamefont {P.}~\bibnamefont {{Tiwari}}}\ and\ \bibinfo {author} {\bibfnamefont {A.}~\bibnamefont {{Pai}}},\ }\bibfield  {title} {\bibinfo {title} {{Deep TOV to characterize Neutron Stars}},\ }\href {https://doi.org/10.48550/arXiv.2405.08163} {\bibfield  {journal} {\bibinfo  {journal} {arXiv e-prints}\ ,\ \bibinfo {eid} {arXiv:2405.08163}} (\bibinfo {year} {2024})},\ \Eprint {https://arxiv.org/abs/2405.08163} {arXiv:2405.08163 [astro-ph.HE]} \BibitemShut {NoStop}%
\bibitem [{\citenamefont {{Reed}}\ \emph {et~al.}(2024)\citenamefont {{Reed}}, \citenamefont {{Somasundaram}}, \citenamefont {{De}}, \citenamefont {{Armstrong}}, \citenamefont {{Giuliani}}, \citenamefont {{Capano}}, \citenamefont {{Brown}},\ and\ \citenamefont {{Tews}}}]{2024Reed+}%
  \BibitemOpen
  \bibfield  {author} {\bibinfo {author} {\bibfnamefont {B.~T.}\ \bibnamefont {{Reed}}}, \bibinfo {author} {\bibfnamefont {R.}~\bibnamefont {{Somasundaram}}}, \bibinfo {author} {\bibfnamefont {S.}~\bibnamefont {{De}}}, \bibinfo {author} {\bibfnamefont {C.~L.}\ \bibnamefont {{Armstrong}}}, \bibinfo {author} {\bibfnamefont {P.}~\bibnamefont {{Giuliani}}}, \bibinfo {author} {\bibfnamefont {C.}~\bibnamefont {{Capano}}}, \bibinfo {author} {\bibfnamefont {D.~A.}\ \bibnamefont {{Brown}}},\ and\ \bibinfo {author} {\bibfnamefont {I.}~\bibnamefont {{Tews}}},\ }\bibfield  {title} {\bibinfo {title} {{Towards accelerated nuclear-physics parameter estimation from binary neutron star mergers: Emulators for the Tolman-Oppenheimer-Volkoff equations}},\ }\href {https://doi.org/10.48550/arXiv.2405.20558} {\bibfield  {journal} {\bibinfo  {journal} {arXiv e-prints}\ ,\ \bibinfo {eid} {arXiv:2405.20558}} (\bibinfo {year} {2024})},\ \Eprint {https://arxiv.org/abs/2405.20558} {arXiv:2405.20558 [astro-ph.HE]} \BibitemShut {NoStop}%
\bibitem [{\citenamefont {{Ashton}}\ \emph {et~al.}(2019)\citenamefont {{Ashton}}, \citenamefont {{H{\"u}bner}}, \citenamefont {{Lasky}}, \citenamefont {{Talbot}}, \citenamefont {{Ackley}}, \citenamefont {{Biscoveanu}}, \citenamefont {{Chu}}, \citenamefont {{Divakarla}}, \citenamefont {{Easter}}, \citenamefont {{Goncharov}}, \citenamefont {{Hernandez Vivanco}}, \citenamefont {{Harms}}, \citenamefont {{Lower}}, \citenamefont {{Meadors}}, \citenamefont {{Melchor}}, \citenamefont {{Payne}}, \citenamefont {{Pitkin}}, \citenamefont {{Powell}}, \citenamefont {{Sarin}}, \citenamefont {{Smith}},\ and\ \citenamefont {{Thrane}}}]{Bilby_Ashton}%
  \BibitemOpen
  \bibfield  {author} {\bibinfo {author} {\bibfnamefont {G.}~\bibnamefont {{Ashton}}}, \bibinfo {author} {\bibfnamefont {M.}~\bibnamefont {{H{\"u}bner}}}, \bibinfo {author} {\bibfnamefont {P.~D.}\ \bibnamefont {{Lasky}}}, \bibinfo {author} {\bibfnamefont {C.}~\bibnamefont {{Talbot}}}, \bibinfo {author} {\bibfnamefont {K.}~\bibnamefont {{Ackley}}}, \bibinfo {author} {\bibfnamefont {S.}~\bibnamefont {{Biscoveanu}}}, \bibinfo {author} {\bibfnamefont {Q.}~\bibnamefont {{Chu}}}, \bibinfo {author} {\bibfnamefont {A.}~\bibnamefont {{Divakarla}}}, \bibinfo {author} {\bibfnamefont {P.~J.}\ \bibnamefont {{Easter}}}, \bibinfo {author} {\bibfnamefont {B.}~\bibnamefont {{Goncharov}}}, \bibinfo {author} {\bibfnamefont {F.}~\bibnamefont {{Hernandez Vivanco}}}, \bibinfo {author} {\bibfnamefont {J.}~\bibnamefont {{Harms}}}, \bibinfo {author} {\bibfnamefont {M.~E.}\ \bibnamefont {{Lower}}}, \bibinfo {author} {\bibfnamefont {G.~D.}\ \bibnamefont {{Meadors}}}, \bibinfo {author} {\bibfnamefont {D.}~\bibnamefont {{Melchor}}},
  \bibinfo {author} {\bibfnamefont {E.}~\bibnamefont {{Payne}}}, \bibinfo {author} {\bibfnamefont {M.~D.}\ \bibnamefont {{Pitkin}}}, \bibinfo {author} {\bibfnamefont {J.}~\bibnamefont {{Powell}}}, \bibinfo {author} {\bibfnamefont {N.}~\bibnamefont {{Sarin}}}, \bibinfo {author} {\bibfnamefont {R.~J.~E.}\ \bibnamefont {{Smith}}},\ and\ \bibinfo {author} {\bibfnamefont {E.}~\bibnamefont {{Thrane}}},\ }\bibfield  {title} {\bibinfo {title} {{BILBY: A User-friendly Bayesian Inference Library for Gravitational-wave Astronomy}},\ }\href {https://doi.org/10.3847/1538-4365/ab06fc} {\bibfield  {journal} {\bibinfo  {journal} {\apjs}\ }\textbf {\bibinfo {volume} {241}},\ \bibinfo {eid} {27} (\bibinfo {year} {2019})},\ \Eprint {https://arxiv.org/abs/1811.02042} {arXiv:1811.02042 [astro-ph.IM]} \BibitemShut {NoStop}%
\bibitem [{\citenamefont {{Romero-Shaw}}\ \emph {et~al.}(2020)\citenamefont {{Romero-Shaw}}, \citenamefont {{Talbot}}, \citenamefont {{Biscoveanu}}, \citenamefont {{D'Emilio}}, \citenamefont {{Ashton}}, \citenamefont {{Berry}}, \citenamefont {{Coughlin}}, \citenamefont {{Galaudage}}, \citenamefont {{Hoy}}, \citenamefont {{H{\"u}bner}}, \citenamefont {{Phukon}}, \citenamefont {{Pitkin}}, \citenamefont {{Rizzo}}, \citenamefont {{Sarin}}, \citenamefont {{Smith}}, \citenamefont {{Stevenson}}, \citenamefont {{Vajpeyi}}, \citenamefont {{Ar{\`e}ne}}, \citenamefont {{Athar}}, \citenamefont {{Banagiri}}, \citenamefont {{Bose}}, \citenamefont {{Carney}}, \citenamefont {{Chatziioannou}}, \citenamefont {{Clark}}, \citenamefont {{Colleoni}}, \citenamefont {{Cotesta}}, \citenamefont {{Edelman}}, \citenamefont {{Estell{\'e}s}}, \citenamefont {{Garc{\'\i}a-Quir{\'o}s}}, \citenamefont {{Ghosh}}, \citenamefont {{Green}}, \citenamefont {{Haster}}, \citenamefont {{Husa}}, \citenamefont {{Keitel}}, \citenamefont {{Kim}},
  \citenamefont {{Hernandez-Vivanco}}, \citenamefont {{Maga{\~n}a Hernandez}}, \citenamefont {{Karathanasis}}, \citenamefont {{Lasky}}, \citenamefont {{De Lillo}}, \citenamefont {{Lower}}, \citenamefont {{Macleod}}, \citenamefont {{Mateu-Lucena}}, \citenamefont {{Miller}}, \citenamefont {{Millhouse}}, \citenamefont {{Morisaki}}, \citenamefont {{Oh}}, \citenamefont {{Ossokine}}, \citenamefont {{Payne}}, \citenamefont {{Powell}}, \citenamefont {{Pratten}}, \citenamefont {{P{\"u}rrer}}, \citenamefont {{Ramos-Buades}}, \citenamefont {{Raymond}}, \citenamefont {{Thrane}}, \citenamefont {{Veitch}}, \citenamefont {{Williams}}, \citenamefont {{Williams}},\ and\ \citenamefont {{Xiao}}}]{Bilby_Romero-Shaw}%
  \BibitemOpen
  \bibfield  {author} {\bibinfo {author} {\bibfnamefont {I.~M.}\ \bibnamefont {{Romero-Shaw}}}, \bibinfo {author} {\bibfnamefont {C.}~\bibnamefont {{Talbot}}}, \bibinfo {author} {\bibfnamefont {S.}~\bibnamefont {{Biscoveanu}}}, \bibinfo {author} {\bibfnamefont {V.}~\bibnamefont {{D'Emilio}}}, \bibinfo {author} {\bibfnamefont {G.}~\bibnamefont {{Ashton}}}, \bibinfo {author} {\bibfnamefont {C.~P.~L.}\ \bibnamefont {{Berry}}}, \bibinfo {author} {\bibfnamefont {S.}~\bibnamefont {{Coughlin}}}, \bibinfo {author} {\bibfnamefont {S.}~\bibnamefont {{Galaudage}}}, \bibinfo {author} {\bibfnamefont {C.}~\bibnamefont {{Hoy}}}, \bibinfo {author} {\bibfnamefont {M.}~\bibnamefont {{H{\"u}bner}}}, \bibinfo {author} {\bibfnamefont {K.~S.}\ \bibnamefont {{Phukon}}}, \bibinfo {author} {\bibfnamefont {M.}~\bibnamefont {{Pitkin}}}, \bibinfo {author} {\bibfnamefont {M.}~\bibnamefont {{Rizzo}}}, \bibinfo {author} {\bibfnamefont {N.}~\bibnamefont {{Sarin}}}, \bibinfo {author} {\bibfnamefont {R.}~\bibnamefont {{Smith}}}, \bibinfo
  {author} {\bibfnamefont {S.}~\bibnamefont {{Stevenson}}}, \bibinfo {author} {\bibfnamefont {A.}~\bibnamefont {{Vajpeyi}}}, \bibinfo {author} {\bibfnamefont {M.}~\bibnamefont {{Ar{\`e}ne}}}, \bibinfo {author} {\bibfnamefont {K.}~\bibnamefont {{Athar}}}, \bibinfo {author} {\bibfnamefont {S.}~\bibnamefont {{Banagiri}}}, \bibinfo {author} {\bibfnamefont {N.}~\bibnamefont {{Bose}}}, \bibinfo {author} {\bibfnamefont {M.}~\bibnamefont {{Carney}}}, \bibinfo {author} {\bibfnamefont {K.}~\bibnamefont {{Chatziioannou}}}, \bibinfo {author} {\bibfnamefont {J.~A.}\ \bibnamefont {{Clark}}}, \bibinfo {author} {\bibfnamefont {M.}~\bibnamefont {{Colleoni}}}, \bibinfo {author} {\bibfnamefont {R.}~\bibnamefont {{Cotesta}}}, \bibinfo {author} {\bibfnamefont {B.}~\bibnamefont {{Edelman}}}, \bibinfo {author} {\bibfnamefont {H.}~\bibnamefont {{Estell{\'e}s}}}, \bibinfo {author} {\bibfnamefont {C.}~\bibnamefont {{Garc{\'\i}a-Quir{\'o}s}}}, \bibinfo {author} {\bibfnamefont {A.}~\bibnamefont {{Ghosh}}}, \bibinfo {author}
  {\bibfnamefont {R.}~\bibnamefont {{Green}}}, \bibinfo {author} {\bibfnamefont {C.~J.}\ \bibnamefont {{Haster}}}, \bibinfo {author} {\bibfnamefont {S.}~\bibnamefont {{Husa}}}, \bibinfo {author} {\bibfnamefont {D.}~\bibnamefont {{Keitel}}}, \bibinfo {author} {\bibfnamefont {A.~X.}\ \bibnamefont {{Kim}}}, \bibinfo {author} {\bibfnamefont {F.}~\bibnamefont {{Hernandez-Vivanco}}}, \bibinfo {author} {\bibfnamefont {I.}~\bibnamefont {{Maga{\~n}a Hernandez}}}, \bibinfo {author} {\bibfnamefont {C.}~\bibnamefont {{Karathanasis}}}, \bibinfo {author} {\bibfnamefont {P.~D.}\ \bibnamefont {{Lasky}}}, \bibinfo {author} {\bibfnamefont {N.}~\bibnamefont {{De Lillo}}}, \bibinfo {author} {\bibfnamefont {M.~E.}\ \bibnamefont {{Lower}}}, \bibinfo {author} {\bibfnamefont {D.}~\bibnamefont {{Macleod}}}, \bibinfo {author} {\bibfnamefont {M.}~\bibnamefont {{Mateu-Lucena}}}, \bibinfo {author} {\bibfnamefont {A.}~\bibnamefont {{Miller}}}, \bibinfo {author} {\bibfnamefont {M.}~\bibnamefont {{Millhouse}}}, \bibinfo {author}
  {\bibfnamefont {S.}~\bibnamefont {{Morisaki}}}, \bibinfo {author} {\bibfnamefont {S.~H.}\ \bibnamefont {{Oh}}}, \bibinfo {author} {\bibfnamefont {S.}~\bibnamefont {{Ossokine}}}, \bibinfo {author} {\bibfnamefont {E.}~\bibnamefont {{Payne}}}, \bibinfo {author} {\bibfnamefont {J.}~\bibnamefont {{Powell}}}, \bibinfo {author} {\bibfnamefont {G.}~\bibnamefont {{Pratten}}}, \bibinfo {author} {\bibfnamefont {M.}~\bibnamefont {{P{\"u}rrer}}}, \bibinfo {author} {\bibfnamefont {A.}~\bibnamefont {{Ramos-Buades}}}, \bibinfo {author} {\bibfnamefont {V.}~\bibnamefont {{Raymond}}}, \bibinfo {author} {\bibfnamefont {E.}~\bibnamefont {{Thrane}}}, \bibinfo {author} {\bibfnamefont {J.}~\bibnamefont {{Veitch}}}, \bibinfo {author} {\bibfnamefont {D.}~\bibnamefont {{Williams}}}, \bibinfo {author} {\bibfnamefont {M.~J.}\ \bibnamefont {{Williams}}},\ and\ \bibinfo {author} {\bibfnamefont {L.}~\bibnamefont {{Xiao}}},\ }\bibfield  {title} {\bibinfo {title} {{Bayesian inference for compact binary coalescences with BILBY: validation
  and application to the first LIGO-Virgo gravitational-wave transient catalogue}},\ }\href {https://doi.org/10.1093/mnras/staa2850} {\bibfield  {journal} {\bibinfo  {journal} {\mnras}\ }\textbf {\bibinfo {volume} {499}},\ \bibinfo {pages} {3295} (\bibinfo {year} {2020})},\ \Eprint {https://arxiv.org/abs/2006.00714} {arXiv:2006.00714 [astro-ph.IM]} \BibitemShut {NoStop}%
\bibitem [{\citenamefont {{Smith}}\ \emph {et~al.}(2016)\citenamefont {{Smith}}, \citenamefont {{Field}}, \citenamefont {{Blackburn}}, \citenamefont {{Haster}}, \citenamefont {{P{\"u}rrer}}, \citenamefont {{Raymond}},\ and\ \citenamefont {{Schmidt}}}]{Smith+2016}%
  \BibitemOpen
  \bibfield  {author} {\bibinfo {author} {\bibfnamefont {R.}~\bibnamefont {{Smith}}}, \bibinfo {author} {\bibfnamefont {S.~E.}\ \bibnamefont {{Field}}}, \bibinfo {author} {\bibfnamefont {K.}~\bibnamefont {{Blackburn}}}, \bibinfo {author} {\bibfnamefont {C.-J.}\ \bibnamefont {{Haster}}}, \bibinfo {author} {\bibfnamefont {M.}~\bibnamefont {{P{\"u}rrer}}}, \bibinfo {author} {\bibfnamefont {V.}~\bibnamefont {{Raymond}}},\ and\ \bibinfo {author} {\bibfnamefont {P.}~\bibnamefont {{Schmidt}}},\ }\bibfield  {title} {\bibinfo {title} {{Fast and accurate inference on gravitational waves from precessing compact binaries}},\ }\href {https://doi.org/10.1103/PhysRevD.94.044031} {\bibfield  {journal} {\bibinfo  {journal} {\prd}\ }\textbf {\bibinfo {volume} {94}},\ \bibinfo {eid} {044031} (\bibinfo {year} {2016})},\ \Eprint {https://arxiv.org/abs/1604.08253} {arXiv:1604.08253 [gr-qc]} \BibitemShut {NoStop}%
\bibitem [{\citenamefont {{Read}}\ \emph {et~al.}(2009)\citenamefont {{Read}}, \citenamefont {{Lackey}}, \citenamefont {{Owen}},\ and\ \citenamefont {{Friedman}}}]{Read2009}%
  \BibitemOpen
  \bibfield  {author} {\bibinfo {author} {\bibfnamefont {J.~S.}\ \bibnamefont {{Read}}}, \bibinfo {author} {\bibfnamefont {B.~D.}\ \bibnamefont {{Lackey}}}, \bibinfo {author} {\bibfnamefont {B.~J.}\ \bibnamefont {{Owen}}},\ and\ \bibinfo {author} {\bibfnamefont {J.~L.}\ \bibnamefont {{Friedman}}},\ }\bibfield  {title} {\bibinfo {title} {{Constraints on a phenomenologically parametrized neutron-star equation of state}},\ }\href {https://doi.org/10.1103/PhysRevD.79.124032} {\bibfield  {journal} {\bibinfo  {journal} {\prd}\ }\textbf {\bibinfo {volume} {79}},\ \bibinfo {eid} {124032} (\bibinfo {year} {2009})},\ \Eprint {https://arxiv.org/abs/0812.2163} {arXiv:0812.2163 [astro-ph]} \BibitemShut {NoStop}%
\bibitem [{\citenamefont {{Lindblom}}(2010)}]{Lindbolm2010}%
  \BibitemOpen
  \bibfield  {author} {\bibinfo {author} {\bibfnamefont {L.}~\bibnamefont {{Lindblom}}},\ }\bibfield  {title} {\bibinfo {title} {{Spectral representations of neutron-star equations of state}},\ }\href {https://doi.org/10.1103/PhysRevD.82.103011} {\bibfield  {journal} {\bibinfo  {journal} {\prd}\ }\textbf {\bibinfo {volume} {82}},\ \bibinfo {eid} {103011} (\bibinfo {year} {2010})},\ \Eprint {https://arxiv.org/abs/1009.0738} {arXiv:1009.0738 [astro-ph.HE]} \BibitemShut {NoStop}%
\bibitem [{\citenamefont {{Landry}}\ and\ \citenamefont {{Essick}}(2019)}]{2019Landry}%
  \BibitemOpen
  \bibfield  {author} {\bibinfo {author} {\bibfnamefont {P.}~\bibnamefont {{Landry}}}\ and\ \bibinfo {author} {\bibfnamefont {R.}~\bibnamefont {{Essick}}},\ }\bibfield  {title} {\bibinfo {title} {{Nonparametric inference of the neutron star equation of state from gravitational wave observations}},\ }\href {https://doi.org/10.1103/PhysRevD.99.084049} {\bibfield  {journal} {\bibinfo  {journal} {\prd}\ }\textbf {\bibinfo {volume} {99}},\ \bibinfo {eid} {084049} (\bibinfo {year} {2019})},\ \Eprint {https://arxiv.org/abs/1811.12529} {arXiv:1811.12529 [gr-qc]} \BibitemShut {NoStop}%
\bibitem [{\citenamefont {{Essick}}\ \emph {et~al.}(2020)\citenamefont {{Essick}}, \citenamefont {{Landry}},\ and\ \citenamefont {{Holz}}}]{2020Essick}%
  \BibitemOpen
  \bibfield  {author} {\bibinfo {author} {\bibfnamefont {R.}~\bibnamefont {{Essick}}}, \bibinfo {author} {\bibfnamefont {P.}~\bibnamefont {{Landry}}},\ and\ \bibinfo {author} {\bibfnamefont {D.~E.}\ \bibnamefont {{Holz}}},\ }\bibfield  {title} {\bibinfo {title} {{Nonparametric inference of neutron star composition, equation of state, and maximum mass with GW170817}},\ }\href {https://doi.org/10.1103/PhysRevD.101.063007} {\bibfield  {journal} {\bibinfo  {journal} {\prd}\ }\textbf {\bibinfo {volume} {101}},\ \bibinfo {eid} {063007} (\bibinfo {year} {2020})},\ \Eprint {https://arxiv.org/abs/1910.09740} {arXiv:1910.09740 [astro-ph.HE]} \BibitemShut {NoStop}%
\bibitem [{\citenamefont {{Legred}}\ \emph {et~al.}(2022{\natexlab{a}})\citenamefont {{Legred}}, \citenamefont {{Chatziioannou}}, \citenamefont {{Essick}},\ and\ \citenamefont {{Landry}}}]{2022Legred}%
  \BibitemOpen
  \bibfield  {author} {\bibinfo {author} {\bibfnamefont {I.}~\bibnamefont {{Legred}}}, \bibinfo {author} {\bibfnamefont {K.}~\bibnamefont {{Chatziioannou}}}, \bibinfo {author} {\bibfnamefont {R.}~\bibnamefont {{Essick}}},\ and\ \bibinfo {author} {\bibfnamefont {P.}~\bibnamefont {{Landry}}},\ }\bibfield  {title} {\bibinfo {title} {{Implicit correlations within phenomenological parametric models of the neutron star equation of state}},\ }\href {https://doi.org/10.1103/PhysRevD.105.043016} {\bibfield  {journal} {\bibinfo  {journal} {\prd}\ }\textbf {\bibinfo {volume} {105}},\ \bibinfo {eid} {043016} (\bibinfo {year} {2022}{\natexlab{a}})},\ \Eprint {https://arxiv.org/abs/2201.06791} {arXiv:2201.06791 [astro-ph.HE]} \BibitemShut {NoStop}%
\bibitem [{\citenamefont {{Sarin}}\ \emph {et~al.}(2023)\citenamefont {{Sarin}}, \citenamefont {{Peiris}}, \citenamefont {{Mortlock}}, \citenamefont {{Alsing}}, \citenamefont {{Nissanke}},\ and\ \citenamefont {{Feeney}}}]{Sarin2023}%
  \BibitemOpen
  \bibfield  {author} {\bibinfo {author} {\bibfnamefont {N.}~\bibnamefont {{Sarin}}}, \bibinfo {author} {\bibfnamefont {H.~V.}\ \bibnamefont {{Peiris}}}, \bibinfo {author} {\bibfnamefont {D.~J.}\ \bibnamefont {{Mortlock}}}, \bibinfo {author} {\bibfnamefont {J.}~\bibnamefont {{Alsing}}}, \bibinfo {author} {\bibfnamefont {S.~M.}\ \bibnamefont {{Nissanke}}},\ and\ \bibinfo {author} {\bibfnamefont {S.~M.}\ \bibnamefont {{Feeney}}},\ }\bibfield  {title} {\bibinfo {title} {{Measuring the nuclear equation of state with neutron star-black hole mergers}},\ }\href {https://doi.org/10.48550/arXiv.2311.05689} {\bibfield  {journal} {\bibinfo  {journal} {arXiv e-prints}\ ,\ \bibinfo {eid} {arXiv:2311.05689}} (\bibinfo {year} {2023})},\ \Eprint {https://arxiv.org/abs/2311.05689} {arXiv:2311.05689 [gr-qc]} \BibitemShut {NoStop}%
\bibitem [{\citenamefont {{Raaijmakers}}\ \emph {et~al.}(2018)\citenamefont {{Raaijmakers}}, \citenamefont {{Riley}},\ and\ \citenamefont {{Watts}}}]{Geert2018}%
  \BibitemOpen
  \bibfield  {author} {\bibinfo {author} {\bibfnamefont {G.}~\bibnamefont {{Raaijmakers}}}, \bibinfo {author} {\bibfnamefont {T.~E.}\ \bibnamefont {{Riley}}},\ and\ \bibinfo {author} {\bibfnamefont {A.~L.}\ \bibnamefont {{Watts}}},\ }\bibfield  {title} {\bibinfo {title} {{A pitfall of piecewise-polytropic equation of state inference}},\ }\href {https://doi.org/10.1093/mnras/sty1052} {\bibfield  {journal} {\bibinfo  {journal} {\mnras}\ }\textbf {\bibinfo {volume} {478}},\ \bibinfo {pages} {2177} (\bibinfo {year} {2018})},\ \Eprint {https://arxiv.org/abs/1804.09087} {arXiv:1804.09087 [astro-ph.HE]} \BibitemShut {NoStop}%
\bibitem [{\citenamefont {{Legred}}\ \emph {et~al.}(2022{\natexlab{b}})\citenamefont {{Legred}}, \citenamefont {{Chatziioannou}}, \citenamefont {{Essick}},\ and\ \citenamefont {{Landry}}}]{Legred2022}%
  \BibitemOpen
  \bibfield  {author} {\bibinfo {author} {\bibfnamefont {I.}~\bibnamefont {{Legred}}}, \bibinfo {author} {\bibfnamefont {K.}~\bibnamefont {{Chatziioannou}}}, \bibinfo {author} {\bibfnamefont {R.}~\bibnamefont {{Essick}}},\ and\ \bibinfo {author} {\bibfnamefont {P.}~\bibnamefont {{Landry}}},\ }\bibfield  {title} {\bibinfo {title} {{Implicit correlations within phenomenological parametric models of the neutron star equation of state}},\ }\href {https://doi.org/10.1103/PhysRevD.105.043016} {\bibfield  {journal} {\bibinfo  {journal} {\prd}\ }\textbf {\bibinfo {volume} {105}},\ \bibinfo {eid} {043016} (\bibinfo {year} {2022}{\natexlab{b}})},\ \Eprint {https://arxiv.org/abs/2201.06791} {arXiv:2201.06791 [astro-ph.HE]} \BibitemShut {NoStop}%
\bibitem [{\citenamefont {{Douchin}}\ and\ \citenamefont {{Haensel}}(2001{\natexlab{a}})}]{SLY2001}%
  \BibitemOpen
  \bibfield  {author} {\bibinfo {author} {\bibfnamefont {F.}~\bibnamefont {{Douchin}}}\ and\ \bibinfo {author} {\bibfnamefont {P.}~\bibnamefont {{Haensel}}},\ }\bibfield  {title} {\bibinfo {title} {{A unified equation of state of dense matter and neutron star structure}},\ }\href {https://doi.org/10.1051/0004-6361:20011402} {\bibfield  {journal} {\bibinfo  {journal} {\aap}\ }\textbf {\bibinfo {volume} {380}},\ \bibinfo {pages} {151} (\bibinfo {year} {2001}{\natexlab{a}})},\ \Eprint {https://arxiv.org/abs/astro-ph/0111092} {arXiv:astro-ph/0111092 [astro-ph]} \BibitemShut {NoStop}%
\bibitem [{\citenamefont {{Haensel}}\ and\ \citenamefont {{Potekhin}}(2004)}]{SLY2004}%
  \BibitemOpen
  \bibfield  {author} {\bibinfo {author} {\bibfnamefont {P.}~\bibnamefont {{Haensel}}}\ and\ \bibinfo {author} {\bibfnamefont {A.~Y.}\ \bibnamefont {{Potekhin}}},\ }\bibfield  {title} {\bibinfo {title} {{Analytical representations of unified equations of state of neutron-star matter}},\ }\href {https://doi.org/10.1051/0004-6361:20041722} {\bibfield  {journal} {\bibinfo  {journal} {\aap}\ }\textbf {\bibinfo {volume} {428}},\ \bibinfo {pages} {191} (\bibinfo {year} {2004})},\ \Eprint {https://arxiv.org/abs/astro-ph/0408324} {arXiv:astro-ph/0408324 [astro-ph]} \BibitemShut {NoStop}%
\bibitem [{\citenamefont {{Antoniadis}}\ \emph {et~al.}(2013)\citenamefont {{Antoniadis}}, \citenamefont {{Freire}}, \citenamefont {{Wex}}, \citenamefont {{Tauris}}, \citenamefont {{Lynch}}, \citenamefont {{van Kerkwijk}}, \citenamefont {{Kramer}}, \citenamefont {{Bassa}}, \citenamefont {{Dhillon}}, \citenamefont {{Driebe}}, \citenamefont {{Hessels}}, \citenamefont {{Kaspi}}, \citenamefont {{Kondratiev}}, \citenamefont {{Langer}}, \citenamefont {{Marsh}}, \citenamefont {{McLaughlin}}, \citenamefont {{Pennucci}}, \citenamefont {{Ransom}}, \citenamefont {{Stairs}}, \citenamefont {{van Leeuwen}}, \citenamefont {{Verbiest}},\ and\ \citenamefont {{Whelan}}}]{J0384+0432}%
  \BibitemOpen
  \bibfield  {author} {\bibinfo {author} {\bibfnamefont {J.}~\bibnamefont {{Antoniadis}}}, \bibinfo {author} {\bibfnamefont {P.~C.~C.}\ \bibnamefont {{Freire}}}, \bibinfo {author} {\bibfnamefont {N.}~\bibnamefont {{Wex}}}, \bibinfo {author} {\bibfnamefont {T.~M.}\ \bibnamefont {{Tauris}}}, \bibinfo {author} {\bibfnamefont {R.~S.}\ \bibnamefont {{Lynch}}}, \bibinfo {author} {\bibfnamefont {M.~H.}\ \bibnamefont {{van Kerkwijk}}}, \bibinfo {author} {\bibfnamefont {M.}~\bibnamefont {{Kramer}}}, \bibinfo {author} {\bibfnamefont {C.}~\bibnamefont {{Bassa}}}, \bibinfo {author} {\bibfnamefont {V.~S.}\ \bibnamefont {{Dhillon}}}, \bibinfo {author} {\bibfnamefont {T.}~\bibnamefont {{Driebe}}}, \bibinfo {author} {\bibfnamefont {J.~W.~T.}\ \bibnamefont {{Hessels}}}, \bibinfo {author} {\bibfnamefont {V.~M.}\ \bibnamefont {{Kaspi}}}, \bibinfo {author} {\bibfnamefont {V.~I.}\ \bibnamefont {{Kondratiev}}}, \bibinfo {author} {\bibfnamefont {N.}~\bibnamefont {{Langer}}}, \bibinfo {author} {\bibfnamefont {T.~R.}\ \bibnamefont
  {{Marsh}}}, \bibinfo {author} {\bibfnamefont {M.~A.}\ \bibnamefont {{McLaughlin}}}, \bibinfo {author} {\bibfnamefont {T.~T.}\ \bibnamefont {{Pennucci}}}, \bibinfo {author} {\bibfnamefont {S.~M.}\ \bibnamefont {{Ransom}}}, \bibinfo {author} {\bibfnamefont {I.~H.}\ \bibnamefont {{Stairs}}}, \bibinfo {author} {\bibfnamefont {J.}~\bibnamefont {{van Leeuwen}}}, \bibinfo {author} {\bibfnamefont {J.~P.~W.}\ \bibnamefont {{Verbiest}}},\ and\ \bibinfo {author} {\bibfnamefont {D.~G.}\ \bibnamefont {{Whelan}}},\ }\bibfield  {title} {\bibinfo {title} {{A Massive Pulsar in a Compact Relativistic Binary}},\ }\href {https://doi.org/10.1126/science.1233232} {\bibfield  {journal} {\bibinfo  {journal} {Science}\ }\textbf {\bibinfo {volume} {340}},\ \bibinfo {pages} {448} (\bibinfo {year} {2013})},\ \Eprint {https://arxiv.org/abs/1304.6875} {arXiv:1304.6875 [astro-ph.HE]} \BibitemShut {NoStop}%
\bibitem [{\citenamefont {{LIGO Scientific Collaboration}}\ \emph {et~al.}(2018)\citenamefont {{LIGO Scientific Collaboration}}, \citenamefont {{Virgo Collaboration}},\ and\ \citenamefont {{KAGRA Collaboration}}}]{lalsuite}%
  \BibitemOpen
  \bibfield  {author} {\bibinfo {author} {\bibnamefont {{LIGO Scientific Collaboration}}}, \bibinfo {author} {\bibnamefont {{Virgo Collaboration}}},\ and\ \bibinfo {author} {\bibnamefont {{KAGRA Collaboration}}},\ }\href {https://doi.org/10.7935/GT1W-FZ16} {\bibinfo {title} {{LVK} {A}lgorithm {L}ibrary - {LALS}uite}},\ \bibinfo {howpublished} {Free software (GPL)} (\bibinfo {year} {2018})\BibitemShut {NoStop}%
\bibitem [{\citenamefont {{Hinderer}}(2008)}]{2008Hinderer}%
  \BibitemOpen
  \bibfield  {author} {\bibinfo {author} {\bibfnamefont {T.}~\bibnamefont {{Hinderer}}},\ }\bibfield  {title} {\bibinfo {title} {{Tidal Love Numbers of Neutron Stars}},\ }\href {https://doi.org/10.1086/533487} {\bibfield  {journal} {\bibinfo  {journal} {\apj}\ }\textbf {\bibinfo {volume} {677}},\ \bibinfo {pages} {1216} (\bibinfo {year} {2008})},\ \Eprint {https://arxiv.org/abs/0711.2420} {arXiv:0711.2420 [astro-ph]} \BibitemShut {NoStop}%
\bibitem [{\citenamefont {Abadi}\ \emph {et~al.}(2015)\citenamefont {Abadi}, \citenamefont {Agarwal}, \citenamefont {Barham}, \citenamefont {Brevdo}, \citenamefont {Chen}, \citenamefont {Citro}, \citenamefont {Corrado}, \citenamefont {Davis}, \citenamefont {Dean}, \citenamefont {Devin}, \citenamefont {Ghemawat}, \citenamefont {Goodfellow}, \citenamefont {Harp}, \citenamefont {Irving}, \citenamefont {Isard}, \citenamefont {Jia}, \citenamefont {Jozefowicz}, \citenamefont {Kaiser}, \citenamefont {Kudlur}, \citenamefont {Levenberg}, \citenamefont {Man\'{e}}, \citenamefont {Monga}, \citenamefont {Moore}, \citenamefont {Murray}, \citenamefont {Olah}, \citenamefont {Schuster}, \citenamefont {Shlens}, \citenamefont {Steiner}, \citenamefont {Sutskever}, \citenamefont {Talwar}, \citenamefont {Tucker}, \citenamefont {Vanhoucke}, \citenamefont {Vasudevan}, \citenamefont {Vi\'{e}gas}, \citenamefont {Vinyals}, \citenamefont {Warden}, \citenamefont {Wattenberg}, \citenamefont {Wicke}, \citenamefont {Yu},\ and\ \citenamefont
  {Zheng}}]{tensorflow2015-whitepaper}%
  \BibitemOpen
  \bibfield  {author} {\bibinfo {author} {\bibfnamefont {M.}~\bibnamefont {Abadi}}, \bibinfo {author} {\bibfnamefont {A.}~\bibnamefont {Agarwal}}, \bibinfo {author} {\bibfnamefont {P.}~\bibnamefont {Barham}}, \bibinfo {author} {\bibfnamefont {E.}~\bibnamefont {Brevdo}}, \bibinfo {author} {\bibfnamefont {Z.}~\bibnamefont {Chen}}, \bibinfo {author} {\bibfnamefont {C.}~\bibnamefont {Citro}}, \bibinfo {author} {\bibfnamefont {G.~S.}\ \bibnamefont {Corrado}}, \bibinfo {author} {\bibfnamefont {A.}~\bibnamefont {Davis}}, \bibinfo {author} {\bibfnamefont {J.}~\bibnamefont {Dean}}, \bibinfo {author} {\bibfnamefont {M.}~\bibnamefont {Devin}}, \bibinfo {author} {\bibfnamefont {S.}~\bibnamefont {Ghemawat}}, \bibinfo {author} {\bibfnamefont {I.}~\bibnamefont {Goodfellow}}, \bibinfo {author} {\bibfnamefont {A.}~\bibnamefont {Harp}}, \bibinfo {author} {\bibfnamefont {G.}~\bibnamefont {Irving}}, \bibinfo {author} {\bibfnamefont {M.}~\bibnamefont {Isard}}, \bibinfo {author} {\bibfnamefont {Y.}~\bibnamefont {Jia}}, \bibinfo
  {author} {\bibfnamefont {R.}~\bibnamefont {Jozefowicz}}, \bibinfo {author} {\bibfnamefont {L.}~\bibnamefont {Kaiser}}, \bibinfo {author} {\bibfnamefont {M.}~\bibnamefont {Kudlur}}, \bibinfo {author} {\bibfnamefont {J.}~\bibnamefont {Levenberg}}, \bibinfo {author} {\bibfnamefont {D.}~\bibnamefont {Man\'{e}}}, \bibinfo {author} {\bibfnamefont {R.}~\bibnamefont {Monga}}, \bibinfo {author} {\bibfnamefont {S.}~\bibnamefont {Moore}}, \bibinfo {author} {\bibfnamefont {D.}~\bibnamefont {Murray}}, \bibinfo {author} {\bibfnamefont {C.}~\bibnamefont {Olah}}, \bibinfo {author} {\bibfnamefont {M.}~\bibnamefont {Schuster}}, \bibinfo {author} {\bibfnamefont {J.}~\bibnamefont {Shlens}}, \bibinfo {author} {\bibfnamefont {B.}~\bibnamefont {Steiner}}, \bibinfo {author} {\bibfnamefont {I.}~\bibnamefont {Sutskever}}, \bibinfo {author} {\bibfnamefont {K.}~\bibnamefont {Talwar}}, \bibinfo {author} {\bibfnamefont {P.}~\bibnamefont {Tucker}}, \bibinfo {author} {\bibfnamefont {V.}~\bibnamefont {Vanhoucke}}, \bibinfo {author}
  {\bibfnamefont {V.}~\bibnamefont {Vasudevan}}, \bibinfo {author} {\bibfnamefont {F.}~\bibnamefont {Vi\'{e}gas}}, \bibinfo {author} {\bibfnamefont {O.}~\bibnamefont {Vinyals}}, \bibinfo {author} {\bibfnamefont {P.}~\bibnamefont {Warden}}, \bibinfo {author} {\bibfnamefont {M.}~\bibnamefont {Wattenberg}}, \bibinfo {author} {\bibfnamefont {M.}~\bibnamefont {Wicke}}, \bibinfo {author} {\bibfnamefont {Y.}~\bibnamefont {Yu}},\ and\ \bibinfo {author} {\bibfnamefont {X.}~\bibnamefont {Zheng}},\ }\href {https://www.tensorflow.org/} {\bibinfo {title} {{TensorFlow}: Large-scale machine learning on heterogeneous systems}} (\bibinfo {year} {2015}),\ \bibinfo {note} {software available from tensorflow.org}\BibitemShut {NoStop}%
\bibitem [{\citenamefont {{Smith}}\ \emph {et~al.}(2021)\citenamefont {{Smith}}, \citenamefont {{Borhanian}}, \citenamefont {{Sathyaprakash}}, \citenamefont {{Hernandez Vivanco}}, \citenamefont {{Field}}, \citenamefont {{Lasky}}, \citenamefont {{Mandel}}, \citenamefont {{Morisaki}}, \citenamefont {{Ottaway}}, \citenamefont {{Slagmolen}}, \citenamefont {{Thrane}}, \citenamefont {{T{\"o}yr{\"a}}},\ and\ \citenamefont {{Vitale}}}]{Smith+2021}%
  \BibitemOpen
  \bibfield  {author} {\bibinfo {author} {\bibfnamefont {R.}~\bibnamefont {{Smith}}}, \bibinfo {author} {\bibfnamefont {S.}~\bibnamefont {{Borhanian}}}, \bibinfo {author} {\bibfnamefont {B.}~\bibnamefont {{Sathyaprakash}}}, \bibinfo {author} {\bibfnamefont {F.}~\bibnamefont {{Hernandez Vivanco}}}, \bibinfo {author} {\bibfnamefont {S.~E.}\ \bibnamefont {{Field}}}, \bibinfo {author} {\bibfnamefont {P.}~\bibnamefont {{Lasky}}}, \bibinfo {author} {\bibfnamefont {I.}~\bibnamefont {{Mandel}}}, \bibinfo {author} {\bibfnamefont {S.}~\bibnamefont {{Morisaki}}}, \bibinfo {author} {\bibfnamefont {D.}~\bibnamefont {{Ottaway}}}, \bibinfo {author} {\bibfnamefont {B.~J.~J.}\ \bibnamefont {{Slagmolen}}}, \bibinfo {author} {\bibfnamefont {E.}~\bibnamefont {{Thrane}}}, \bibinfo {author} {\bibfnamefont {D.}~\bibnamefont {{T{\"o}yr{\"a}}}},\ and\ \bibinfo {author} {\bibfnamefont {S.}~\bibnamefont {{Vitale}}},\ }\bibfield  {title} {\bibinfo {title} {{Bayesian Inference for Gravitational Waves from Binary Neutron Star Mergers in
  Third Generation Observatories}},\ }\href {https://doi.org/10.1103/PhysRevLett.127.081102} {\bibfield  {journal} {\bibinfo  {journal} {\prl}\ }\textbf {\bibinfo {volume} {127}},\ \bibinfo {eid} {081102} (\bibinfo {year} {2021})},\ \Eprint {https://arxiv.org/abs/2103.12274} {arXiv:2103.12274 [gr-qc]} \BibitemShut {NoStop}%
\bibitem [{\citenamefont {Smith}\ \emph {et~al.}(2020)\citenamefont {Smith}, \citenamefont {Ashton}, \citenamefont {Vajpeyi},\ and\ \citenamefont {Talbot}}]{pbilby_paper}%
  \BibitemOpen
  \bibfield  {author} {\bibinfo {author} {\bibfnamefont {R.~J.~E.}\ \bibnamefont {Smith}}, \bibinfo {author} {\bibfnamefont {G.}~\bibnamefont {Ashton}}, \bibinfo {author} {\bibfnamefont {A.}~\bibnamefont {Vajpeyi}},\ and\ \bibinfo {author} {\bibfnamefont {C.}~\bibnamefont {Talbot}},\ }\bibfield  {title} {\bibinfo {title} {{Massively parallel Bayesian inference for transient gravitational-wave astronomy}},\ }\href {https://doi.org/10.1093/mnras/staa2483} {\bibfield  {journal} {\bibinfo  {journal} {Mon. Not. Roy. Astron. Soc.}\ }\textbf {\bibinfo {volume} {498}},\ \bibinfo {pages} {4492} (\bibinfo {year} {2020})},\ \Eprint {https://arxiv.org/abs/1909.11873} {arXiv:1909.11873 [gr-qc]} \BibitemShut {NoStop}%
\bibitem [{\citenamefont {{Veitch}}\ \emph {et~al.}(2015)\citenamefont {{Veitch}}, \citenamefont {{Raymond}}, \citenamefont {{Farr}}, \citenamefont {{Farr}}, \citenamefont {{Graff}}, \citenamefont {{Vitale}}, \citenamefont {{Aylott}}, \citenamefont {{Blackburn}}, \citenamefont {{Christensen}}, \citenamefont {{Coughlin}}, \citenamefont {{Del Pozzo}}, \citenamefont {{Feroz}}, \citenamefont {{Gair}}, \citenamefont {{Haster}}, \citenamefont {{Kalogera}}, \citenamefont {{Littenberg}}, \citenamefont {{Mandel}}, \citenamefont {{O'Shaughnessy}}, \citenamefont {{Pitkin}}, \citenamefont {{Rodriguez}}, \citenamefont {{R{\"o}ver}}, \citenamefont {{Sidery}}, \citenamefont {{Smith}}, \citenamefont {{Van Der Sluys}}, \citenamefont {{Vecchio}}, \citenamefont {{Vousden}},\ and\ \citenamefont {{Wade}}}]{veitch2015}%
  \BibitemOpen
  \bibfield  {author} {\bibinfo {author} {\bibfnamefont {J.}~\bibnamefont {{Veitch}}}, \bibinfo {author} {\bibfnamefont {V.}~\bibnamefont {{Raymond}}}, \bibinfo {author} {\bibfnamefont {B.}~\bibnamefont {{Farr}}}, \bibinfo {author} {\bibfnamefont {W.}~\bibnamefont {{Farr}}}, \bibinfo {author} {\bibfnamefont {P.}~\bibnamefont {{Graff}}}, \bibinfo {author} {\bibfnamefont {S.}~\bibnamefont {{Vitale}}}, \bibinfo {author} {\bibfnamefont {B.}~\bibnamefont {{Aylott}}}, \bibinfo {author} {\bibfnamefont {K.}~\bibnamefont {{Blackburn}}}, \bibinfo {author} {\bibfnamefont {N.}~\bibnamefont {{Christensen}}}, \bibinfo {author} {\bibfnamefont {M.}~\bibnamefont {{Coughlin}}}, \bibinfo {author} {\bibfnamefont {W.}~\bibnamefont {{Del Pozzo}}}, \bibinfo {author} {\bibfnamefont {F.}~\bibnamefont {{Feroz}}}, \bibinfo {author} {\bibfnamefont {J.}~\bibnamefont {{Gair}}}, \bibinfo {author} {\bibfnamefont {C.~J.}\ \bibnamefont {{Haster}}}, \bibinfo {author} {\bibfnamefont {V.}~\bibnamefont {{Kalogera}}}, \bibinfo {author}
  {\bibfnamefont {T.}~\bibnamefont {{Littenberg}}}, \bibinfo {author} {\bibfnamefont {I.}~\bibnamefont {{Mandel}}}, \bibinfo {author} {\bibfnamefont {R.}~\bibnamefont {{O'Shaughnessy}}}, \bibinfo {author} {\bibfnamefont {M.}~\bibnamefont {{Pitkin}}}, \bibinfo {author} {\bibfnamefont {C.}~\bibnamefont {{Rodriguez}}}, \bibinfo {author} {\bibfnamefont {C.}~\bibnamefont {{R{\"o}ver}}}, \bibinfo {author} {\bibfnamefont {T.}~\bibnamefont {{Sidery}}}, \bibinfo {author} {\bibfnamefont {R.}~\bibnamefont {{Smith}}}, \bibinfo {author} {\bibfnamefont {M.}~\bibnamefont {{Van Der Sluys}}}, \bibinfo {author} {\bibfnamefont {A.}~\bibnamefont {{Vecchio}}}, \bibinfo {author} {\bibfnamefont {W.}~\bibnamefont {{Vousden}}},\ and\ \bibinfo {author} {\bibfnamefont {L.}~\bibnamefont {{Wade}}},\ }\bibfield  {title} {\bibinfo {title} {{Parameter estimation for compact binaries with ground-based gravitational-wave observations using the LALInference software library}},\ }\href {https://doi.org/10.1103/PhysRevD.91.042003} {\bibfield
  {journal} {\bibinfo  {journal} {\prd}\ }\textbf {\bibinfo {volume} {91}},\ \bibinfo {eid} {042003} (\bibinfo {year} {2015})},\ \Eprint {https://arxiv.org/abs/1409.7215} {arXiv:1409.7215 [gr-qc]} \BibitemShut {NoStop}%
\bibitem [{\citenamefont {{Dietrich}}\ \emph {et~al.}(2019)\citenamefont {{Dietrich}}, \citenamefont {{Samajdar}}, \citenamefont {{Khan}}, \citenamefont {{Johnson-McDaniel}}, \citenamefont {{Dudi}},\ and\ \citenamefont {{Tichy}}}]{Dietrich2019_NRtidal2}%
  \BibitemOpen
  \bibfield  {author} {\bibinfo {author} {\bibfnamefont {T.}~\bibnamefont {{Dietrich}}}, \bibinfo {author} {\bibfnamefont {A.}~\bibnamefont {{Samajdar}}}, \bibinfo {author} {\bibfnamefont {S.}~\bibnamefont {{Khan}}}, \bibinfo {author} {\bibfnamefont {N.~K.}\ \bibnamefont {{Johnson-McDaniel}}}, \bibinfo {author} {\bibfnamefont {R.}~\bibnamefont {{Dudi}}},\ and\ \bibinfo {author} {\bibfnamefont {W.}~\bibnamefont {{Tichy}}},\ }\bibfield  {title} {\bibinfo {title} {{Improving the NRTidal model for binary neutron star systems}},\ }\href {https://doi.org/10.1103/PhysRevD.100.044003} {\bibfield  {journal} {\bibinfo  {journal} {\prd}\ }\textbf {\bibinfo {volume} {100}},\ \bibinfo {eid} {044003} (\bibinfo {year} {2019})},\ \Eprint {https://arxiv.org/abs/1905.06011} {arXiv:1905.06011 [gr-qc]} \BibitemShut {NoStop}%
\bibitem [{\citenamefont {{Speagle}}(2020)}]{Dynesty}%
  \BibitemOpen
  \bibfield  {author} {\bibinfo {author} {\bibfnamefont {J.~S.}\ \bibnamefont {{Speagle}}},\ }\bibfield  {title} {\bibinfo {title} {{DYNESTY: a dynamic nested sampling package for estimating Bayesian posteriors and evidences}},\ }\href {https://doi.org/10.1093/mnras/staa278} {\bibfield  {journal} {\bibinfo  {journal} {\mnras}\ }\textbf {\bibinfo {volume} {493}},\ \bibinfo {pages} {3132} (\bibinfo {year} {2020})},\ \Eprint {https://arxiv.org/abs/1904.02180} {arXiv:1904.02180 [astro-ph.IM]} \BibitemShut {NoStop}%
\bibitem [{\citenamefont {{Broekgaarden}}\ \emph {et~al.}(2023)\citenamefont {{Broekgaarden}}, \citenamefont {{Banagiri}},\ and\ \citenamefont {{Payne}}}]{Floor_A+_Asharp_rates}%
  \BibitemOpen
  \bibfield  {author} {\bibinfo {author} {\bibfnamefont {F.~S.}\ \bibnamefont {{Broekgaarden}}}, \bibinfo {author} {\bibfnamefont {S.}~\bibnamefont {{Banagiri}}},\ and\ \bibinfo {author} {\bibfnamefont {E.}~\bibnamefont {{Payne}}},\ }\bibfield  {title} {\bibinfo {title} {{Visualizing the Number of Existing and Future Gravitational-Wave Detections from Merging Double Compact Objects}},\ }\href {https://doi.org/10.48550/arXiv.2303.17628} {\bibfield  {journal} {\bibinfo  {journal} {arXiv e-prints}\ ,\ \bibinfo {eid} {arXiv:2303.17628}} (\bibinfo {year} {2023})},\ \Eprint {https://arxiv.org/abs/2303.17628} {arXiv:2303.17628 [astro-ph.HE]} \BibitemShut {NoStop}%
\bibitem [{\citenamefont {{Alsing}}\ \emph {et~al.}(2018)\citenamefont {{Alsing}}, \citenamefont {{Silva}},\ and\ \citenamefont {{Berti}}}]{2018Alsing}%
  \BibitemOpen
  \bibfield  {author} {\bibinfo {author} {\bibfnamefont {J.}~\bibnamefont {{Alsing}}}, \bibinfo {author} {\bibfnamefont {H.~O.}\ \bibnamefont {{Silva}}},\ and\ \bibinfo {author} {\bibfnamefont {E.}~\bibnamefont {{Berti}}},\ }\bibfield  {title} {\bibinfo {title} {{Evidence for a maximum mass cut-off in the neutron star mass distribution and constraints on the equation of state}},\ }\href {https://doi.org/10.1093/mnras/sty1065} {\bibfield  {journal} {\bibinfo  {journal} {\mnras}\ }\textbf {\bibinfo {volume} {478}},\ \bibinfo {pages} {1377} (\bibinfo {year} {2018})},\ \Eprint {https://arxiv.org/abs/1709.07889} {arXiv:1709.07889 [astro-ph.HE]} \BibitemShut {NoStop}%
\bibitem [{\citenamefont {{Douchin}}\ and\ \citenamefont {{Haensel}}(2001{\natexlab{b}})}]{2001SLy}%
  \BibitemOpen
  \bibfield  {author} {\bibinfo {author} {\bibfnamefont {F.}~\bibnamefont {{Douchin}}}\ and\ \bibinfo {author} {\bibfnamefont {P.}~\bibnamefont {{Haensel}}},\ }\bibfield  {title} {\bibinfo {title} {{A unified equation of state of dense matter and neutron star structure}},\ }\href {https://doi.org/10.1051/0004-6361:20011402} {\bibfield  {journal} {\bibinfo  {journal} {\aap}\ }\textbf {\bibinfo {volume} {380}},\ \bibinfo {pages} {151} (\bibinfo {year} {2001}{\natexlab{b}})},\ \Eprint {https://arxiv.org/abs/astro-ph/0111092} {arXiv:astro-ph/0111092 [astro-ph]} \BibitemShut {NoStop}%
\bibitem [{\citenamefont {{Planck Collaboration}}\ \emph {et~al.}(2020)\citenamefont {{Planck Collaboration}}, \citenamefont {{Aghanim}}, \citenamefont {{Akrami}}, \citenamefont {{Ashdown}}, \citenamefont {{Aumont}}, \citenamefont {{Baccigalupi}}, \citenamefont {{Ballardini}}, \citenamefont {{Banday}}, \citenamefont {{Barreiro}}, \citenamefont {{Bartolo}}, \citenamefont {{Basak}}, \citenamefont {{Battye}}, \citenamefont {{Benabed}}, \citenamefont {{Bernard}}, \citenamefont {{Bersanelli}}, \citenamefont {{Bielewicz}}, \citenamefont {{Bock}}, \citenamefont {{Bond}}, \citenamefont {{Borrill}}, \citenamefont {{Bouchet}}, \citenamefont {{Boulanger}}, \citenamefont {{Bucher}}, \citenamefont {{Burigana}}, \citenamefont {{Butler}}, \citenamefont {{Calabrese}}, \citenamefont {{Cardoso}}, \citenamefont {{Carron}}, \citenamefont {{Challinor}}, \citenamefont {{Chiang}}, \citenamefont {{Chluba}}, \citenamefont {{Colombo}}, \citenamefont {{Combet}}, \citenamefont {{Contreras}}, \citenamefont {{Crill}}, \citenamefont
  {{Cuttaia}}, \citenamefont {{de Bernardis}}, \citenamefont {{de Zotti}}, \citenamefont {{Delabrouille}}, \citenamefont {{Delouis}}, \citenamefont {{Di Valentino}}, \citenamefont {{Diego}}, \citenamefont {{Dor{\'e}}}, \citenamefont {{Douspis}}, \citenamefont {{Ducout}}, \citenamefont {{Dupac}}, \citenamefont {{Dusini}}, \citenamefont {{Efstathiou}}, \citenamefont {{Elsner}}, \citenamefont {{En{\ss}lin}}, \citenamefont {{Eriksen}}, \citenamefont {{Fantaye}}, \citenamefont {{Farhang}}, \citenamefont {{Fergusson}}, \citenamefont {{Fernandez-Cobos}}, \citenamefont {{Finelli}}, \citenamefont {{Forastieri}}, \citenamefont {{Frailis}}, \citenamefont {{Fraisse}}, \citenamefont {{Franceschi}}, \citenamefont {{Frolov}}, \citenamefont {{Galeotta}}, \citenamefont {{Galli}}, \citenamefont {{Ganga}}, \citenamefont {{G{\'e}nova-Santos}}, \citenamefont {{Gerbino}}, \citenamefont {{Ghosh}}, \citenamefont {{Gonz{\'a}lez-Nuevo}}, \citenamefont {{G{\'o}rski}}, \citenamefont {{Gratton}}, \citenamefont {{Gruppuso}}, \citenamefont
  {{Gudmundsson}}, \citenamefont {{Hamann}}, \citenamefont {{Handley}}, \citenamefont {{Hansen}}, \citenamefont {{Herranz}}, \citenamefont {{Hildebrandt}}, \citenamefont {{Hivon}}, \citenamefont {{Huang}}, \citenamefont {{Jaffe}}, \citenamefont {{Jones}}, \citenamefont {{Karakci}}, \citenamefont {{Keih{\"a}nen}}, \citenamefont {{Keskitalo}}, \citenamefont {{Kiiveri}}, \citenamefont {{Kim}}, \citenamefont {{Kisner}}, \citenamefont {{Knox}}, \citenamefont {{Krachmalnicoff}}, \citenamefont {{Kunz}}, \citenamefont {{Kurki-Suonio}}, \citenamefont {{Lagache}}, \citenamefont {{Lamarre}}, \citenamefont {{Lasenby}}, \citenamefont {{Lattanzi}}, \citenamefont {{Lawrence}}, \citenamefont {{Le Jeune}}, \citenamefont {{Lemos}}, \citenamefont {{Lesgourgues}}, \citenamefont {{Levrier}}, \citenamefont {{Lewis}}, \citenamefont {{Liguori}}, \citenamefont {{Lilje}}, \citenamefont {{Lilley}}, \citenamefont {{Lindholm}}, \citenamefont {{L{\'o}pez-Caniego}}, \citenamefont {{Lubin}}, \citenamefont {{Ma}}, \citenamefont
  {{Mac{\'\i}as-P{\'e}rez}}, \citenamefont {{Maggio}}, \citenamefont {{Maino}}, \citenamefont {{Mandolesi}}, \citenamefont {{Mangilli}}, \citenamefont {{Marcos-Caballero}}, \citenamefont {{Maris}}, \citenamefont {{Martin}}, \citenamefont {{Martinelli}}, \citenamefont {{Mart{\'\i}nez-Gonz{\'a}lez}}, \citenamefont {{Matarrese}}, \citenamefont {{Mauri}}, \citenamefont {{McEwen}}, \citenamefont {{Meinhold}}, \citenamefont {{Melchiorri}}, \citenamefont {{Mennella}}, \citenamefont {{Migliaccio}}, \citenamefont {{Millea}}, \citenamefont {{Mitra}}, \citenamefont {{Miville-Desch{\^e}nes}}, \citenamefont {{Molinari}}, \citenamefont {{Montier}}, \citenamefont {{Morgante}}, \citenamefont {{Moss}}, \citenamefont {{Natoli}}, \citenamefont {{N{\o}rgaard-Nielsen}}, \citenamefont {{Pagano}}, \citenamefont {{Paoletti}}, \citenamefont {{Partridge}}, \citenamefont {{Patanchon}}, \citenamefont {{Peiris}}, \citenamefont {{Perrotta}}, \citenamefont {{Pettorino}}, \citenamefont {{Piacentini}}, \citenamefont {{Polastri}},
  \citenamefont {{Polenta}}, \citenamefont {{Puget}}, \citenamefont {{Rachen}}, \citenamefont {{Reinecke}}, \citenamefont {{Remazeilles}}, \citenamefont {{Renzi}}, \citenamefont {{Rocha}}, \citenamefont {{Rosset}}, \citenamefont {{Roudier}}, \citenamefont {{Rubi{\~n}o-Mart{\'\i}n}}, \citenamefont {{Ruiz-Granados}}, \citenamefont {{Salvati}}, \citenamefont {{Sandri}}, \citenamefont {{Savelainen}}, \citenamefont {{Scott}}, \citenamefont {{Shellard}}, \citenamefont {{Sirignano}}, \citenamefont {{Sirri}}, \citenamefont {{Spencer}}, \citenamefont {{Sunyaev}}, \citenamefont {{Suur-Uski}}, \citenamefont {{Tauber}}, \citenamefont {{Tavagnacco}}, \citenamefont {{Tenti}}, \citenamefont {{Toffolatti}}, \citenamefont {{Tomasi}}, \citenamefont {{Trombetti}}, \citenamefont {{Valenziano}}, \citenamefont {{Valiviita}}, \citenamefont {{Van Tent}}, \citenamefont {{Vibert}}, \citenamefont {{Vielva}}, \citenamefont {{Villa}}, \citenamefont {{Vittorio}}, \citenamefont {{Wandelt}}, \citenamefont {{Wehus}}, \citenamefont {{White}},
  \citenamefont {{White}}, \citenamefont {{Zacchei}},\ and\ \citenamefont {{Zonca}}}]{2020A&A...641A...6P}%
  \BibitemOpen
  \bibfield  {author} {\bibinfo {author} {\bibnamefont {{Planck Collaboration}}}, \bibinfo {author} {\bibfnamefont {N.}~\bibnamefont {{Aghanim}}}, \bibinfo {author} {\bibfnamefont {Y.}~\bibnamefont {{Akrami}}}, \bibinfo {author} {\bibfnamefont {M.}~\bibnamefont {{Ashdown}}}, \bibinfo {author} {\bibfnamefont {J.}~\bibnamefont {{Aumont}}}, \bibinfo {author} {\bibfnamefont {C.}~\bibnamefont {{Baccigalupi}}}, \bibinfo {author} {\bibfnamefont {M.}~\bibnamefont {{Ballardini}}}, \bibinfo {author} {\bibfnamefont {A.~J.}\ \bibnamefont {{Banday}}}, \bibinfo {author} {\bibfnamefont {R.~B.}\ \bibnamefont {{Barreiro}}}, \bibinfo {author} {\bibfnamefont {N.}~\bibnamefont {{Bartolo}}}, \bibinfo {author} {\bibfnamefont {S.}~\bibnamefont {{Basak}}}, \bibinfo {author} {\bibfnamefont {R.}~\bibnamefont {{Battye}}}, \bibinfo {author} {\bibfnamefont {K.}~\bibnamefont {{Benabed}}}, \bibinfo {author} {\bibfnamefont {J.~P.}\ \bibnamefont {{Bernard}}}, \bibinfo {author} {\bibfnamefont {M.}~\bibnamefont {{Bersanelli}}}, \bibinfo
  {author} {\bibfnamefont {P.}~\bibnamefont {{Bielewicz}}}, \bibinfo {author} {\bibfnamefont {J.~J.}\ \bibnamefont {{Bock}}}, \bibinfo {author} {\bibfnamefont {J.~R.}\ \bibnamefont {{Bond}}}, \bibinfo {author} {\bibfnamefont {J.}~\bibnamefont {{Borrill}}}, \bibinfo {author} {\bibfnamefont {F.~R.}\ \bibnamefont {{Bouchet}}}, \bibinfo {author} {\bibfnamefont {F.}~\bibnamefont {{Boulanger}}}, \bibinfo {author} {\bibfnamefont {M.}~\bibnamefont {{Bucher}}}, \bibinfo {author} {\bibfnamefont {C.}~\bibnamefont {{Burigana}}}, \bibinfo {author} {\bibfnamefont {R.~C.}\ \bibnamefont {{Butler}}}, \bibinfo {author} {\bibfnamefont {E.}~\bibnamefont {{Calabrese}}}, \bibinfo {author} {\bibfnamefont {J.~F.}\ \bibnamefont {{Cardoso}}}, \bibinfo {author} {\bibfnamefont {J.}~\bibnamefont {{Carron}}}, \bibinfo {author} {\bibfnamefont {A.}~\bibnamefont {{Challinor}}}, \bibinfo {author} {\bibfnamefont {H.~C.}\ \bibnamefont {{Chiang}}}, \bibinfo {author} {\bibfnamefont {J.}~\bibnamefont {{Chluba}}}, \bibinfo {author} {\bibfnamefont
  {L.~P.~L.}\ \bibnamefont {{Colombo}}}, \bibinfo {author} {\bibfnamefont {C.}~\bibnamefont {{Combet}}}, \bibinfo {author} {\bibfnamefont {D.}~\bibnamefont {{Contreras}}}, \bibinfo {author} {\bibfnamefont {B.~P.}\ \bibnamefont {{Crill}}}, \bibinfo {author} {\bibfnamefont {F.}~\bibnamefont {{Cuttaia}}}, \bibinfo {author} {\bibfnamefont {P.}~\bibnamefont {{de Bernardis}}}, \bibinfo {author} {\bibfnamefont {G.}~\bibnamefont {{de Zotti}}}, \bibinfo {author} {\bibfnamefont {J.}~\bibnamefont {{Delabrouille}}}, \bibinfo {author} {\bibfnamefont {J.~M.}\ \bibnamefont {{Delouis}}}, \bibinfo {author} {\bibfnamefont {E.}~\bibnamefont {{Di Valentino}}}, \bibinfo {author} {\bibfnamefont {J.~M.}\ \bibnamefont {{Diego}}}, \bibinfo {author} {\bibfnamefont {O.}~\bibnamefont {{Dor{\'e}}}}, \bibinfo {author} {\bibfnamefont {M.}~\bibnamefont {{Douspis}}}, \bibinfo {author} {\bibfnamefont {A.}~\bibnamefont {{Ducout}}}, \bibinfo {author} {\bibfnamefont {X.}~\bibnamefont {{Dupac}}}, \bibinfo {author} {\bibfnamefont {S.}~\bibnamefont
  {{Dusini}}}, \bibinfo {author} {\bibfnamefont {G.}~\bibnamefont {{Efstathiou}}}, \bibinfo {author} {\bibfnamefont {F.}~\bibnamefont {{Elsner}}}, \bibinfo {author} {\bibfnamefont {T.~A.}\ \bibnamefont {{En{\ss}lin}}}, \bibinfo {author} {\bibfnamefont {H.~K.}\ \bibnamefont {{Eriksen}}}, \bibinfo {author} {\bibfnamefont {Y.}~\bibnamefont {{Fantaye}}}, \bibinfo {author} {\bibfnamefont {M.}~\bibnamefont {{Farhang}}}, \bibinfo {author} {\bibfnamefont {J.}~\bibnamefont {{Fergusson}}}, \bibinfo {author} {\bibfnamefont {R.}~\bibnamefont {{Fernandez-Cobos}}}, \bibinfo {author} {\bibfnamefont {F.}~\bibnamefont {{Finelli}}}, \bibinfo {author} {\bibfnamefont {F.}~\bibnamefont {{Forastieri}}}, \bibinfo {author} {\bibfnamefont {M.}~\bibnamefont {{Frailis}}}, \bibinfo {author} {\bibfnamefont {A.~A.}\ \bibnamefont {{Fraisse}}}, \bibinfo {author} {\bibfnamefont {E.}~\bibnamefont {{Franceschi}}}, \bibinfo {author} {\bibfnamefont {A.}~\bibnamefont {{Frolov}}}, \bibinfo {author} {\bibfnamefont {S.}~\bibnamefont {{Galeotta}}},
  \bibinfo {author} {\bibfnamefont {S.}~\bibnamefont {{Galli}}}, \bibinfo {author} {\bibfnamefont {K.}~\bibnamefont {{Ganga}}}, \bibinfo {author} {\bibfnamefont {R.~T.}\ \bibnamefont {{G{\'e}nova-Santos}}}, \bibinfo {author} {\bibfnamefont {M.}~\bibnamefont {{Gerbino}}}, \bibinfo {author} {\bibfnamefont {T.}~\bibnamefont {{Ghosh}}}, \bibinfo {author} {\bibfnamefont {J.}~\bibnamefont {{Gonz{\'a}lez-Nuevo}}}, \bibinfo {author} {\bibfnamefont {K.~M.}\ \bibnamefont {{G{\'o}rski}}}, \bibinfo {author} {\bibfnamefont {S.}~\bibnamefont {{Gratton}}}, \bibinfo {author} {\bibfnamefont {A.}~\bibnamefont {{Gruppuso}}}, \bibinfo {author} {\bibfnamefont {J.~E.}\ \bibnamefont {{Gudmundsson}}}, \bibinfo {author} {\bibfnamefont {J.}~\bibnamefont {{Hamann}}}, \bibinfo {author} {\bibfnamefont {W.}~\bibnamefont {{Handley}}}, \bibinfo {author} {\bibfnamefont {F.~K.}\ \bibnamefont {{Hansen}}}, \bibinfo {author} {\bibfnamefont {D.}~\bibnamefont {{Herranz}}}, \bibinfo {author} {\bibfnamefont {S.~R.}\ \bibnamefont {{Hildebrandt}}},
  \bibinfo {author} {\bibfnamefont {E.}~\bibnamefont {{Hivon}}}, \bibinfo {author} {\bibfnamefont {Z.}~\bibnamefont {{Huang}}}, \bibinfo {author} {\bibfnamefont {A.~H.}\ \bibnamefont {{Jaffe}}}, \bibinfo {author} {\bibfnamefont {W.~C.}\ \bibnamefont {{Jones}}}, \bibinfo {author} {\bibfnamefont {A.}~\bibnamefont {{Karakci}}}, \bibinfo {author} {\bibfnamefont {E.}~\bibnamefont {{Keih{\"a}nen}}}, \bibinfo {author} {\bibfnamefont {R.}~\bibnamefont {{Keskitalo}}}, \bibinfo {author} {\bibfnamefont {K.}~\bibnamefont {{Kiiveri}}}, \bibinfo {author} {\bibfnamefont {J.}~\bibnamefont {{Kim}}}, \bibinfo {author} {\bibfnamefont {T.~S.}\ \bibnamefont {{Kisner}}}, \bibinfo {author} {\bibfnamefont {L.}~\bibnamefont {{Knox}}}, \bibinfo {author} {\bibfnamefont {N.}~\bibnamefont {{Krachmalnicoff}}}, \bibinfo {author} {\bibfnamefont {M.}~\bibnamefont {{Kunz}}}, \bibinfo {author} {\bibfnamefont {H.}~\bibnamefont {{Kurki-Suonio}}}, \bibinfo {author} {\bibfnamefont {G.}~\bibnamefont {{Lagache}}}, \bibinfo {author} {\bibfnamefont
  {J.~M.}\ \bibnamefont {{Lamarre}}}, \bibinfo {author} {\bibfnamefont {A.}~\bibnamefont {{Lasenby}}}, \bibinfo {author} {\bibfnamefont {M.}~\bibnamefont {{Lattanzi}}}, \bibinfo {author} {\bibfnamefont {C.~R.}\ \bibnamefont {{Lawrence}}}, \bibinfo {author} {\bibfnamefont {M.}~\bibnamefont {{Le Jeune}}}, \bibinfo {author} {\bibfnamefont {P.}~\bibnamefont {{Lemos}}}, \bibinfo {author} {\bibfnamefont {J.}~\bibnamefont {{Lesgourgues}}}, \bibinfo {author} {\bibfnamefont {F.}~\bibnamefont {{Levrier}}}, \bibinfo {author} {\bibfnamefont {A.}~\bibnamefont {{Lewis}}}, \bibinfo {author} {\bibfnamefont {M.}~\bibnamefont {{Liguori}}}, \bibinfo {author} {\bibfnamefont {P.~B.}\ \bibnamefont {{Lilje}}}, \bibinfo {author} {\bibfnamefont {M.}~\bibnamefont {{Lilley}}}, \bibinfo {author} {\bibfnamefont {V.}~\bibnamefont {{Lindholm}}}, \bibinfo {author} {\bibfnamefont {M.}~\bibnamefont {{L{\'o}pez-Caniego}}}, \bibinfo {author} {\bibfnamefont {P.~M.}\ \bibnamefont {{Lubin}}}, \bibinfo {author} {\bibfnamefont {Y.~Z.}\ \bibnamefont
  {{Ma}}}, \bibinfo {author} {\bibfnamefont {J.~F.}\ \bibnamefont {{Mac{\'\i}as-P{\'e}rez}}}, \bibinfo {author} {\bibfnamefont {G.}~\bibnamefont {{Maggio}}}, \bibinfo {author} {\bibfnamefont {D.}~\bibnamefont {{Maino}}}, \bibinfo {author} {\bibfnamefont {N.}~\bibnamefont {{Mandolesi}}}, \bibinfo {author} {\bibfnamefont {A.}~\bibnamefont {{Mangilli}}}, \bibinfo {author} {\bibfnamefont {A.}~\bibnamefont {{Marcos-Caballero}}}, \bibinfo {author} {\bibfnamefont {M.}~\bibnamefont {{Maris}}}, \bibinfo {author} {\bibfnamefont {P.~G.}\ \bibnamefont {{Martin}}}, \bibinfo {author} {\bibfnamefont {M.}~\bibnamefont {{Martinelli}}}, \bibinfo {author} {\bibfnamefont {E.}~\bibnamefont {{Mart{\'\i}nez-Gonz{\'a}lez}}}, \bibinfo {author} {\bibfnamefont {S.}~\bibnamefont {{Matarrese}}}, \bibinfo {author} {\bibfnamefont {N.}~\bibnamefont {{Mauri}}}, \bibinfo {author} {\bibfnamefont {J.~D.}\ \bibnamefont {{McEwen}}}, \bibinfo {author} {\bibfnamefont {P.~R.}\ \bibnamefont {{Meinhold}}}, \bibinfo {author} {\bibfnamefont
  {A.}~\bibnamefont {{Melchiorri}}}, \bibinfo {author} {\bibfnamefont {A.}~\bibnamefont {{Mennella}}}, \bibinfo {author} {\bibfnamefont {M.}~\bibnamefont {{Migliaccio}}}, \bibinfo {author} {\bibfnamefont {M.}~\bibnamefont {{Millea}}}, \bibinfo {author} {\bibfnamefont {S.}~\bibnamefont {{Mitra}}}, \bibinfo {author} {\bibfnamefont {M.~A.}\ \bibnamefont {{Miville-Desch{\^e}nes}}}, \bibinfo {author} {\bibfnamefont {D.}~\bibnamefont {{Molinari}}}, \bibinfo {author} {\bibfnamefont {L.}~\bibnamefont {{Montier}}}, \bibinfo {author} {\bibfnamefont {G.}~\bibnamefont {{Morgante}}}, \bibinfo {author} {\bibfnamefont {A.}~\bibnamefont {{Moss}}}, \bibinfo {author} {\bibfnamefont {P.}~\bibnamefont {{Natoli}}}, \bibinfo {author} {\bibfnamefont {H.~U.}\ \bibnamefont {{N{\o}rgaard-Nielsen}}}, \bibinfo {author} {\bibfnamefont {L.}~\bibnamefont {{Pagano}}}, \bibinfo {author} {\bibfnamefont {D.}~\bibnamefont {{Paoletti}}}, \bibinfo {author} {\bibfnamefont {B.}~\bibnamefont {{Partridge}}}, \bibinfo {author} {\bibfnamefont
  {G.}~\bibnamefont {{Patanchon}}}, \bibinfo {author} {\bibfnamefont {H.~V.}\ \bibnamefont {{Peiris}}}, \bibinfo {author} {\bibfnamefont {F.}~\bibnamefont {{Perrotta}}}, \bibinfo {author} {\bibfnamefont {V.}~\bibnamefont {{Pettorino}}}, \bibinfo {author} {\bibfnamefont {F.}~\bibnamefont {{Piacentini}}}, \bibinfo {author} {\bibfnamefont {L.}~\bibnamefont {{Polastri}}}, \bibinfo {author} {\bibfnamefont {G.}~\bibnamefont {{Polenta}}}, \bibinfo {author} {\bibfnamefont {J.~L.}\ \bibnamefont {{Puget}}}, \bibinfo {author} {\bibfnamefont {J.~P.}\ \bibnamefont {{Rachen}}}, \bibinfo {author} {\bibfnamefont {M.}~\bibnamefont {{Reinecke}}}, \bibinfo {author} {\bibfnamefont {M.}~\bibnamefont {{Remazeilles}}}, \bibinfo {author} {\bibfnamefont {A.}~\bibnamefont {{Renzi}}}, \bibinfo {author} {\bibfnamefont {G.}~\bibnamefont {{Rocha}}}, \bibinfo {author} {\bibfnamefont {C.}~\bibnamefont {{Rosset}}}, \bibinfo {author} {\bibfnamefont {G.}~\bibnamefont {{Roudier}}}, \bibinfo {author} {\bibfnamefont {J.~A.}\ \bibnamefont
  {{Rubi{\~n}o-Mart{\'\i}n}}}, \bibinfo {author} {\bibfnamefont {B.}~\bibnamefont {{Ruiz-Granados}}}, \bibinfo {author} {\bibfnamefont {L.}~\bibnamefont {{Salvati}}}, \bibinfo {author} {\bibfnamefont {M.}~\bibnamefont {{Sandri}}}, \bibinfo {author} {\bibfnamefont {M.}~\bibnamefont {{Savelainen}}}, \bibinfo {author} {\bibfnamefont {D.}~\bibnamefont {{Scott}}}, \bibinfo {author} {\bibfnamefont {E.~P.~S.}\ \bibnamefont {{Shellard}}}, \bibinfo {author} {\bibfnamefont {C.}~\bibnamefont {{Sirignano}}}, \bibinfo {author} {\bibfnamefont {G.}~\bibnamefont {{Sirri}}}, \bibinfo {author} {\bibfnamefont {L.~D.}\ \bibnamefont {{Spencer}}}, \bibinfo {author} {\bibfnamefont {R.}~\bibnamefont {{Sunyaev}}}, \bibinfo {author} {\bibfnamefont {A.~S.}\ \bibnamefont {{Suur-Uski}}}, \bibinfo {author} {\bibfnamefont {J.~A.}\ \bibnamefont {{Tauber}}}, \bibinfo {author} {\bibfnamefont {D.}~\bibnamefont {{Tavagnacco}}}, \bibinfo {author} {\bibfnamefont {M.}~\bibnamefont {{Tenti}}}, \bibinfo {author} {\bibfnamefont {L.}~\bibnamefont
  {{Toffolatti}}}, \bibinfo {author} {\bibfnamefont {M.}~\bibnamefont {{Tomasi}}}, \bibinfo {author} {\bibfnamefont {T.}~\bibnamefont {{Trombetti}}}, \bibinfo {author} {\bibfnamefont {L.}~\bibnamefont {{Valenziano}}}, \bibinfo {author} {\bibfnamefont {J.}~\bibnamefont {{Valiviita}}}, \bibinfo {author} {\bibfnamefont {B.}~\bibnamefont {{Van Tent}}}, \bibinfo {author} {\bibfnamefont {L.}~\bibnamefont {{Vibert}}}, \bibinfo {author} {\bibfnamefont {P.}~\bibnamefont {{Vielva}}}, \bibinfo {author} {\bibfnamefont {F.}~\bibnamefont {{Villa}}}, \bibinfo {author} {\bibfnamefont {N.}~\bibnamefont {{Vittorio}}}, \bibinfo {author} {\bibfnamefont {B.~D.}\ \bibnamefont {{Wandelt}}}, \bibinfo {author} {\bibfnamefont {I.~K.}\ \bibnamefont {{Wehus}}}, \bibinfo {author} {\bibfnamefont {M.}~\bibnamefont {{White}}}, \bibinfo {author} {\bibfnamefont {S.~D.~M.}\ \bibnamefont {{White}}}, \bibinfo {author} {\bibfnamefont {A.}~\bibnamefont {{Zacchei}}},\ and\ \bibinfo {author} {\bibfnamefont {A.}~\bibnamefont {{Zonca}}},\ }\bibfield
  {title} {\bibinfo {title} {{Planck 2018 results. VI. Cosmological parameters}},\ }\href {https://doi.org/10.1051/0004-6361/201833910} {\bibfield  {journal} {\bibinfo  {journal} {\aap}\ }\textbf {\bibinfo {volume} {641}},\ \bibinfo {eid} {A6} (\bibinfo {year} {2020})},\ \Eprint {https://arxiv.org/abs/1807.06209} {arXiv:1807.06209 [astro-ph.CO]} \BibitemShut {NoStop}%
\bibitem [{\citenamefont {{Morisaki}}\ \emph {et~al.}(2023)\citenamefont {{Morisaki}}, \citenamefont {{Smith}}, \citenamefont {{Tsukada}}, \citenamefont {{Sachdev}}, \citenamefont {{Stevenson}}, \citenamefont {{Talbot}},\ and\ \citenamefont {{Zimmerman}}}]{2023MorisakiROQ}%
  \BibitemOpen
  \bibfield  {author} {\bibinfo {author} {\bibfnamefont {S.}~\bibnamefont {{Morisaki}}}, \bibinfo {author} {\bibfnamefont {R.}~\bibnamefont {{Smith}}}, \bibinfo {author} {\bibfnamefont {L.}~\bibnamefont {{Tsukada}}}, \bibinfo {author} {\bibfnamefont {S.}~\bibnamefont {{Sachdev}}}, \bibinfo {author} {\bibfnamefont {S.}~\bibnamefont {{Stevenson}}}, \bibinfo {author} {\bibfnamefont {C.}~\bibnamefont {{Talbot}}},\ and\ \bibinfo {author} {\bibfnamefont {A.}~\bibnamefont {{Zimmerman}}},\ }\bibfield  {title} {\bibinfo {title} {{Rapid localization and inference on compact binary coalescences with the Advanced LIGO-Virgo-KAGRA gravitational-wave detector network}},\ }\href {https://doi.org/10.1103/PhysRevD.108.123040} {\bibfield  {journal} {\bibinfo  {journal} {\prd}\ }\textbf {\bibinfo {volume} {108}},\ \bibinfo {eid} {123040} (\bibinfo {year} {2023})},\ \Eprint {https://arxiv.org/abs/2307.13380} {arXiv:2307.13380 [gr-qc]} \BibitemShut {NoStop}%
\bibitem [{\citenamefont {{Gair}}\ \emph {et~al.}(2023)\citenamefont {{Gair}}, \citenamefont {{Ghosh}}, \citenamefont {{Gray}}, \citenamefont {{Holz}}, \citenamefont {{Mastrogiovanni}}, \citenamefont {{Mukherjee}}, \citenamefont {{Palmese}}, \citenamefont {{Tamanini}}, \citenamefont {{Baker}}, \citenamefont {{Beirnaert}}, \citenamefont {{Bilicki}}, \citenamefont {{Chen}}, \citenamefont {{D{\'a}lya}}, \citenamefont {{Ezquiaga}}, \citenamefont {{Farr}}, \citenamefont {{Fishbach}}, \citenamefont {{Garcia-Bellido}}, \citenamefont {{Ghosh}}, \citenamefont {{Huang}}, \citenamefont {{Karathanasis}}, \citenamefont {{Leyde}}, \citenamefont {{Maga{\~n}a Hernandez}}, \citenamefont {{Noller}}, \citenamefont {{Pierra}}, \citenamefont {{Raffai}}, \citenamefont {{Romano}}, \citenamefont {{Seglar-Arroyo}}, \citenamefont {{Steer}}, \citenamefont {{Turski}}, \citenamefont {{Vaccaro}},\ and\ \citenamefont {{Vallejo-Pe{\~n}a}}}]{2023Galaxy_Catalogue}%
  \BibitemOpen
  \bibfield  {author} {\bibinfo {author} {\bibfnamefont {J.~R.}\ \bibnamefont {{Gair}}}, \bibinfo {author} {\bibfnamefont {A.}~\bibnamefont {{Ghosh}}}, \bibinfo {author} {\bibfnamefont {R.}~\bibnamefont {{Gray}}}, \bibinfo {author} {\bibfnamefont {D.~E.}\ \bibnamefont {{Holz}}}, \bibinfo {author} {\bibfnamefont {S.}~\bibnamefont {{Mastrogiovanni}}}, \bibinfo {author} {\bibfnamefont {S.}~\bibnamefont {{Mukherjee}}}, \bibinfo {author} {\bibfnamefont {A.}~\bibnamefont {{Palmese}}}, \bibinfo {author} {\bibfnamefont {N.}~\bibnamefont {{Tamanini}}}, \bibinfo {author} {\bibfnamefont {T.}~\bibnamefont {{Baker}}}, \bibinfo {author} {\bibfnamefont {F.}~\bibnamefont {{Beirnaert}}}, \bibinfo {author} {\bibfnamefont {M.}~\bibnamefont {{Bilicki}}}, \bibinfo {author} {\bibfnamefont {H.-Y.}\ \bibnamefont {{Chen}}}, \bibinfo {author} {\bibfnamefont {G.}~\bibnamefont {{D{\'a}lya}}}, \bibinfo {author} {\bibfnamefont {J.~M.}\ \bibnamefont {{Ezquiaga}}}, \bibinfo {author} {\bibfnamefont {W.~M.}\ \bibnamefont {{Farr}}}, \bibinfo
  {author} {\bibfnamefont {M.}~\bibnamefont {{Fishbach}}}, \bibinfo {author} {\bibfnamefont {J.}~\bibnamefont {{Garcia-Bellido}}}, \bibinfo {author} {\bibfnamefont {T.}~\bibnamefont {{Ghosh}}}, \bibinfo {author} {\bibfnamefont {H.-Y.}\ \bibnamefont {{Huang}}}, \bibinfo {author} {\bibfnamefont {C.}~\bibnamefont {{Karathanasis}}}, \bibinfo {author} {\bibfnamefont {K.}~\bibnamefont {{Leyde}}}, \bibinfo {author} {\bibfnamefont {I.}~\bibnamefont {{Maga{\~n}a Hernandez}}}, \bibinfo {author} {\bibfnamefont {J.}~\bibnamefont {{Noller}}}, \bibinfo {author} {\bibfnamefont {G.}~\bibnamefont {{Pierra}}}, \bibinfo {author} {\bibfnamefont {P.}~\bibnamefont {{Raffai}}}, \bibinfo {author} {\bibfnamefont {A.~E.}\ \bibnamefont {{Romano}}}, \bibinfo {author} {\bibfnamefont {M.}~\bibnamefont {{Seglar-Arroyo}}}, \bibinfo {author} {\bibfnamefont {D.~A.}\ \bibnamefont {{Steer}}}, \bibinfo {author} {\bibfnamefont {C.}~\bibnamefont {{Turski}}}, \bibinfo {author} {\bibfnamefont {M.~P.}\ \bibnamefont {{Vaccaro}}},\ and\ \bibinfo
  {author} {\bibfnamefont {S.~A.}\ \bibnamefont {{Vallejo-Pe{\~n}a}}},\ }\bibfield  {title} {\bibinfo {title} {{The Hitchhiker's Guide to the Galaxy Catalog Approach for Dark Siren Gravitational-wave Cosmology}},\ }\href {https://doi.org/10.3847/1538-3881/acca78} {\bibfield  {journal} {\bibinfo  {journal} {\aj}\ }\textbf {\bibinfo {volume} {166}},\ \bibinfo {eid} {22} (\bibinfo {year} {2023})},\ \Eprint {https://arxiv.org/abs/2212.08694} {arXiv:2212.08694 [gr-qc]} \BibitemShut {NoStop}%
\bibitem [{\citenamefont {{Pang}}\ \emph {et~al.}(2023)\citenamefont {{Pang}}, \citenamefont {{Dietrich}}, \citenamefont {{Coughlin}}, \citenamefont {{Bulla}}, \citenamefont {{Tews}}, \citenamefont {{Almualla}}, \citenamefont {{Barna}}, \citenamefont {{Kiendrebeogo}}, \citenamefont {{Kunert}}, \citenamefont {{Mansingh}}, \citenamefont {{Reed}}, \citenamefont {{Sravan}}, \citenamefont {{Toivonen}}, \citenamefont {{Antier}}, \citenamefont {{VandenBerg}}, \citenamefont {{Heinzel}}, \citenamefont {{Nedora}}, \citenamefont {{Salehi}}, \citenamefont {{Sharma}}, \citenamefont {{Somasundaram}},\ and\ \citenamefont {{Van Den Broeck}}}]{2023Pang}%
  \BibitemOpen
  \bibfield  {author} {\bibinfo {author} {\bibfnamefont {P.~T.~H.}\ \bibnamefont {{Pang}}}, \bibinfo {author} {\bibfnamefont {T.}~\bibnamefont {{Dietrich}}}, \bibinfo {author} {\bibfnamefont {M.~W.}\ \bibnamefont {{Coughlin}}}, \bibinfo {author} {\bibfnamefont {M.}~\bibnamefont {{Bulla}}}, \bibinfo {author} {\bibfnamefont {I.}~\bibnamefont {{Tews}}}, \bibinfo {author} {\bibfnamefont {M.}~\bibnamefont {{Almualla}}}, \bibinfo {author} {\bibfnamefont {T.}~\bibnamefont {{Barna}}}, \bibinfo {author} {\bibfnamefont {R.~W.}\ \bibnamefont {{Kiendrebeogo}}}, \bibinfo {author} {\bibfnamefont {N.}~\bibnamefont {{Kunert}}}, \bibinfo {author} {\bibfnamefont {G.}~\bibnamefont {{Mansingh}}}, \bibinfo {author} {\bibfnamefont {B.}~\bibnamefont {{Reed}}}, \bibinfo {author} {\bibfnamefont {N.}~\bibnamefont {{Sravan}}}, \bibinfo {author} {\bibfnamefont {A.}~\bibnamefont {{Toivonen}}}, \bibinfo {author} {\bibfnamefont {S.}~\bibnamefont {{Antier}}}, \bibinfo {author} {\bibfnamefont {R.~O.}\ \bibnamefont {{VandenBerg}}}, \bibinfo
  {author} {\bibfnamefont {J.}~\bibnamefont {{Heinzel}}}, \bibinfo {author} {\bibfnamefont {V.}~\bibnamefont {{Nedora}}}, \bibinfo {author} {\bibfnamefont {P.}~\bibnamefont {{Salehi}}}, \bibinfo {author} {\bibfnamefont {R.}~\bibnamefont {{Sharma}}}, \bibinfo {author} {\bibfnamefont {R.}~\bibnamefont {{Somasundaram}}},\ and\ \bibinfo {author} {\bibfnamefont {C.}~\bibnamefont {{Van Den Broeck}}},\ }\bibfield  {title} {\bibinfo {title} {{An updated nuclear-physics and multi-messenger astrophysics framework for binary neutron star mergers}},\ }\href {https://doi.org/10.1038/s41467-023-43932-6} {\bibfield  {journal} {\bibinfo  {journal} {Nature Communications}\ }\textbf {\bibinfo {volume} {14}},\ \bibinfo {eid} {8352} (\bibinfo {year} {2023})},\ \Eprint {https://arxiv.org/abs/2205.08513} {arXiv:2205.08513 [astro-ph.HE]} \BibitemShut {NoStop}%
\bibitem [{\citenamefont {{Riley}}\ \emph {et~al.}(2019)\citenamefont {{Riley}}, \citenamefont {{Watts}}, \citenamefont {{Bogdanov}}, \citenamefont {{Ray}}, \citenamefont {{Ludlam}}, \citenamefont {{Guillot}}, \citenamefont {{Arzoumanian}}, \citenamefont {{Baker}}, \citenamefont {{Bilous}}, \citenamefont {{Chakrabarty}}, \citenamefont {{Gendreau}}, \citenamefont {{Harding}}, \citenamefont {{Ho}}, \citenamefont {{Lattimer}}, \citenamefont {{Morsink}},\ and\ \citenamefont {{Strohmayer}}}]{2019RileyNICERView}%
  \BibitemOpen
  \bibfield  {author} {\bibinfo {author} {\bibfnamefont {T.~E.}\ \bibnamefont {{Riley}}}, \bibinfo {author} {\bibfnamefont {A.~L.}\ \bibnamefont {{Watts}}}, \bibinfo {author} {\bibfnamefont {S.}~\bibnamefont {{Bogdanov}}}, \bibinfo {author} {\bibfnamefont {P.~S.}\ \bibnamefont {{Ray}}}, \bibinfo {author} {\bibfnamefont {R.~M.}\ \bibnamefont {{Ludlam}}}, \bibinfo {author} {\bibfnamefont {S.}~\bibnamefont {{Guillot}}}, \bibinfo {author} {\bibfnamefont {Z.}~\bibnamefont {{Arzoumanian}}}, \bibinfo {author} {\bibfnamefont {C.~L.}\ \bibnamefont {{Baker}}}, \bibinfo {author} {\bibfnamefont {A.~V.}\ \bibnamefont {{Bilous}}}, \bibinfo {author} {\bibfnamefont {D.}~\bibnamefont {{Chakrabarty}}}, \bibinfo {author} {\bibfnamefont {K.~C.}\ \bibnamefont {{Gendreau}}}, \bibinfo {author} {\bibfnamefont {A.~K.}\ \bibnamefont {{Harding}}}, \bibinfo {author} {\bibfnamefont {W.~C.~G.}\ \bibnamefont {{Ho}}}, \bibinfo {author} {\bibfnamefont {J.~M.}\ \bibnamefont {{Lattimer}}}, \bibinfo {author} {\bibfnamefont {S.~M.}\ \bibnamefont
  {{Morsink}}},\ and\ \bibinfo {author} {\bibfnamefont {T.~E.}\ \bibnamefont {{Strohmayer}}},\ }\bibfield  {title} {\bibinfo {title} {{A NICER View of PSR J0030+0451: Millisecond Pulsar Parameter Estimation}},\ }\href {https://doi.org/10.3847/2041-8213/ab481c} {\bibfield  {journal} {\bibinfo  {journal} {\apjl}\ }\textbf {\bibinfo {volume} {887}},\ \bibinfo {eid} {L21} (\bibinfo {year} {2019})},\ \Eprint {https://arxiv.org/abs/1912.05702} {arXiv:1912.05702 [astro-ph.HE]} \BibitemShut {NoStop}%
\bibitem [{\citenamefont {{Biscoveanu}}\ \emph {et~al.}(2023)\citenamefont {{Biscoveanu}}, \citenamefont {{Landry}},\ and\ \citenamefont {{Vitale}}}]{2023Biscoveanu}%
  \BibitemOpen
  \bibfield  {author} {\bibinfo {author} {\bibfnamefont {S.}~\bibnamefont {{Biscoveanu}}}, \bibinfo {author} {\bibfnamefont {P.}~\bibnamefont {{Landry}}},\ and\ \bibinfo {author} {\bibfnamefont {S.}~\bibnamefont {{Vitale}}},\ }\bibfield  {title} {\bibinfo {title} {{Population properties and multimessenger prospects of neutron star-black hole mergers following GWTC-3}},\ }\href {https://doi.org/10.1093/mnras/stac3052} {\bibfield  {journal} {\bibinfo  {journal} {\mnras}\ }\textbf {\bibinfo {volume} {518}},\ \bibinfo {pages} {5298} (\bibinfo {year} {2023})},\ \Eprint {https://arxiv.org/abs/2207.01568} {arXiv:2207.01568 [astro-ph.HE]} \BibitemShut {NoStop}%
\bibitem [{\citenamefont {{Clarke}}\ \emph {et~al.}(2023)\citenamefont {{Clarke}}, \citenamefont {{Chastain}}, \citenamefont {{Lasky}},\ and\ \citenamefont {{Thrane}}}]{2023Clarke}%
  \BibitemOpen
  \bibfield  {author} {\bibinfo {author} {\bibfnamefont {T.~A.}\ \bibnamefont {{Clarke}}}, \bibinfo {author} {\bibfnamefont {L.}~\bibnamefont {{Chastain}}}, \bibinfo {author} {\bibfnamefont {P.~D.}\ \bibnamefont {{Lasky}}},\ and\ \bibinfo {author} {\bibfnamefont {E.}~\bibnamefont {{Thrane}}},\ }\bibfield  {title} {\bibinfo {title} {{Nuclear Physics with Gravitational Waves from Neutron Stars Disrupted by Black Holes}},\ }\href {https://doi.org/10.3847/2041-8213/acd33b} {\bibfield  {journal} {\bibinfo  {journal} {\apjl}\ }\textbf {\bibinfo {volume} {949}},\ \bibinfo {eid} {L6} (\bibinfo {year} {2023})},\ \Eprint {https://arxiv.org/abs/2302.09711} {arXiv:2302.09711 [gr-qc]} \BibitemShut {NoStop}%
\bibitem [{\citenamefont {{Landry}}\ \emph {et~al.}(2020)\citenamefont {{Landry}}, \citenamefont {{Essick}},\ and\ \citenamefont {{Chatziioannou}}}]{2020Landry}%
  \BibitemOpen
  \bibfield  {author} {\bibinfo {author} {\bibfnamefont {P.}~\bibnamefont {{Landry}}}, \bibinfo {author} {\bibfnamefont {R.}~\bibnamefont {{Essick}}},\ and\ \bibinfo {author} {\bibfnamefont {K.}~\bibnamefont {{Chatziioannou}}},\ }\bibfield  {title} {\bibinfo {title} {{Nonparametric constraints on neutron star matter with existing and upcoming gravitational wave and pulsar observations}},\ }\href {https://doi.org/10.1103/PhysRevD.101.123007} {\bibfield  {journal} {\bibinfo  {journal} {\prd}\ }\textbf {\bibinfo {volume} {101}},\ \bibinfo {eid} {123007} (\bibinfo {year} {2020})},\ \Eprint {https://arxiv.org/abs/2003.04880} {arXiv:2003.04880 [astro-ph.HE]} \BibitemShut {NoStop}%
\bibitem [{\citenamefont {Fukushima}(1969)}]{Fukushima1969VisualFE}%
  \BibitemOpen
  \bibfield  {author} {\bibinfo {author} {\bibfnamefont {K.}~\bibnamefont {Fukushima}},\ }\bibfield  {title} {\bibinfo {title} {Visual feature extraction by a multilayered network of analog threshold elements},\ }\href {https://api.semanticscholar.org/CorpusID:206799280} {\bibfield  {journal} {\bibinfo  {journal} {IEEE Trans. Syst. Sci. Cybern.}\ }\textbf {\bibinfo {volume} {5}},\ \bibinfo {pages} {322} (\bibinfo {year} {1969})}\BibitemShut {NoStop}%
\bibitem [{\citenamefont {{Kingma}}\ and\ \citenamefont {{Ba}}(2014)}]{ADAM}%
  \BibitemOpen
  \bibfield  {author} {\bibinfo {author} {\bibfnamefont {D.~P.}\ \bibnamefont {{Kingma}}}\ and\ \bibinfo {author} {\bibfnamefont {J.}~\bibnamefont {{Ba}}},\ }\bibfield  {title} {\bibinfo {title} {{Adam: A Method for Stochastic Optimization}},\ }\href {https://doi.org/10.48550/arXiv.1412.6980} {\bibfield  {journal} {\bibinfo  {journal} {arXiv e-prints}\ ,\ \bibinfo {eid} {arXiv:1412.6980}} (\bibinfo {year} {2014})},\ \Eprint {https://arxiv.org/abs/1412.6980} {arXiv:1412.6980 [cs.LG]} \BibitemShut {NoStop}%
\bibitem [{\citenamefont {{Lattimer}}\ and\ \citenamefont {{Prakash}}(2001)}]{2001Lattimer_Prakash}%
  \BibitemOpen
  \bibfield  {author} {\bibinfo {author} {\bibfnamefont {J.~M.}\ \bibnamefont {{Lattimer}}}\ and\ \bibinfo {author} {\bibfnamefont {M.}~\bibnamefont {{Prakash}}},\ }\bibfield  {title} {\bibinfo {title} {{Neutron Star Structure and the Equation of State}},\ }\href {https://doi.org/10.1086/319702} {\bibfield  {journal} {\bibinfo  {journal} {\apj}\ }\textbf {\bibinfo {volume} {550}},\ \bibinfo {pages} {426} (\bibinfo {year} {2001})},\ \Eprint {https://arxiv.org/abs/astro-ph/0002232} {arXiv:astro-ph/0002232 [astro-ph]} \BibitemShut {NoStop}%
\end{thebibliography}

%
\end{document}